\documentclass{article}
\usepackage[utf8]{inputenc}
\usepackage{url}
\usepackage{hyperref}
\usepackage{color,comment}

\usepackage{graphicx}
\usepackage{amsmath,amssymb}
\allowdisplaybreaks
\usepackage{float}
\usepackage{graphicx}
\usepackage{appendix}
\usepackage{listings}
\usepackage{subfigure}
\usepackage{cite}
\usepackage{bm}
\usepackage{makecell}
\usepackage{multirow}
\usepackage{algorithm}
\usepackage{algorithmicx}
\usepackage{algpseudocode}
\usepackage[T1]{fontenc}
\usepackage[utf8]{inputenc}
\usepackage{authblk}

\usepackage{multicol}
\newcommand\keywords[1]{\textbf{Keywords}:#1}

\newtheorem{lemma}{Lemma}[section]



\newcommand{\bn}{\bm n}

\newcommand{\bx}{\bm x}

\newcommand{\bX}{\bm X}

\newcommand{\bu}{\bm u}
\newcommand{\bU}{\bm U}
\newcommand{\bv}{\bm v}
\newcommand{\bA}{\bm A}

\newcommand{\bV}{\bm V}

\newcommand{\bS}{\bm \sigma}

\newcommand{\bP}{\bm P}

\newcommand{\bF}{\bm F}

\newcommand{\bpsi}{\bm \psi}

\newcommand{\bxi}{\bm \xi}

\title{A Projection-Based Time-Segmented Reduced Order Model for Fluid-Structure Interactions}

\author[a]{Qijia Zhai \thanks{zhaiqijia@stu.scu.edu.cn}}
\author[a]{Shiquan Zhang \thanks{shiquanzhang@scu.edu.cn}}
\author[b]{Pengtao Sun \thanks{Corresponding author: pengtao.sun@unlv.edu}}
\author[a]{Xiaoping Xie \thanks{xpxie@scu.edu.cn}}
\affil[a]{School of Mathematics, Sichuan University, Chengdu 610064, China}
\affil[b]{Department of Mathematical Sciences, University of Nevada Las Vegas, Las Vegas, Nevada 89154, USA}

\date{ }

\begin{document}

\maketitle

\begin{abstract}
In this paper, a type of novel projection-based, time-segmented
reduced order model (ROM) is proposed for dynamic fluid-structure
interaction (FSI) problems based upon the arbitrary
Lagrangian--Eulerian
(ALE)-finite element method (FEM) in a monolithic frame, where 
spatially, each variable is separated from others in terms of their
attribution (fluid/structure), category (velocity/pressure) and
component (horizontal/vertical) while
temporally, the proper orthogonal decomposition (POD) bases are
constructed in some deliberately partitioned time segments tailored
through extensive numerical trials. By the combination of spatial
and temporal decompositions, the developed ROM approach enables
prolonged simulations under prescribed accuracy thresholds.
Numerical experiments are carried out to compare numerical
performances of the proposed ROM with corresponding full-order model
(FOM) by solving a two-dimensional FSI benchmark problem that
involves a vibrating elastic beam in the fluid, where the
performance of offline ROM on perturbed physical parameters in the
online phase is investigated as well. Extensive numerical results
demonstrate that the proposed ROM has a comparable accuracy to while
much higher efficiency than the FOM. The developed ROM approach is
dimension-independent and can be seamlessly extended to solve high
dimensional FSI problems.
\end{abstract}

\section{Introduction}
Fluid-structure interaction (FSI) phenomena are encountered in many
engineering systems and biological processes, such as the vibration
of turbine blades impacted by the fluid flow, the response of
bridges and tall buildings to winds, the floating parachute wafted
by the air current, the rotating mechanical parts driven by the
pressurized liquid, the blood fluid through the cardiovascular
system, and etc. Simulating FSI problems enables analyzing relevant
engineering/biological system performance and durability under
complex operating conditions and with an accurate fashion by
monolithically considering the interactional effects between the
fluid and structure through subtle interface conditions (which have
two kinds: the dynamic one and the kinematic one). However,
high-fidelity FSI modeling presents immense computational challenges
due to the need for interfacing distinct fluid and structural
solvers. FSI simulations must couple solutions of the unsteady
Navier-Stokes equations in the fluid domain, which is usually
defined in Eulerian description, with solutions of potentially
nonlinear structural dynamics in the Lagrangian-based structural
domain. This requires exchanging the interface information at each
time step while accounting for the moving interface thus moving
subdomains and even moving meshes if a body-fitted mesh method is
adopted.

When addressing numerical methods for solving FSI problems, two
different categories always exist, as demonstrated below:
\begin{enumerate}
    \item The body-fitted/unfitted mesh method in terms of the mesh conformity between the fluidic and structural meshes.

The most representative body-fitted mesh method is the arbitrary
Lagrangian--Eulerian (ALE) method (see e.g.,
\cite{Hughes1981,Huerta1988,Nitikitpaiboon1993,Souli2010})
which adapts the fluid mesh to accommodate the
deformations/displacements of structure on the interface, thus
interface conditions are naturally satisfied therein. On the other
hand, the fictitious domain/immersed boundary method (FD/IBM) (see
e.g.,
\cite{Peskin1972,Peskin2002,LeVeque1994,ZLi2001})
represents the body-unfitted mesh method the most, where the fluid
is extended into the structural domain as the fictitious fluid thus
a fixed background Eulerian fluidic mesh fills up the entire domain,
while the foreground Lagrangian structural mesh changes with time
and thus does not conform with the fluidic mesh, simultaneously, the
kinematic interface condition is also reinforced throughout the
structural domain besides the interface, either strongly or weakly.
    \item The monolithic/partitioned approach in terms of the coupling strategy
    between the fluid and structural solver.

With the monolithic approach
\cite{WANG;Yan;ZHANG2012,Ryzhakov2010a,langer2016robust,Barker.A;Cai.X2009a},
the fluid and structural equations are solved simultaneously, where
the dynamic interface condition naturally vanishes in the
variational/weak form of FSI while the kinematic interface condition
is reinforced on the interface. It owns unconditional stability and
the immunity of any systematic error in the implementation of
interface conditions for moving interface (e.g., FSI) problems
without doing an alternating iteration by subdomains.
Whereas, the partitioned approach
\cite{grandmont_2001_numerical,causin2005added,tuprints254,Hou2012}
solves the fluid and structure equations, separately, by a
Dirichlet/Robin-Neumann alternating iteration between the fluid and
structural solver, where the kinematic interface condition is taken
as Dirichlet boundary condition of fluid equations and the dynamic
interface condition as Neumann boundary condition of structural
equations. It is conditionally stable and conditionally convergent
under a particular range of the physical parameters of FSI model.
For instance, it fails to converge when fluid and structural
densities are of the same magnitude due to the so-called added-mass
effect \cite{Idelsohn_Pin2009,causin2005added}.
\end{enumerate}

In regard to the above numerical methodologies, the full-order model
(FOM) approach that we adopt to numerically solve FSI problems in
this paper is the ALE-finite element method (ALE-FEM) within the
monolithic frame, as the author Sun et al. do in
\cite{Wang;Sun2016,HaoSun2021,SunLan2020,SunLan2019}. The FOM
approach may undergo a large-scale computation over a long term
evolvement, leading to a time-consuming FSI simulation with high
computational costs in practice.
As a contrast, the reduced-order model (ROM) approach
\cite{yang2016active,li2019numerical,kalashnikova_2013_a,ballarin_2016_podgalerkin,ballarin2017reduced20,nonino2019overcoming18},
which constructs low-dimensional surrogate models by projecting
governing equations onto a compact modal basis, can provide a
computationally inexpensive possibility to perform the same
computations as the FOM does but with minimum complexity while
keeping the essential features of the system intact. Thus the ROM
may radically accelerate detailed FSI simulations and save a great
deal of computational costs while remain a comparable accuracy to
the FOM.

Traditionally, utilizing the ROM to solve fluid dynamics is to apply
the Galerkin projection to Navier-Stokes equations within the space
spanned by the time-invariant, proper orthogonal decomposition (POD)
bases, which induces reduced ODE systems with respect to spatial
variables, and then the spatial integration is performed to obtain
space-invariant POD coefficients
\cite{alfioquarteroni_2014_reduced}. As for FSI problems, ROM is
studied with either the partitioned or monolithic approach. For
instance, in the partitioned approach,
two distinct ROMs are normally adopted to solve fluid and structural
equations in their respective domains first, and then are coupled
together to form a ROM for FSI \cite{beran_2004_reducedorder,
vierendeels_2007_implicit,
kalashnikova_2013_a, alfioquarteroni_2014_reduced}.
Additionally, the moving interface of fluid and structure can also be brought into the formation of ROM for FSI 
\cite{vierendeels_2007_implicit}. When the ALE-based body-fitted
mesh method is applied, the fluidic mesh moves along with structural
deformations, which sways over POD modes and thus induces a
multi-POD approach in \cite{anttonen_2003_podbased} by selecting
bases on grid shifts. The aforementioned Galerkin POD-ROM approach
is also proposed to account for modest deformation cases when
capturing the transonic flow's capricious nature
\cite{bourguet_2011_reducedorder} and when solving generic FSI
problems \cite{Nonino2022ProjectionBS}, which shows that in the
online phase of ROM,
the dimension of online FSI system can still be reduced further with
minor modifications like variable changes and careful selections of
interface coupling. ROM-ROM and ROM-FOM coupling strategies are also
actively developed for interface problems in the field of model
order reduction \cite{khacchihoang_2021_domaindecomposition,
decastro2022novel,bergmann_2018_a}, and the current research into
coupled ROM systems promises additional performance gains
\cite{astorino_2010_robin}, showing that the ROM is an extremely
interesting approach that could benefit many FSI-related
applications.


Because FSI problems are parameter-dependent problems, an extra
assumption, i.e., all operators in FSI problems hold an affine
parametric dependence, needs to be made in order to enable fast
online parametric queries. The empirical interpolation method can
recover this affine dependence, but with many parametrized
functions, which adds major offline costs. This tradeoff is
necessary to decouple expensive, one-time, parameter-independent
high-fidelity data structures from inexpensive, parameter-dependent
online queries. For instance, domains of varying shapes are
considered in \cite{rozza_2006_reduced} that are parametrized by
affine and non-affine maps related to a reference domain, where the
proposed method is well-suited for repeated and rapid evaluations
needed for parameter estimation, design, optimization, and real-time
control. In \cite{cho_2022_efficient}, a projection-based ROM using
POD and discrete empirical interpolation method, together with a
characteristic-based split scheme, is applied to the ALE-based
Navier-Stokes equations on dynamic grids.
A monolithic approach of ROM for parametrized FSI problems is put
forth in \cite{ballarin_2016_podgalerkin}, where a detailed
parametrized formulation of FSI and its components is provided to
demonstrate how an efficient offline-online computation by
approximating parametrized nonlinear tensors is achieved. In
addition, the monolithic POD-Galerkin method is also presented
therein to show how the fluid velocity, pressure and structural
displacement of FSI problems are efficiently computed during the
online phase.

In this paper, we intend to develop a monolithic ALE-FEM based,
novel POD-ROM to solve dynamic FSI problems undergoing a long-term
evolvement. It is well known that the traditional POD-ROM approach
may blow up for extended simulations after a long run, as revealed
by e.g., \cite{tello2014fluid} or the authors Zhai et al.
\cite{zhai2023new} in which
the comparison error between ROM and FOM for parabolic equations
under fixed POD bases increases with time, and the error can only be
controlled within $O(\tau^{-\frac{1}{2}})$ time steps, where $\tau$
is the time step size. This inspires us to first consider a
construction of unsteady POD bases that vary in each time segment,
where all time segments are divided from the entire time interval as
a coarse partition, then to conduct the online POD-ROM computation
following these time segments. Our numerical results in Section
\ref{sec:notimedivide} also show that if the time dimension is not
segmented and the POD bases are independent of time, then the total
relative $L^{2}$ error between ROM and FOM reaches up to the
magnitude of $10^{11}$, causing an error blowup for the ROM to FSI
simulation. To the best of our knowledge, so far there has not such
a time-segmented idea being proposed for ROM yet on solutions to
time-dependent problems including FSI.

Based upon the above insights, our commitments in this paper are as
follows:
\begin{itemize}
 \item In an innovative fashion, treat time as a non-reduced variable and design a new
ROM approach by dividing the entire time interval into some typical
time segments, then carrying out the classical POD method following
these time segments.
 \item Apply the developed ROM to a FSI benchmark problem involving a
vibrating elastic beam in the fluid, and suppress the increasing
error of the beam vibration's amplitude at the tail part after a
long-term simulation.
 \item Numerically demonstrate that the developed ROM not only remains a comparable
 accuracy to the FOM but also adapts to perturbed model parameters.
\end{itemize}


%
%

The structure of this paper is organized as follows. In
{\color{red}Section \ref{S2}}, we introduce the generic FSI model
and its fully discrete ALE-FEM. The ROM approach in both offline and
online phases are proposed in {\color{red}Section \ref{S3}.} We
conduct numerical experiments in {\color{red}Section \ref{S4}} to
validate the proposed ROM by comparing with FOM on solutions to a
FSI benchmark problem. Finally, the concluding remarks are given in
{\color{red}Section \ref{S5}}.

\section{Model description and finite element method for FSI}\label{S2}
\begin{figure}[H]
    \centering
    \includegraphics[width = 7cm, height = 3cm]{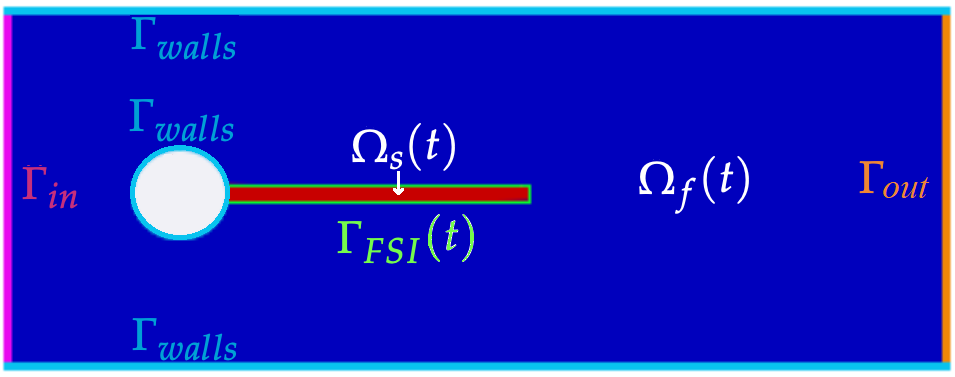}
    \caption{A schematic domain of FSI benchmark problem, where the fluid
    channel $\Omega_f(t)$ is in blue, the red-colored structural domain $\Omega_s(t)$
    represents a elastic beam behind a rigid cylinder.
    From the left to right, lines in purple, cyan, green and orange
    represent the inlet $\Gamma_{in}$, the fluidic channel wall $\Gamma_{walls}$,
    the fluid-structure interface $\Gamma_{FSI}(t)$, and the outlet
    $\Gamma_{out}$, respectively.}    \label{domainsetting}
\end{figure}

Let $\Omega \subset \mathbb{R}^d\ (d=2,3)$ be an open bounded domain
of interest with a polygonal boundary $\partial\Omega$, which is
divided by the interface $\Gamma_{FSI}(t)$ into two subdomains:
$\Omega=\Omega_f(t) \cup \Omega_s(t)$ and
$\Gamma_{FSI}(t)=\bar{\Omega}_f(t) \cap \bar{\Omega}_s(t)\ (t\in
[0,T])$, where $\Omega_f(t)$ is the fluid domain and $\Omega_s(t)$
the structural domain, and $\Omega_f(t) \cap
\Omega_s(t)=\varnothing$. Figure \ref{domainsetting} illustrates a
domain of a FSI benchmark problem \cite{turek2006proposal,
turek2010numerical}, where
$\partial\Omega=\Gamma_{in}\cup\Gamma_{walls}\cup\Gamma_{out}$,
$\Gamma_{in}$, $\Gamma_{walls}$ and $\Gamma_{out}$ represent the
inlet, walls and outlet of the fluidic channel, respectively.


\subsection{Model of dynamic FSI problems}
In the coupled multiphysics system of FSI, the fluid is assumed to
be Newtonian and incompressible, therefore, its behavior can be
modeled by incompressible Navier-Stokes equations in terms of the
fluid velocity $\boldsymbol{u}_f$ and fluid pressure $p_f$: find
$(\boldsymbol{u}_f,p_f)\in H^1(0,T;H^2(\Omega_f(t))^d)\times
L^2(0,T;H^1(\Omega_f(t)))$ such that
\begin{equation}\label{nseq}
    \left\{\begin{array}{rcll}
\rho_f\left(\frac{\partial \boldsymbol{u}_f}{\partial t}+\left(\boldsymbol{u}_f \cdot \nabla\right) \boldsymbol{u}_f\right)
-\operatorname{div} \boldsymbol{\sigma}_f\left(\boldsymbol{u}_f, p_f\right)&=&\boldsymbol{b}_f,& \text {in } \Omega_f(t) \times(0, T], \\
\operatorname{div} \boldsymbol{u}_f&=&0,& \text {in } \Omega_f(t)
\times(0, T],
\end{array}\right.
\end{equation}
subjecting to the following boundary conditions and initial
condition:
\begin{equation}\label{fluid-BICs}
\begin{array}{rclll}
\boldsymbol{u}_f&=&\boldsymbol{u}_{in}, & \text{on } \Gamma_{in},&(\text{incoming flow condition})\\
\boldsymbol{u}_f&=&\boldsymbol{0}, & \text{on } \Gamma_{walls},&(\text{no-slip condition})\\
\boldsymbol{\sigma}_f\boldsymbol{n}_f&=&\boldsymbol{0}, & \text{on } \Gamma_{out},&(\text{do-nothing condition})\\
\boldsymbol{u}_f(\boldsymbol{x},0)&=&\boldsymbol{u}_{f}^0, &
\text{in } \hat\Omega_f,&(\text{initial condition})
\end{array}
\end{equation}
where $\rho_f$ is the fluid density, $\boldsymbol{b}_f$ the fluid
volume external force, $\boldsymbol{n}_f$ the outward normal unit
vector on $\Gamma_{out}$, and
$\boldsymbol{\sigma}_f\left(\boldsymbol{u}_f, p_f\right)$ the fluid
stress tensor that is expressed as follows:
$$
\boldsymbol{\sigma}_f\left(\boldsymbol{u}_f,
p_f\right)=\rho_f\nu_f\left(\nabla
\boldsymbol{u}_f+(\nabla\boldsymbol{u}_f)^T\right)-p_f
\boldsymbol{I},
$$
here $\nu_f$ is the fluid kinematic viscosity, and $\boldsymbol{I}$
the identity matrix.

We note that the gradient operator ``$\nabla$'' and divergence
operator ``$\operatorname{div}$'' in (\ref{nseq}) and in what
follows refer to differentiations with respect to the time-dependent
spatial coordinates,
$\boldsymbol{x}_f=\hat{\boldsymbol{X}}_f(\hat{\boldsymbol{x}}_f,t)$,
where $\boldsymbol{x}_f$ denotes the current/Eulerian coordinates in
$\Omega_f(t)$ while $\hat{\boldsymbol{x}}_f$ denotes the
initial/reference/Lagrangian coordinates in
$\hat\Omega_i=\Omega_i(0)$, $i=f$ or $s$. In fact,
$\boldsymbol{x}_i=\hat{\boldsymbol{X}}_i(\hat{\boldsymbol{x}}_i,t)
=\hat{\boldsymbol{x}}_i+\hat{\boldsymbol{d}}_i$ defines a bijective
flow map, $\hat{\boldsymbol{X}}_i:\hat\Omega_i\mapsto\Omega_i(t)$,
where $\hat{\boldsymbol{d}}_i$ denotes the material displacement in
$\hat\Omega_i\ (i=f,s)$. Without further indication, we use ``
$\hat\cdot$ '' to denote an object `` $\cdot$ '' that is associated
with the initial/reference domain of $\Omega_i(t)$, $\hat\Omega_i\
(i=f,s)$, in the rest of this paper.

As for the structural material, we adopt the linear elasticity to
model the structural constitution relation in this paper, which can
be generalized to more complicated nonlinear structural materials.
The structural dynamics can be defined below in terms of the
structural displacement $\hat{\boldsymbol{d}}_s$: find
$\hat{\boldsymbol{d}}_s(t)\in H^2(0,T;H^2(\hat{\Omega}_s)^d)$ such
that:
\begin{equation}\label{soliddisplacement}
\hat\rho_s\frac{\partial^2 \hat{\boldsymbol{d}}_s}{\partial
t^2}-\widehat{\operatorname{div}}
\hat{\boldsymbol{P}}\left(\hat{\boldsymbol{d}}_s\right)=\hat{\boldsymbol{b}}_s,
\quad \text { in } \hat{\Omega}_s \times(0, T],
\end{equation}
where $\hat\rho_s$ is the structural density,
$\hat{\boldsymbol{b}}_s$ is the external force acting on the
structure, and
$\hat{\boldsymbol{P}}\left(\hat{\boldsymbol{d}}_s\right)$ is the
first Piola-Kirchhoff stress tensor expressed as
$$
\hat{\boldsymbol{P}}\left(\hat{\boldsymbol{d}}_s\right)=2 \mu_s
\hat{\boldsymbol{\varepsilon}}\left(\hat{\boldsymbol{d}}_s\right)+\lambda_s
\operatorname{tr}
\hat{\boldsymbol{\varepsilon}}\left(\hat{\boldsymbol{d}}_s\right)
\boldsymbol{I},
$$
where $\mu_s$ and $\lambda_s$ denote the Lam\'e constants, and
$\hat{\boldsymbol{\varepsilon}}\left(\hat{\boldsymbol{d}}_s\right)$
is the linearized strain operator defined as
$$
\hat{\boldsymbol{\varepsilon}}\left(\hat{\boldsymbol{d}}_s\right):=\frac{1}{2}\left(\hat{\nabla}
\hat{\boldsymbol{d}}_s+(\hat{\nabla}
\hat{\boldsymbol{d}}_s)^T\right).
$$
Here the gradient operator ``$\hat{\nabla}$'' and divergence
operator ``$\widehat{\operatorname{div}}$'' represent
differentiations with respect to Lagrangian coordinates,
$\hat{\boldsymbol{x}}_s$, in the reference configuration.

The following boundary condition and initial condition are defined
for (\ref{soliddisplacement}):
\begin{equation}\label{solid-BICs}
\begin{array}{rclll}
\hat{\boldsymbol{d}}_s&=&0,& \text{on } \hat\Gamma_{walls}\cap \hat\Gamma_{FSI},&\text{(fixed support condition)}\\
\hat{\boldsymbol{d}}_s(\hat{\boldsymbol{x}},0)&=&\hat{\boldsymbol{d}}_{s}^0,&
\text{in } \hat\Omega_s.&\text{(initial condition)}
\end{array}
\end{equation}

To close the definition of FSI model, we need to introduce the
following no-slip type interface conditions that can be applied to
most cases of FSI problems,
\begin{equation}\label{interface}
\begin{array}{rclll}
\boldsymbol{u}_f(\boldsymbol{x}(\hat{\boldsymbol{x}},t),t)&=&\frac{\partial\hat{\boldsymbol{d}}_s}{\partial t}, & \text{on } \hat\Gamma_{FSI}\times[0,T],&\text{(kinematic condition)}\\
\hat{J}\boldsymbol{\sigma}_f(\boldsymbol{x}(\hat{\boldsymbol{x}},t),t)
\hat{\boldsymbol{F}}^{-T}\hat{\boldsymbol{n}}_s&=&\hat{\boldsymbol{P}}\hat{\boldsymbol{n}}_s,
& \text{on } \hat\Gamma_{FSI}\times[0,T],&\text{(dynamic condition)}
\end{array}
\end{equation}
which describe the continuity of velocity and of normal stress
across the interface, respectively, where the Jacobian matrix,
$\hat{\boldsymbol{F}}=\hat\nabla_{\hat{\boldsymbol{x}}_s}\boldsymbol{x}_s
=\hat\nabla_{\hat{\boldsymbol{x}}_s}\hat{\boldsymbol{X}}_s
=\boldsymbol{I}+\hat\nabla_{\hat{\boldsymbol{x}}_s}\hat{\boldsymbol{d}}_s$,
denotes the deformation gradient tensor of structure,
$\hat{J}=\operatorname{det}(\hat{\boldsymbol{F}})$ is the Jacobian,
and $\hat{\boldsymbol{n}}_s$ denotes the outward normal unit vector
on $\hat\Gamma_{FSI}$ pointing into the fluid domain from the
structural domain.

\subsection{ALE mapping}
Instead of using the material/Lagrangian mapping of fluid,
$\hat{\boldsymbol{X}}_f$, to move the fluidic mesh that may suffer
grid distortions due to the large displacement of fluidic material
point, we introduce the following invertible ALE mapping:
\begin{equation*}
   \begin{aligned}
\mathcal{\hat{\boldsymbol{A}}}_f: \hat{\Omega}_f & \mapsto \Omega_f(t) \\
\hat{\boldsymbol{x}}_f & \mapsto
\boldsymbol{x}_m=\hat{\boldsymbol{x}}_f+\hat{\boldsymbol{m}}_f,
\end{aligned}
\end{equation*}
where $\boldsymbol{x}_m\in\Omega_f(t)$ denotes the current position
of fluidic mesh, $\hat{\boldsymbol{m}}_f$ denotes the fluidic mesh
displacement that is defined as an extension of the structural
displacement $\hat{\boldsymbol{d}}_s$ into $\hat{\Omega}_f$. This
extension can be defined in different ways. Here we adopt a harmonic
extension to define the ALE mapping:
\begin{equation}\label{alemap}
\left\{\begin{array}{rcll}
-\operatorname{div}\left(\hat{\nabla} \hat{\boldsymbol{m}}_f\right)&=&0,& \text { in } \hat{\Omega}_f, \\
\hat{\boldsymbol{m}}_f&=&\hat{\boldsymbol{d}}_s, & \text { on }
\hat{\Gamma}_{FSI},\\
\hat{\boldsymbol{m}}_f&=&0, & \text { on }
\partial\hat{\Omega}_f\backslash\hat{\Gamma}_{FSI}.
\end{array}\right.
\end{equation}
Thus, the moving fluidic mesh is obtained by $\bm x_m=\hat{\bm
x}_f+\hat{\bm m}_f$, $\forall \hat{\bm x}_f\in \hat\Omega_f$. Let
${\bm v}_m$ denote the velocity of fluidic mesh, defined as
\begin{equation}\label{mesh_velocity}
{\bm v}_{m}({\bm x_m},t)=\frac{\partial\mathcal{\hat{\bm
A}}_f}{\partial t} \circ\mathcal{{\hat{\bm A}}}_f^{-1}({\bm x_m})
=\frac{\partial(\hat{\bm x}_f+\hat{\bm m}_f)}{\partial
t}=\frac{\partial\hat{\bm m}_f}{\partial t}(\hat{\bm x}_f,t).
\end{equation}
Then we have the material derivative defined on the moving fluidic
mesh as follows
\begin{equation}
D_t {\bm u}_f = \partial_{t}^{\mathcal{\hat{\bm A}}_f} {\bm u}_f +
\big(({\bm u}_f - {\bm v}_m) \cdot\nabla\big){\bm u}_f,
\end{equation}
where $\partial_t^{\mathcal{\hat{\bm A}}_f} {\bm u}_f=\frac{\partial
{\bm u}_f}{\partial t}+({\bm v}_m\cdot\nabla) {\bm u}_f$ denotes the
ALE material derivative. Hence, fluid equations (\ref{nseq}) can be
reformulated as follows in ALE description:
\begin{equation}\label{ALE-momentum}
\left\{
\begin{aligned}
\rho_f \big(\partial_{t}^{\mathcal{\hat{\bm A}}_f} {\bm u}_f +
(({\bm u}_f - {\bm v}_m) \cdot\nabla){\bm u}_f\big) -
\operatorname{div} {\bm
\sigma}_f &= {\bm b}_f,& \hbox{~~~in~} \Omega_f(t)\times (0,T],\\
\operatorname{div}{\bm u}_f&=0,&\hbox{~in~}\Omega_f(t)\times (0,T].
\end{aligned}
\right.
\end{equation}

The following Lemma \ref{prop1} shows that the ALE mapping,
$\mathcal{\hat{\bm A}}_f$, holds $H^1$-invariance all the time for
any $\bu_f(\bx(\hat\bx,t),t)$ and its ALE material derivative
$\partial_t^{\mathcal{\hat\bA}_f} {\bm u}_f$.
\begin{lemma}\label{prop1}\cite{Gastaldi2001,Nobile;Formaggia1999}
For any $t\in(0,T]$, $\bu_f(\bx,t)\in H^1(\Omega_f(t))^d$ and
$\partial_t^{\mathcal{\hat\bA}_f} {\bm u}_f(\bx,t)\in
H^1(\Omega_f(t))^d$ if and only if $\hat
\bu(\hat\bx,t)=\bu(\bx,t)\circ \mathcal{\hat\bA}_f(\hat\bx,t)\in
H^1(\hat\Omega_f)^d$.
\end{lemma}

\subsection{Monolithic ALE-FEM for FSI}
In this section, we depict the monolithic, fully discrete ALE-FEM
for the presented FSI model (\ref{ALE-momentum}),
(\ref{fluid-BICs})-(\ref{alemap}) as the author Sun et al. do in,
e.g., \cite{HaoSun2021,Wang;Sun2016}. We first introduce the
following Sobolev spaces that are adopted to define the ALE weak
form of FSI model:
\begin{equation*}
\begin{array}{rcl}
\bV^f&:=&\{\bpsi_f\in H^1(\Omega_f(t))^d:
\bpsi_f=\hat\bpsi_f\circ\mathcal{\hat\bA}_f^{-1},\hat\bpsi_f\in
H^1(\hat\Omega_f)^d\},\\\vspace{0.2cm} \bV^f_{0}&:=&\{\bpsi_f\in
\bV^f:\bpsi_f=\bm 0 \text{ on
}\Gamma_{in}\cup\Gamma_{walls}\},\\\vspace{0.2cm}
\bV^f_{D}&:=&\{\bpsi_f\in \bV^f:\bpsi_f=\bu_{in} \text{ on
}\Gamma_{in},\bpsi_f=\bm 0 \text{ on
}\Gamma_{walls}\},\\\vspace{0.2cm} \hat\bV^s&:=&\{\hat\bpsi_s\in
H^1(\hat\Omega_s)^d:\frac{\partial\hat\bpsi_s}{\partial
t}=\bpsi_f\circ\mathcal{\bA}_f \text{ on
}\hat\Gamma_{FSI},\bpsi_f\in \bV^f\cap
L^2(\Gamma_{FSI})\},\\\vspace{0.2cm}
\hat\bV^s_0&:=&\{\hat\bpsi_s\in\hat\bV^s:\hat\bpsi_s=\bm 0 \text{ on
}\hat\Gamma_{walls}\cap \hat\Gamma_{FSI}\},\\\vspace{0.2cm}
Q^f&:=&\{q_f\in L^2(\Omega_f(t)): q_f=\hat
q_f\circ\mathcal{\bA}_f^{-1},\hat q_f\in
L^2(\hat\Omega_f)\},\\\vspace{0.2cm}
\hat\bV^m&:=&H^1(\hat\Omega_f)^d,\\\vspace{0.2cm}
\hat\bV^m_0&:=&\{\hat\bxi_f\in \hat\bV^m:\hat\bxi_f=\bm 0 \text{ on
 }\hat\partial\hat\Omega_f\},\\\vspace{0.2cm}
\hat\bV^m_D&:=&\{\hat\bxi_f\in \hat\bV^m:\hat\bxi_f=\bm 0 \text{ on
}\hat\partial\hat\Omega_f\backslash\hat\Gamma_{FSI};
\hat\bxi_f=\hat\bpsi_s \text{ on }\hat\Gamma_{FSI},\hat\bpsi_s\in
\hat\bV^s\cap L^2(\hat\Gamma_{FSI})\},
\end{array}
\end{equation*}
where the interface condition (\ref{interface})$_1$ and Lemma
\ref{prop1} are applied.

Then, we can define the following monolithic ALE weak form of FSI:
find
$(\boldsymbol{u}_f,p_f,\hat{\boldsymbol{d}}_s,\hat{\boldsymbol{m}}_f)\in
\bV^f_D\times Q^f\times \hat\bV^s_0\times\hat\bV^m_D$ such that
\begin{align}
(\hat{\rho}_{s} \frac{\partial^2\hat{\bm d}_{s}}{\partial t^2},
\hat{\bpsi}_{s} )_{\hat{\Omega}_{s}} + ({\bm P}(\hat{\bm d}_{s}),
\nabla \hat{\bpsi}_{s})_{\hat{\Omega}_{s}} + ({\rho}_{f}
\partial_{t}^{\mathcal{\hat\bA}_f} {\bm u}_{f},
{\bpsi_{f}})_{{\Omega}_{f}(t)}\notag\\
+ \left(\big(({\bm u}_{f} - \frac{\partial\hat{\bm m}_f}{\partial
t}\circ \mathcal{\hat\bA}_f^{-1})\cdot\nabla\big) {\bm u}_{f},
{\bpsi_{f}} \right)_{{\Omega}_{f}(t)} +({\bm \sigma}_f({\bm
u}_{f},p_{f}), \nabla\bpsi_f)_{{\Omega}_{f}(t)} \notag\\
+ (\nabla \cdot {\bm u}_{f}, q_{f})_{\Omega_{f}(t)} +(\nabla\hat{\bm
m}_f,\nabla\hat\bxi_f)_{\hat\Omega_f}= (\hat{\bm
b}_s,\hat{\bpsi}_{s} )_{\hat{\Omega}_{s}}+(\bm b_f,{\bpsi}_{f}
)_{{\Omega}_{f}(t)},\label{weakform-fsi}\\
\forall \left(\bpsi_{f},q_{f},\hat{\bpsi}_{s},\hat\bxi_f \right) \in
\bV^f_0\times Q^f\times \hat\bV^s_0\times \hat\bV^m_0,\notag
\end{align}
where due to the Piola transformation of surface integrals
\cite{Richter2017} and the dynamic interface condition
(\ref{interface})$_2$, we have
\begin{align} \label{ic-1}
\int_{\Gamma_{FSI}} \bS_f\bn_s \cdot \bpsi_f ds=\int_{\Gamma_{FSI}}
\bn_s\cdot (\bS_f\bpsi_f) ds \notag\\
=\int_{\hat\Gamma_{FSI}}  \hat \bn_s \cdot \left(\hat J\hat
\bF^{-1}\bS_f(\bx(\hat\bx,t),t) \bpsi_f(\bx(\hat\bx,t),t)\right)
d\hat
s\notag\\
=\int_{\hat\Gamma_{FSI}}  \hat J\bS_f(\bx(\hat\bx,t),t) \hat
\bF^{-T}\hat \bn_s \cdot \bpsi_f(\bx(\hat\bx,t),t) d\hat s =
\int_{\hat\Gamma_{FSI}} \hat\bP \hat \bn_s \cdot  \bpsi_f d\hat s,
\end{align}
which helps to remove interface integrals arising from the
integration by parts in (\ref{weakform-fsi}).

To define a fully discrete finite element approximation to
(\ref{weakform-fsi}), we first introduce a uniform partition
$0=t_0<t_1<\cdots<t_{N_{T}}=T$ with the time-step size $\Delta
t=T/{N_{T}}$. Set $t_n=n\Delta t$,
$\varphi^n=\varphi(\bx(\hat\bx,t_n),t_n)$ for $n=0,\cdots,N_{T}$,
$d_t\varphi^{n}=\frac{\varphi^{n}-\varphi^{n-1}}{\Delta t}$ and
$d_{tt}\varphi^{n}=\frac{\varphi^{n}-2\varphi^{n-1}+\varphi^{n-2}}{\Delta
t^2}$ for $n=1,\cdots,N_{T}$, where $\varphi^{-1}=\varphi^{0}$.

Second, we triangulate $\hat\Omega_f$ and $\hat\Omega_s$ to obtain
two quasi-uniform meshes, $\hat{\mathcal{T}}_{f,h}$ in
$\hat\Omega_f$ and $\hat{\mathcal{T}}_{s,h}$ in $\hat\Omega_s\
(0<h<1)$, which are conforming across $\hat\Gamma_{FSI}$. Then for
any $t\in(0,T]$, we numerically solve the ALE mapping (\ref{alemap})
in the finite element space $\hat\bV^m_{h}:=\big\{\hat\bxi_f\in
\hat\bV^m : \hat\bxi_f\big|_K\in P_1(K)^d,\forall
K\in\hat{\mathcal{T}}_{f,h}\big\}$, where $P_k$ denotes the $k$-th
degree piecewise polynomial space, to attain the discrete ALE
mapping $\mathcal{\hat\bA}_{f,h}$ that is smooth and invertible,
representing the moving fluidic mesh, i.e., for any
$\hat\bx_{f,h}\in \hat{\mathcal{T}}_{f,h}$, there exists
${\bx_{m,h}}\in{\mathcal{T}}_{f,h}(t)$ such that
\begin{equation}\label{discretemapping_fluid}
{\bm x}_{m,h}=\mathcal{\hat\bA}_{f,h}(\hat{\bm x}_{f,h},t)=\hat{\bm
x}_{f,h}+\hat{\bm m}_{f,h},
\end{equation}
where $\mathcal{T}_{f,h}(t)$ is the image of
$\hat{\mathcal{T}}_{f,h}$ under the discrete ALE mapping
$\mathcal{\hat\bA}_{f,h}$, i.e.,
$$
\mathcal{T}_{f,h}(t)=\mathcal{\hat\bA}_{f,h}\left(\hat{\mathcal{T}}_{f,h}\right)
=\hat{\mathcal{T}}_{f,h}+\hat{\bm m}_{f,h},
$$
and $\hat{\bm
{m}}_{f,h}\in \hat\bV^m_{D,h}:=\big\{\hat\bxi_f\in
\hat\bV^m_h:\hat\bxi_f=\bm 0 \text{ on
 }\partial\hat\Omega_f\backslash\hat\Gamma_{FSI},
\hat\bxi_f=\hat{\bm{d}}_s \text{ on }\hat\Gamma_{FSI}\big\}$.

Accordingly, the semi-discrete
ALE material derivative is defined as:
$$\partial_t^{\mathcal{\hat\bA}_{f,h}} {\bpsi}_{f,h}=\frac{\partial
{\bpsi}_{f,h}}{\partial t}+\bv_{m,h}\cdot\nabla
{\bpsi}_{f,h}=\frac{\partial {\bpsi}_{f,h}}{\partial
t}+\left(\frac{\partial\hat{\bm m}_{f,h}}{\partial
t}\circ\mathcal{\hat\bA}_{f,h}^{-1}\right)\cdot\nabla
{\bpsi}_{f,h},$$ which is approximated by the following fully
discrete form at $t=t_{n}$:
\begin{equation}\label{BDF2ALE}
d_{t}^{\mathcal{\hat\bA}_{f,h}}{\bpsi}_{f,h}^{n}=\frac{{\bpsi}_{f,h}^{n}
-{\bpsi}_{f,h}^{n-1}\circ\mathcal{\hat\bA}^{n-1}_{f,h}\circ
(\mathcal{\hat\bA}^{n}_{f,h})^{-1}}{\Delta t}.
\end{equation}

More finite element spaces are defined below for $n=1,\cdots,N_{T}$:
\begin{equation*}
    \begin{array}{rcl}
        \bV^{f,n}_{h}&:=&\big\{\bpsi_f\in \bV^f : \bpsi_f\big|_K\in
P_1(K)^d,\forall K\in{\mathcal{T}}_{f,h}(t_n)\big\},\\
Q^{f,n}_{h}&:=&\big\{q_f\in Q^f : q_f\big|_K\in P_1(K),\forall
K\in{\mathcal{T}}_{f,h}(t_n)\big\},\\
\bV^{f,n}_{0,h}&:=&\{\bpsi_f\in \bV^{f,n}_h:\bpsi_f=\bm 0 \text{ on
}\Gamma_{in}\cup\Gamma_{walls}\},\\
\bV^{f,n}_{D,h}&:=& \{\bpsi_f\in \bV^{f,n}_h:\bpsi_f=\bu_{in} \text{
on }\Gamma_{in},\bpsi_f=\bm 0 \text{ on
}\Gamma_{walls}\},\\
\hat\bV^s_{h}&:=&\big\{\hat\bpsi_s\in \hat\bV^s :
\hat\bpsi_s\big|_K\in P_1(K)^d,\forall
K\in\hat{\mathcal{T}}_{s,h}\big\},\\
\hat\bV^s_{0,h}&:=&\{\hat \bpsi_s \in \hat\bV^s_h:\hat\bpsi_s=\bm 0
\text{ on }\hat\Gamma_{walls}\cap
\hat\Gamma_{FSI}\},\\
\hat\bV^m_{0,h}&:=&\{\hat\bxi_f\in \hat\bV^m_h:\hat\bxi_f=\bm 0
\text{ on }\partial\hat\Omega_f\},
\end{array}
\end{equation*}
which indicate that we employ the lowest equal-order mixed finite
element, $P_1/P_1$ element with the pressure stabilization term
\cite{Hughes.T;Franca.L;Balestra.M1986,Tezduyar.T1992} to
approximate the saddle-point problem arising from the FSI's weak
form (\ref{weakform-fsi}) within finite element spaces,
$\left(\bV^{f,n}_{h}\oplus\hat\bV^s_h\right)\times
Q_h^{f,n}\subset\left(\bV^f\oplus\hat\bV^s\right)\times Q^f$.

Finally, the FOM for FSI simulation, i.e., a fully discrete,
monolithic ALE-FEM for (\ref{weakform-fsi}) is thus defined as
follows: find $({\bm{u}^{n}_{f,h}},p^{n}_{f,h}$,
$\hat{\bm{d}}^{n}_{s,h},\hat{\bm{m}}^{n}_{f,h}) \in
\bm{V}^{f,{n}}_{D,h} \times Q^{f,{n}}_h \times \hat\bV^{s}_{0,h}
\times{\hat\bV^{m}_{D,h}}$ such that for $n=1,\cdots,N_{T}$,
\begin{eqnarray}
&&(\hat{\rho}_{s} d_{tt} \hat{\bm d}^{n}_{s,h}, \hat\bpsi_{s}
)_{\hat{\Omega}_{s}} + ({\bm P}(\hat{\bm d}^{n}_{s,h}), \nabla
\hat\bpsi_{s})_{\hat{\Omega}_{s}}+ ({\rho}_{f}
d_{t}^{\mathcal{\hat\bA}_{f,h}} {\bm u}^{n}_{f,h}, {\bpsi_{f}}
)_{{\Omega}_{f}^n}\notag\\
&& + \left(\big({\bm u}^{n}_{f,h} - d_t\hat{\bm
m}^{n}_{f,h}\circ(\mathcal{\hat\bA}^n_{f,h})^{-1}\big)\cdot\nabla
{\bm u}_{f,h}^{n}, {\bpsi_{f}} \right)_{{\Omega}_{f}^n} +({\bm
\sigma}_f({\bm u}^{n}_{f,h},p^{n}_{f,h}),
\nabla{\bpsi_{f}})_{{\Omega}_{f}^n}
\notag\\
&&+ (\nabla \cdot {\bm u}^{n}_{f,h}, q_{f})_{{\Omega}_{f}^n}
+(\nabla\hat{\bm m}^n_{f,h},\nabla\hat\bxi_f)_{\hat{\Omega}_{f}} +
\delta\frac{h^2}{\rho_f\nu_f}(\nabla p^{n}_{f,h},\nabla
q_{f})_{{\Omega}_{f}^n}\label{disALE-FEM}\\
&&
= (\hat{\bm b}_s^{n},\hat\bpsi_{s} )_{\hat{\Omega}_{s}}+(\bm
b_f^{n},\bpsi_{f} )_{{\Omega}_{f}^n},\ \forall
\left(\bpsi_{f},q_{f},\hat{\bpsi}_{s},\hat\bxi_f \right) \in
\bV^{f,n}_{0,h}\times Q^{f,n}_{h}\times
\hat\bV^s_{0,h}\times\hat\bV^m_{0,h},\notag
\end{eqnarray}
where the last term on the left-hand side of (\ref{disALE-FEM}) is
the pressure stabilization term with a well-tuned parameter
$\delta$. Note that (\ref{disALE-FEM}) is a nonlinear system due to
the nonlinearities of fluidic convection term and of fluidic mesh
update through (\ref{discretemapping_fluid}). Thus a linearization
algorithm is needed to numerically implement (\ref{disALE-FEM}).
Briefly speaking, we adopt a fixed-point iteration to update the
fluidic mesh by solving the discrete ALE mapping
$\mathcal{\hat\bA}_{f,h}$, and in each step of the fixed-point
iteration we utilize Newton's method to linearize the fluidic
convection term and then iteratively solve a linear system of
(\ref{disALE-FEM}) until convergence. A detailed algorithm
description can be referred to early works of the author Sun et al.
\cite{Wang;Sun2016,HaoSun2021}.

\section{Reduced order model for FSI}\label{S3}
In this section, based on the FOM (\ref{disALE-FEM}), we attempt to
develop a proper ROM approach for FSI problems by compressing the
snapshots using POD to generate a set of reduced basis functions
during the offline phase, with which we are able to accomplish a
fast FSI simulation during the online phase. We discuss about both
phases below.

\subsection{Offline Phase}\label{sec:offline}
In order to suppress the growth of approximation errors and achieve
the goal of long-term FSI simulation, first, we divide the total
time interval $[0,T]$ into $G$ time segments $[T_{g},T_{g+1}]$, $g =
0,\ldots,G-1$, where $T_0=0$, $T_{G}=T$,
$[0,T]=\bigcup\limits_{g=0}^{G-1}[T_{g},T_{g+1}]$, and $T_g\ (g =
0,\ldots,G)$ coincide with the discrete time points $t_n$ at some
time steps $n\ (0\leq n \leq N_T)$, i.e., all discrete time points,
$t_0, t_1, \ldots, t_{N_T}$, are reallocated into each
$[T_{g},T_{g+1}]\ (g = 0,\ldots,G-1)$. In other words, each time
segment $[T_{g},T_{g+1}]$ is divided into $N_{T_{g}}-1$ subintervals
with the time step size $\Delta t$, where
$N_{T_{g}}=\frac{T_{g+1}-T_g}{\Delta t}+1,\ \sum\limits_{g =
0}^{G-1}(N_{T_{g}}-1) = N_{T}$, and $t_n=T_g+(n-1)\Delta t\in
[T_{g},T_{g+1}]$, $n=1,\cdots,N_{T_g}$.

Next, we perform the POD in each $[T_{g},T_{g+1}]$, $g =
0,\ldots,G-1$, where we start the time numbering for each variable
at $N_{T_g}$ time points labeled in order as $1,\cdots,N_{T_{g}}$.
Then, we need the snapshot matrices as a main ingredient to conduct
POD. To the end, we first introduce 
dimension notations of four finite element spaces used to define the
FOM (\ref{disALE-FEM}):
$\mathcal{N}_h^{\bm{u}_{f}}$ denotes the dimension of $\bm{V}^{f,{n}}_{D,h}$, 
$\mathcal{N}_h^{p_{f}}$ the dimension of $Q^{f,{n}}_h$, 
$\mathcal{N}_h^{\hat{\bm{d}}_s}$ the dimension of $\hat\bV^{s}_{0,h}$, 
and $\mathcal{N}_h^{\hat{\bm{m}}_f}$ the dimension of
$\hat\bV^{m}_{D,h}$, 
with which we can then assemble the needed snapshot matrices to
perform POD by compressing the high-dimensional finite element
solutions down to a low-dimensional space spanned by a reduced
basis.

We begin by constructing $N_{T_{g}}$ snapshot vectors in
$[T_{g},T_{g+1}]$, $\bm{U}^{n}_g \in \mathbb{R}^{\mathcal{N}_h}\ (n
= 1, \cdots, N_{T_{g}})$, where
$\mathcal{N}_h=\mathcal{N}_h^{\bm{u}_{f}}+\mathcal{N}_h^{p_{f}}
+\mathcal{N}_h^{\hat{\bm{d}}_s}+\mathcal{N}_h^{\hat{\bm{m}}_f}$, and
$\bm{U}^{n}_g$ is defined at the $n$-th time step in
$[T_{g},T_{g+1}]$ as follows
\begin{equation}\label{ui}
\bm{U}^{n}_g = \left( \bm{u}^{n}_{f,h},   p^{n}_{f,h},
\hat{\bm{d}}^{n}_{s,h},  \hat{\bm{m}}^{n}_{f,h}\right)^{\top}, \quad
n = 1, \cdots,  N_{T_{g}},
\end{equation}
resulting in 
the following corresponding snapshot matrix $\bm{U}_g \in
\mathbb{R}^{\mathcal{N}_h \times N_{T_{g}}}$,
\begin{equation}\label{u}
\bm{U}_g = \left[ \bm{U}_g^{1}, \bm{U}_g^{2}, \cdots,
\bm{U}_g^{N_{T_{g}}} \right].
\end{equation}
%
By doing this way, we sort 
all finite element solutions in order in each $[T_{g},T_{g+1}]$
according to their attribution (fluid/structure), category
(velocity/pressure) and component (horizontal/vertical) properties.

Then, we introduce notations of four sub-matrices of $\bU_g$,
$\bX_{\bm{u}_{f}} \in \mathbb{R}^{\mathcal{N}_h^{\bm{u}_{f}} \times
N_{T_{g}}}$, $\bX_p \in \mathbb{R}^{\mathcal{N}_h^{p_{f}} \times
N_{T_{g}}}$, $\bX_{\hat{\bm{d}}_s} \in
\mathbb{R}^{\mathcal{N}_h^{\hat{\bm{d}}_s} \times N_{T_{g}}}$ and
$\bX_{\hat{\bm{m}}_f} \in \mathbb{R}^{\mathcal{N}_h^{\hat{\bm{m}}_f}
\times N_{T_{g}}}$, to denote
\begin{equation}\label{submatrices}
\begin{aligned}
\bX_{\bm{u}_{f}} & =\left[\bm{u}^{1}_{f,h}, \ldots,
\bm{u}^{N_{T_{g}}}_{f,h}\right], & \bX_p & =\left[ p^{1}_{f,h},
\ldots, p^{N_{T_{g}}}_{f,h}\right], \\
\bX_{\hat{\bm{d}}_s} & =\left[\hat{\bm{d}}^{1}_{s,h}, \ldots,
\hat{\bm{d}}^{N_{T_{g}}}_{s,h}\right], & \bX_{\hat{\bm{m}}_f} &
=\left[\hat{\bm{m}}^{1}_{f,h}, \ldots,
\hat{\bm{m}}^{N_{T_{g}}}_{f,h}\right],
\end{aligned}
\end{equation}
thus
\begin{equation}\label{Ug1}
\bU_g=\left[\begin{array}{l}
\bX_{\bm{u}_{f}}\\
\bX_p\\
\bX_{\hat{\bm{d}}_s}\\
\bX_{\hat{\bm{m}}_f}
\end{array}\right].
\end{equation}

Utilizing (\ref{submatrices}), we are now able to define the
correlation matrices ${\mathcal{C}}_{\bm{u}_{f}}$, $
\mathcal{C}_{p_{f}}$,  $ \mathcal{C}_{\hat{\bm{d}}_s}$ and  $
\mathcal{C}_{\hat{\bm{m}}_f}$ as follows
\begin{equation}\label{correlationmatrices}
\begin{aligned}
\mathcal{C}_{\bm{u}_{f}} & :=\bX_{\bm{u}_{f}}^{\top}
\bX_{\bm{u}_{f}},&
\mathcal{C}_{p_{f}} & := \bX_{p_{f}}^{\top} \bX_{p_{f}},  \\
\mathcal{C}_{\hat{\bm{d}}_s} & :=\bX_{\hat{\bm{d}}_s}^{\top}
\bX_{\hat{\bm{d}}_s},& \mathcal{C}_{\hat{\bm{m}}_f} &
:=\bX_{\hat{\bm{m}}_f}^{\top} \bX_{\hat{\bm{m}}_f},
\end{aligned}
\end{equation}
which all belong to $\mathbb{R}^{{N_{T_{g}}} \times {N_{T_{g}}}}$.

By virtue of the above correlation matrices, we conduct a POD
compression on the snapshot matrices, which involves solving the
following $3d+1$ eigenvalue problems:
\begin{equation}\label{snapmatrix}
    \mathcal{C}_* \mathcal{Q}_*=\mathcal{Q}_* \Lambda_{*},
\end{equation}
where $* \in\{\bm{u}_{f},  p_{f}, \hat{\bm{d}}_s, \hat{\bm{m}}_f\},\
\mathcal{Q}_*$ is the eigenvector matrix, and $\Lambda_*$ is the
diagonal eigenvalue matrix in which $N_{T_g}$ eigenvalues are in a
descending order. The $k$-th ($k = 1,\ldots,N_{T_{g}}$) reduced
basis function that is related to the eigenvalue problem
(\ref{snapmatrix}), $\boldsymbol{\Xi}_k^{*}$, is obtained by
multiplying the snapshot matrix $\bX_*$ by the $k$-th column of the
eigenvector matrix $\mathcal{Q}_*$, $\boldsymbol{v}_k^{*}$.
Therefore, we attain the following basis functions for any $*
\in\{\bm{u}_{f},  p_{f}, \hat{\bm{d}}_s, \hat{\bm{m}}_f\}$:
$$
\boldsymbol{\Xi}_k^{*}:=\frac{1}{\sqrt{\lambda_k^{*}}} \bX_{*}
\boldsymbol{v}_k^{*},
$$
where $\lambda_k^{*}$ is the eigenvalue corresponding to the
eigenvector $\boldsymbol{v}_k^{*}$ of $\bX_{*}$.

Thus, the set of $N$ reduced basis functions,
$\left\{\boldsymbol{\Phi}_1, \ldots, \boldsymbol{\Phi}_{N}\right\}$,
are produced, where each basis function $\boldsymbol{\Phi}_i\
(i=1,\cdots,N)$ is a block function of four components, as shown
below
\begin{equation}\label{reducedbasis}
\begin{array}{rclcrcl}
\boldsymbol{\Phi}_1&=&\left(\begin{array}{c}
\boldsymbol{\Xi}_1^{\bm{u}_{f}} \\
\bm{0} \\
\bm{0} \\
\bm{0}
\end{array}\right),& \ldots, &
\boldsymbol{\Phi}_{N_{\bm{u}_{f}}}&=&\left(\begin{array}{c}
\boldsymbol{\Xi}_{N_{\bm{u}_{f}}}^{\bm{u}_{f}} \\
\bm{0} \\
\bm{0} \\
\bm{0}
\end{array}\right),\\
\boldsymbol{\Phi}_{{N_{\bm{u}_{f}}} + 1}&=&\left(\begin{array}{c}
\bm{0} \\
\boldsymbol{\Xi}_1^{p_{f}} \\
\bm{0} \\
\bm{0}
\end{array}\right),& \ldots,&
\boldsymbol{\Phi}_{{N_{\bm{u}_{f}}} +
{N_{p_{f}}}}&=&\left(\begin{array}{c}
\bm{0} \\
\boldsymbol{\Xi}_{N_{p_{f}}}^{p_{f}} \\
\bm{0} \\
\bm{0}
\end{array}\right),\\
\boldsymbol{\Phi}_{{N_{\bm{u}_{f}}} + N_{p_{f}} +
1}&=&\left(\begin{array}{c}
\bm{0} \\
\bm{0} \\
\boldsymbol{\Xi}_1^{\hat{\bm{d}}_s} \\
\bm{0}
\end{array}\right),& \ldots,&
\boldsymbol{\Phi}_{{N_{\bm{u}_{f}}} + {N_{p_{f}}} +
N_{\hat{\bm{d}}_s}}&=&\left(\begin{array}{c}
\bm{0} \\
\bm{0}\\
\boldsymbol{\Xi}_{N_{\hat{\bm{d}}_s}}^{\hat{\bm{d}}_s}\\
\bm{0}
\end{array}\right),\\
\boldsymbol{\Phi}_{{N_{\bm{u}_{f}}} + N_{p_{f}} +N_{\hat{\bm{d}}_s}+
1}&=&\left(\begin{array}{c}
\bm{0} \\
\bm{0} \\
\bm{0} \\
\boldsymbol{\Xi}_1^{\hat{\bm{m}}_f}
\end{array}\right),& \ldots,& \boldsymbol{\Phi}_{N}&=&\left(\begin{array}{c}
\bm{0} \\
\bm{0} \\
\bm{0} \\
\boldsymbol{\Xi}_{N_{\hat{\bm{m}}_f}}^{\hat{\bm{m}}_f}
\end{array}\right),
\end{array}
\end{equation}
where $N_*\ (*=\bm{u}_{f}, p_{f}, \hat{\bm{d}}_s, \hat{\bm{m}}_f)$
denotes the number of reduced basis functions amongst a total of
$N_{T_{g}}$ basis functions for four variables of FOM for FSI,
respectively. Thus $N = N_{\bm{u}_{f}} + N_{p_{f}} +
N_{\hat{\bm{d}}_s} + N_{\hat{\bm{m}}_f}$, and,
$\left\{\boldsymbol{\Phi}_1, \ldots, \boldsymbol{\Phi}_{N}\right\}$
forms the reduced basis of $\bm{U}_g$ defined in (\ref{u}) or
(\ref{Ug1}).

Finally, we introduce the reduced-order finite dimensional spaces at
$t_n\in [T_g,T_{g+1}]$, $n=1, \ldots, N_{T_{g}}$ and $g =
0,\ldots,G-1$:
\begin{equation*}
    \begin{array}{rcl}
\bV^{f,n,rom}_{D,h}&:=&
\operatorname{span}\left\{\boldsymbol{\Xi}_1^{\bm{u}_{f}}, \ldots,
\boldsymbol{\Xi}_{N_{\bm{u}_{f}}}^{\bm{u}_{f}}\right\},\\
Q^{f,n,rom}_{h}&:=&\operatorname{span}\left\{\boldsymbol{\Xi}_1^{p_{f}},
\ldots,
\boldsymbol{\Xi}_{N_{p_{f}}}^{p_{f}}\right\},\\
\hat\bV^{s,rom}_{D,h}&:=&\operatorname{span}\left\{\boldsymbol{\Xi}_1^{\hat{\bm{d}}_s},
\ldots,
\boldsymbol{\Xi}_{N_{\hat{\bm{d}}_s}}^{\hat{\bm{d}}_s}\right\},\\
\hat\bV^{m,rom}_{D,h}&:=&\operatorname{span}\left\{\boldsymbol{\Xi}_1^{\hat{\bm{m}}_f},
\ldots,
\boldsymbol{\Xi}_{N_{\hat{\bm{m}}_f}}^{\hat{\bm{m}}_f}\right\},
\end{array}
\end{equation*}
where $\bV^{f,n,rom}_{D,h}\subset\bV^{f,n}_{D,h}$,
$Q^{f,n,rom}_{h}\subset Q^{f,n}_{h}$, $\hat\bV^{s,rom}_{D,h}\subset
\hat\bV^{s}_{0,h}$, $\hat\bV^{m,rom}_{D,h}\subset
\hat\bV^{m}_{D,h}$.

\subsection{Online Phase}\label{sec:online}
Once we attain the reduced basis functions during the offline phase,
we can utilize them to solve the FOM (\ref{disALE-FEM}) for FSI in
the online phase by defining $(\mathbf{u}_{h}^{n,rom})^{\top}:=$
$\big(\bm{u}^{n,rom}_{f,h}$, $p^{n,rom}_{f,h}$,
$\hat{\bm{d}}^{n,rom}_{s,h}$,$\hat{\bm{m}}^{n,rom}_{f,h} \big)\in
\bV^{f,n,rom}_{D,h} \times Q^{f,n,rom}_{h} \times
\hat\bV^{s,rom}_{D,h} \times\hat\bV^{m,rom}_{D,h}$ as the reduced
solution of the developed ROM for FSI as follows:
\begin{equation}
    \begin{aligned}
\bm{u}^{n,rom}_{f,h} & :=\sum_{k=1}^{N_{\bm{u}_{f}}}  u_{f,k}^{n}
\boldsymbol{\Xi}_k^{\bm{u}_{f}}, &
p^{n,rom}_{f,h} & :=\sum_{k=1}^{N_p} p_{f,k}^{n} \boldsymbol{\Xi}_{k}^{p_{f}} ,\\
\hat{\bm{d}}^{n,rom}_{s,h} & :=\sum_{k=1}^{N_{\hat{\bm{d}}_s}}
\hat{d}_{s, k}^{n} \boldsymbol{\Xi}_k^{\hat{\bm{d}}_s}, &
\hat{\bm{m}}^{n,rom}_{f,h} & :=\sum_{k=1}^{N_{\hat{\bm{m}}_f}}
\hat{m}_{f, k}^{n} \boldsymbol{\Xi}_k^{\hat{\bm{m}}_f}.
\end{aligned}
\end{equation}

Then, the online ROM system for FSI reads as follows: for every
$t_{n}\in[T_g,T_{g+1}]$, $n=2, \ldots, N_{T_{g}}$ and $g =
0,\ldots,G-1$, find
$(\mathbf{u}_{h}^{n,rom})^{\top}=\big(\bm{u}^{n,rom}_{f,h}$,
$p^{n,rom}_{f,h}$,
$\hat{\bm{d}}^{n,rom}_{s,h}$,$\hat{\bm{m}}^{n,rom}_{f,h}\big)$ $\in
\bV^{f,n,rom}_{D,h} \times Q^{f,n,rom}_{h} \times
\hat\bV^{s,rom}_{D,h} \times \hat\bV^{m,rom}_{D,h}$ such that
\begin{eqnarray}
&&(\hat{\rho}_{s} d_{tt} \hat{\bm{d}}^{n,rom}_{s,h},
\hat{\bpsi}^{rom}_{s,h})_{\hat{\Omega}_{s}} + ({\bm
P}(\hat{\bm{d}}^{n,rom}_{s,h}), \nabla
\hat{\bpsi}^{rom}_{s,h})_{\hat{\Omega}_{s}}\notag\\
&&+ ({\rho}_{f}d_{t}^{\mathcal{\hat\bA}_{f,h}}
\bm{u}^{n,rom}_{f,h}, \bpsi^{rom}_{f,h})_{{\Omega}_{f}^n}\notag\\
&& + \left(\big(\bm{u}^{n,rom}_{f,h} -
d_t\hat{\bm{m}}^{n,rom}_{f,h}\circ(\mathcal{\hat\bA}^{n,rom}_{f,h})^{-1}\big)\cdot\nabla
\bm{u}^{n,rom}_{f,h}, \bpsi^{rom}_{f,h} \right)_{{\Omega}_{f}^n}\notag\\
&& +({\bm \sigma}_f(\bm{u}^{n,rom}_{f,h},p^{n,rom}_{f,h}),
\nabla{\bpsi^{rom}_{f,h}})_{{\Omega}_{f}^n} + (\nabla \cdot
\bm{u}^{n,rom}_{f,h}, q^{rom}_{f,h})_{{\Omega}_{f}^n}\notag\\
&&+(\nabla\hat{\bm{m}}^{n,rom}_{f,h},\nabla\hat\bxi^{rom}_{f,h})_{\hat{\Omega}_{f}}
+ \delta\frac{h^2}{\rho_f\nu_f}(\nabla p^{n,rom}_{f,h},\nabla
q^{rom}_{f,h})_{{\Omega}_{f}^n}\notag\\
&& =
(\hat{\bm{b}}_s^{n,rom},\hat{\bpsi}^{rom}_{s,h})_{\hat{\Omega}_{s}}+(\bm
b_f^{n,rom},\bpsi^{rom}_{f,h})_{{\Omega}_{f}^n},\label{romdis}\\
&&\forall
\big(\bpsi^{rom}_{f,{h}},q^{rom}_{f,{h}},
\hat{\bpsi}^{rom}_{s,{h}},\hat\bxi^{rom}_{f,{h}}\big)\in
\bV^{f,n,rom}_{D,h} \times Q^{f,n,rom}_{h} \times
\hat\bV^{s,rom}_{D,h} \times \hat\bV^{m,rom}_{D,h},\notag
\end{eqnarray}
where $\hat{\bm{d}}^{1,rom}_{s,h}$ and $\hat{\bm{d}}^{0,rom}_{s,h}$
in $[T_g,T_{g+1}]$ actually take the values of
$\hat{\bm{d}}^{N_{T_g},rom}_{s,h}$ and
$\hat{\bm{d}}^{N_{T_g-1},rom}_{s,h}$ in $[T_{g-1},T_{g}]$,
respectively, and $\bm{u}^{1,rom}_{f,h}$ in $[T_g,T_{g+1}]$ takes
the value of $\bm{u}^{N_{T_g},rom}_{f,h}$ in $[T_{g-1},T_{g}]$ as
well, for $g = 0,\ldots,G-1$. Particularly, if $g=0$, then
$\hat{\bm{d}}^{1,rom}_{s,h}$ and $\bm{u}^{1,rom}_{f,h}$ take the
interpolation values of $\hat{\bm d}_s^0$ and $\bu_f^0$ in
$\hat\bV^{s,rom}_{D,h}$ and $\bV^{f,1,rom}_{D,h}$ both in
$[T_0,T_{1}]$, respectively, and,
$\hat{\bm{d}}^{0,rom}_{s,h}$=$\hat{\bm{d}}^{1,rom}_{s,h}$ in
$[T_0,T_{1}]$.

\section{Numerical Experiments}\label{S4}
In this section, we carry out some numerical experiments by applying
both existing FOM and the developed ROM to a FSI benchmark problem
initiated by Turek and Hron \cite{turek2006proposal,
turek2010numerical}, and then conduct a comparison study between
them. In addition, we will also illustrate our motivation of
partitioning the spatial and temporal dimensions, separately, for
the developed ROM approach.

The domain of the studied FSI benchmark problem is illustrated in
Figure \ref{domainsetting} and sketched again in Figure
\ref{Computationaldomain}, whose details are depicted below.
\begin{itemize}
    \item Domain length $L=2.5$ and height $H=0.41$;
    \item The center of obstacle cylinder is positioned at $C=(0.2,0.2)$ with the radius
    $r=0.05$;
    \item The elastic structural beam's length $l=0.35$ and height $h=0.02$, whose lower
    right corner is positioned at $(0.6,0.19)$, and whose left end is fully attached to the obstacle
    cylinder;
    \item The control point is $A(t)$ that lies at the midpoint of right end of the beam
    with $A(0)=(0.6,0.2)$, and $B=(0.15,0.2)$.
\end{itemize}
\begin{figure}[H]
    \centering
    \includegraphics[width = 12cm, height = 3cm]{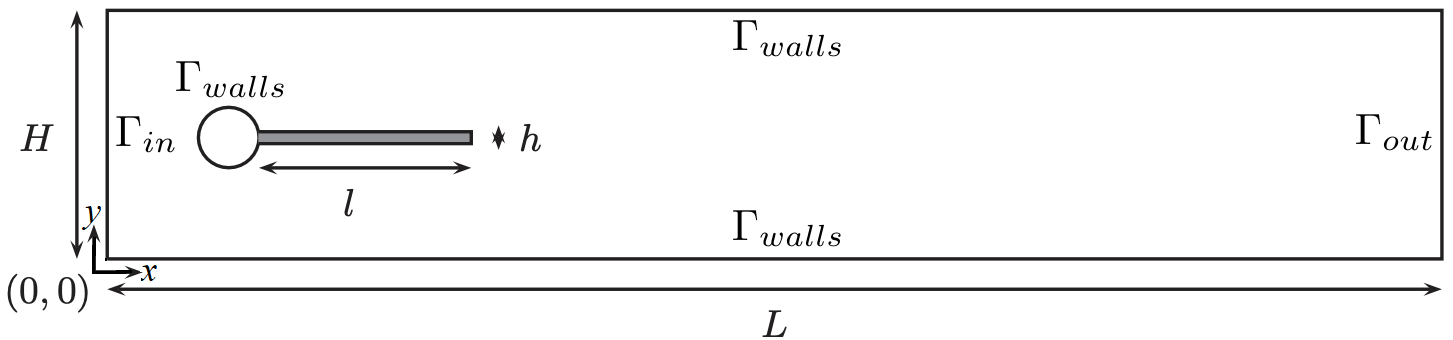}
    \includegraphics[width = 11cm, height = 2.25cm]{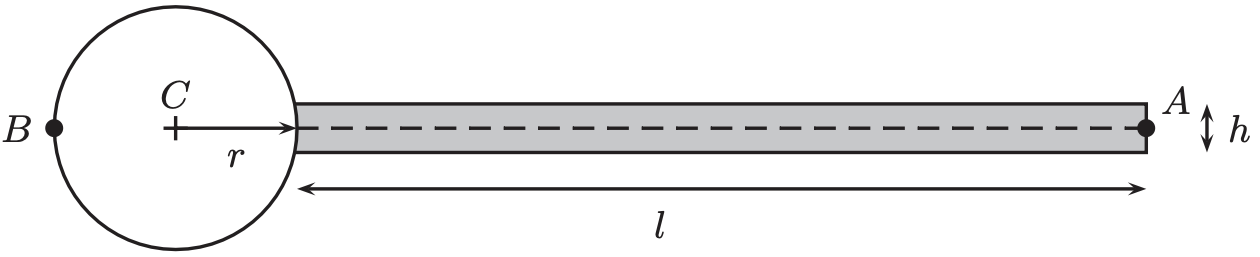}
    \caption{Computational domain of the FSI benchmark problem}
    \label{Computationaldomain}
\end{figure}

We impose the no-slip boundary condition for the fluid velocity on
$\Gamma_{walls}$ that include the surface of obstacle cylinder and
the top and bottom walls of the channel domain, and do-nothing
condition on the outlet $\Gamma_{out}$. In addition, we impose a
parabolic profile of the incoming flow, i.e., the boundary condition
of fluid velocity on the inlet $\Gamma_{in}$ is assigned to
$({u}_{\text {in}},0)^{\top}$ such that
\begin{equation}
{u}_{\text {in}}:=\left\{\begin{array}{ll}
\hat{u}(y) \frac{1-\cos \left(\frac{\pi}{2} t\right)}{2},& \text {if } t<2 s, \\
\hat{u}(y),& \text {otherwise, }
\end{array}\right.
\end{equation}
where $0\leq y\leq 0.41$ is the $y$-coordinate variable, and
\begin{equation}
\hat{u}(y)=\hat{U} \times \frac{4}{0.1681} y(0.41-y),
\end{equation}
here the value of $\hat{U}$ is listed in Table \ref{psforFSI}, in
which all other physical parameters of the FSI problem are reported
as well. We also require the beam to be attached to the obstacle
cylinder, therefore $\hat{\boldsymbol{d}}_s=0$ on $\Gamma_{walls}$.
In what follows, all numerical computations are performed on a
12th-Gen Intel 3.20 GHz Core i9 computer.
\begin{table}[H]
    \centering
    \begin{tabular}{|c|c|c|c|}
\hline Symbol & Description & Value & Unit \\
\hline $\rho_{s}$ & Density of the structure & $10^{4}$ &
$\frac{\mathrm{kg}}{\mathrm{m}^{3}}$\\
\hline $\lambda_{s}$ & Lam\'e constant of the structure & $0.4$ & --- \\
\hline
$\mu_{s}$ & Shear modulus of the structure & $0.5\times 10^{6}$ & $\frac{\mathrm{kg}}{\mathrm{ms}^{2}}$\\
\hline
$\rho_{f}$ & Density of the fluid & $10^{3}$ & $\frac{\mathrm{kg}}{\mathrm{m}^{3}}$ \\
\hline
$\nu_{f}$ & Kinematic viscosity of the fluid & $10^{-3}$ & $\frac{\mathrm{m}^{2}}{\mathrm{s}}$  \\
\hline
$\hat{U}$ & The largest value of incoming fluid velocity & 1.5 & $\frac{\mathrm{m}}{\mathrm{s}}$ \\
\hline
\end{tabular}
    \caption{Parameter settings for the FSI benchmark problem.}
    \label{psforFSI}
\end{table}

\begin{figure}[H]
    \centering
\includegraphics[width = 10cm, height = 1.7cm]{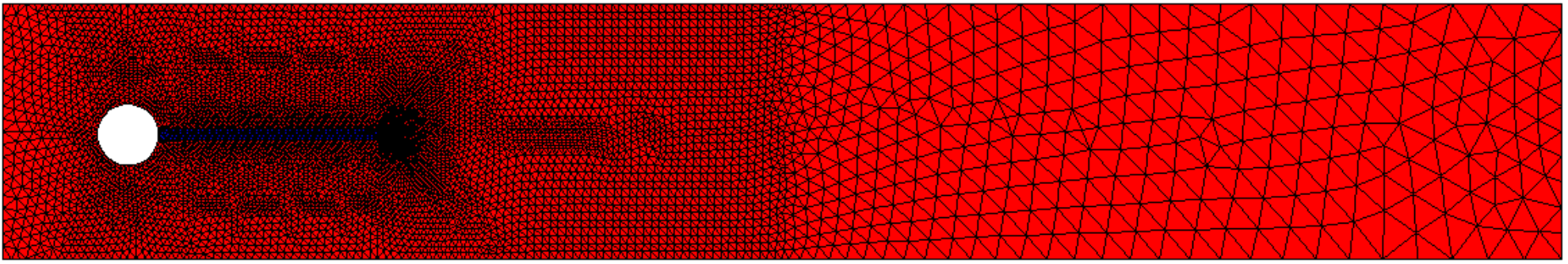}
    \caption{The triangular mesh of benchmark domain.}\label{meshused}
\end{figure}

\begin{figure}[H]
    \centering
    \includegraphics[width = 7cm, height = 5cm]{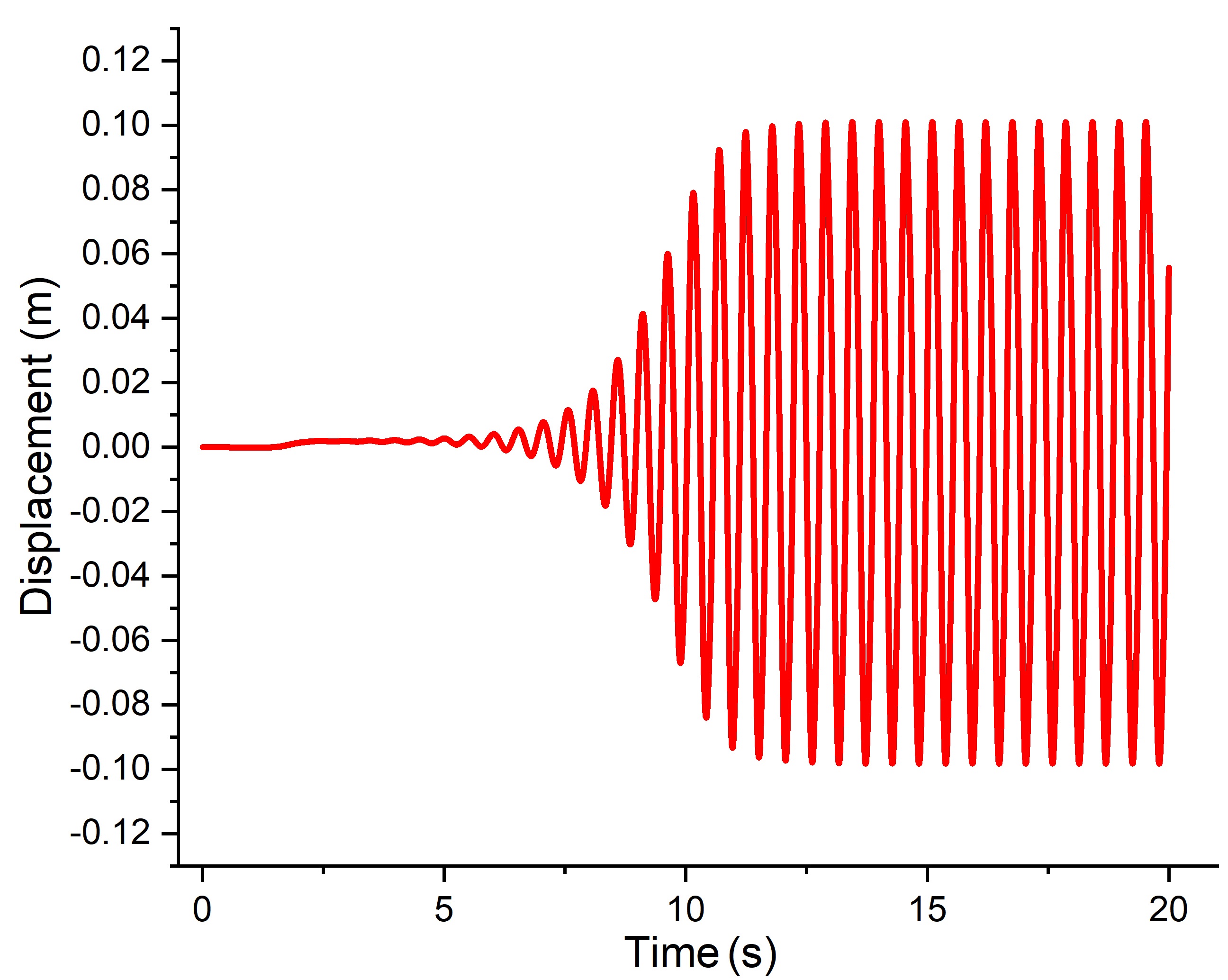}
    \caption{Vibration curve (the $y$-displacement trajectory) of point $A$ obtained from the FOM result.}
    \label{Hfnsdispy18750}
\end{figure}

To verify the effectiveness of our developed ROM by comparing with
FOM, we first perform the FOM computation for this FSI benchmark
problem using the same time step size $\Delta t=0.001s$ as used in
the benchmark problem proposed in
\cite{turek2006proposal,turek2010numerical} on a fine triangular
mesh with $18350$ degree of freedoms (DOFs) in total, as shown in
Figure \ref{meshused}.
%
Figure \ref{Hfnsdispy18750} illustrates the numerical result of
vibration response at the location $A$ (the end of beam tail), where
the vibration amplitude of point $A$ agrees with the benchmark
result in \cite{turek2010numerical} very well. As what we can
observe in Figure \ref{Hfnsdispy18750}, the vibrating frequency and
amplitude stabilize after about 15 seconds. Therefore, we set the
total simulation time $T=15s$ in the following tests.

\subsection{Comparisons of FOM and ROM}\label{sec::comfomrom}
\begin{figure}[H]
    \centering
    \includegraphics[width = 7cm, height = 5cm]{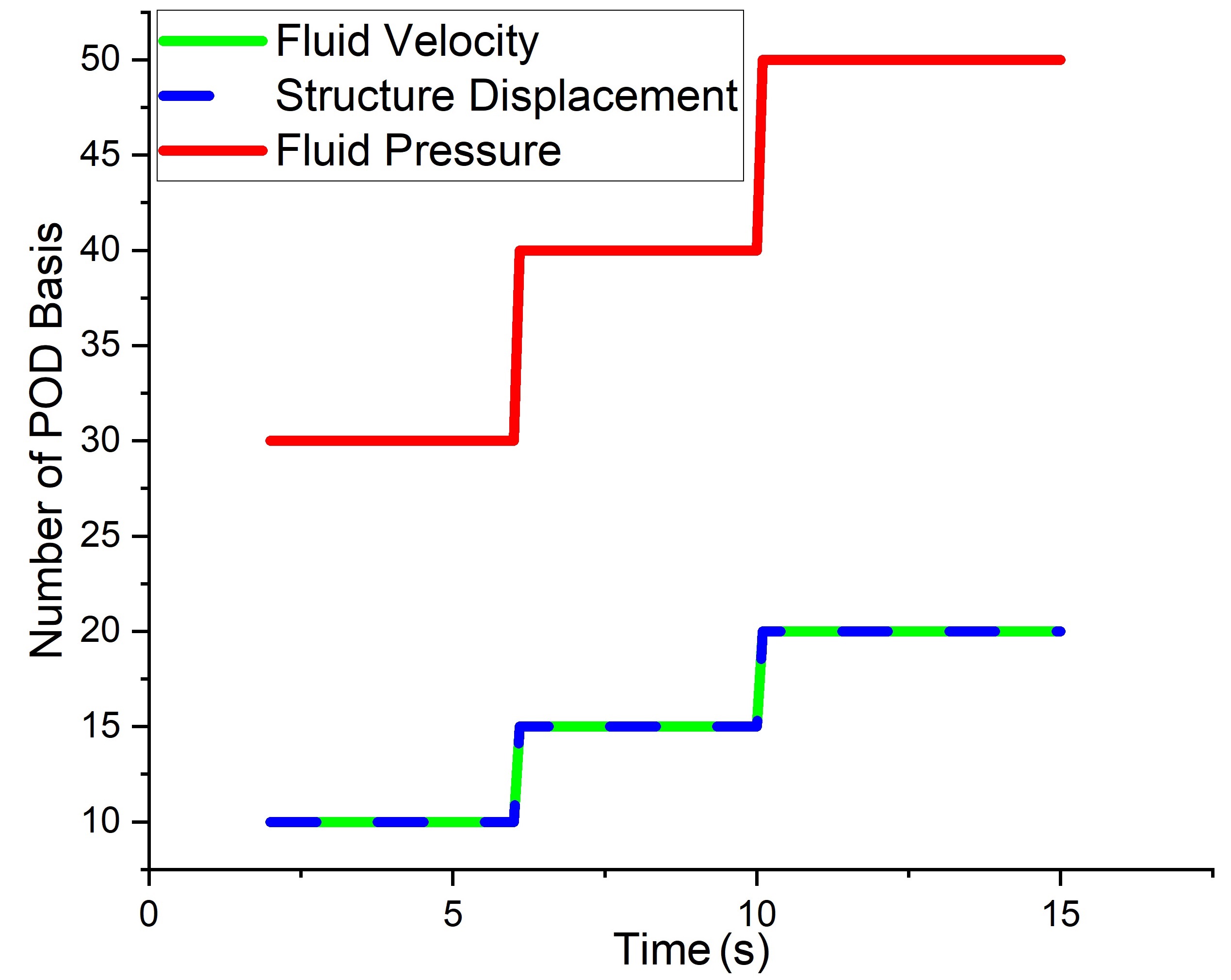}
    \caption{Number of POD bases' selections over the time interval [2s,15s].}
    \label{Numbasis1e3}
\end{figure}

In this subsection, we apply the ROM developed in Section \ref{S3}
to solve the same FSI benchmark problem on the same mesh and with
the same time step size $\Delta t=0.001s$.
We do not reduce the order in the first two seconds, i.e., ROM is
not used during the initial inflow buffer time $(t<2s)$.
Thereafter, we divide the entire time interval $[2s,15s]$ into $130$
time segments with an equal width $0.1s$, and each time segment
contains 100 time steps, i.e., there are $N_{T_g}=101$ time points
in each $[T_g,T_{g+1}]$ with a time step size $\Delta t=0.001s$,
where $T_g=2+g*0.1$, $g=0,1,\cdots,129$. Then during the offline
phase, in three time subintervals
$[2s,6s]=\bigcup\limits_{g=0}^{39}[T_g,T_{g+1}]$,
$[6s,10s]=\bigcup\limits_{g=40}^{79}[T_g,T_{g+1}]$, and
$[10s,15s]=\bigcup\limits_{g=80}^{129}[T_g,T_{g+1}]$ we construct
the POD bases by choosing $N_{\bu_f}=N_{\hat{\bm d}_s}=10, 15$ and
20, and $N_{p_f}=30, 40$ and 50, respectively, as illustrated in
Figure \ref{Numbasis1e3}.
In particular, we let $N_{\hat{\bm m}_f}=N_{T_g}$ in each time
subinterval here, which means we do not utilize the ROM but just the
FOM to solve ALE mapping for the sake of ensuring the generated
moving fluidic mesh shape-regular all the time.
Then, we use the obtained POD basis to carry out the ROM for solving
the FSI benchmark problem in each $[T_g,T_{g+1}]\
(g=0,1,\cdots,129)$ during the online phase, where the ROM solution
obtained by reducing the order at the end of the previous time
segment $[T_{g-1},T_{g}]$ is adopted as the initial value at the
starting point of the current time segment $[T_g,T_{g+1}]$.
Specifically, the case of $g=0$ is referred to the last paragraph of
Section \ref{sec:online}. Finally, we attain the solution of ROM
over the entire time period $[0,15s]$.

\begin{figure}[H]
    \centering
   \subfigure[$2.0 \sim 2.1s$]{\includegraphics[width=3.6cm, height =2.4cm]{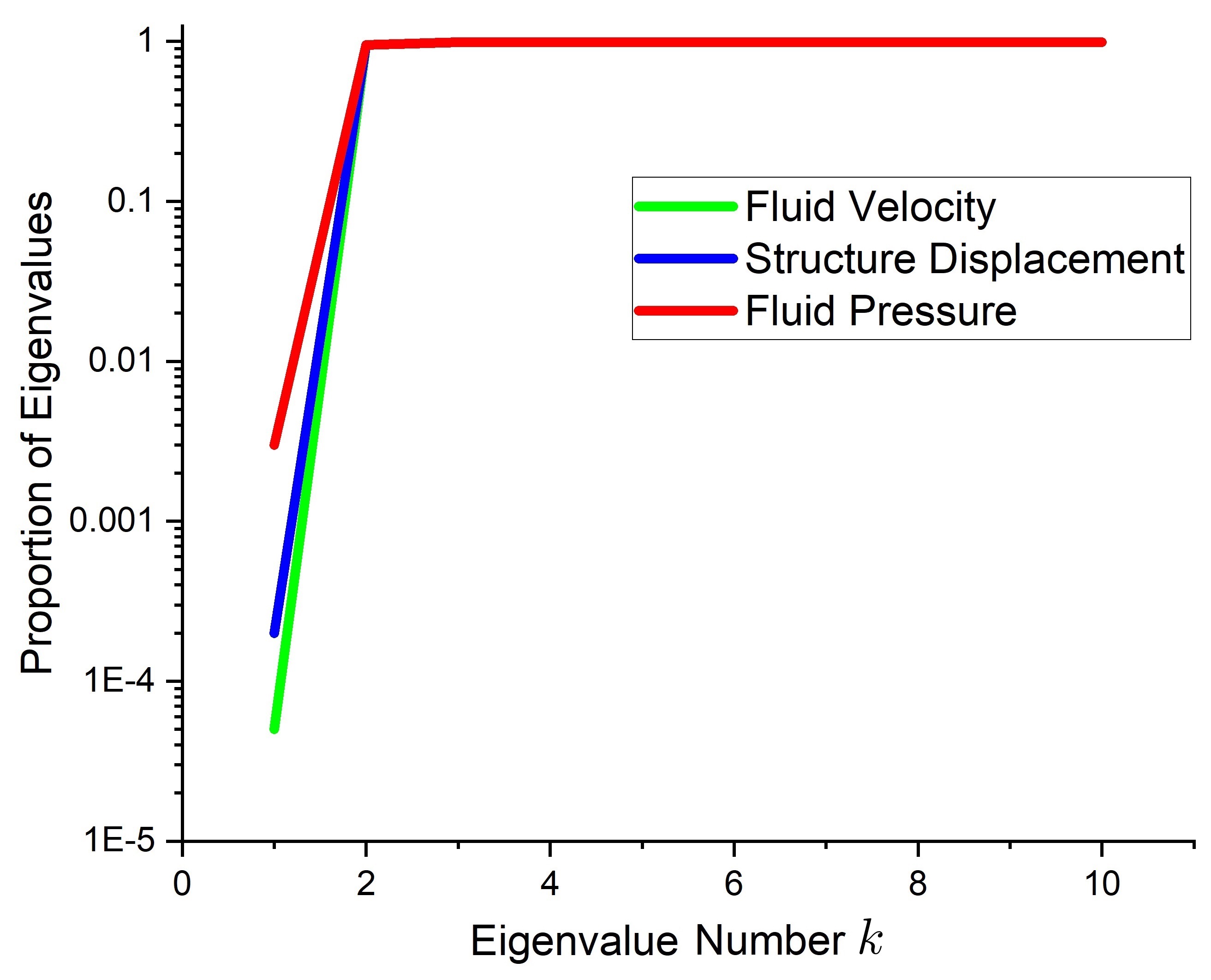}}
    \subfigure[$6.0 \sim 6.1s$]{\includegraphics[width=3.6cm, height =2.4cm]{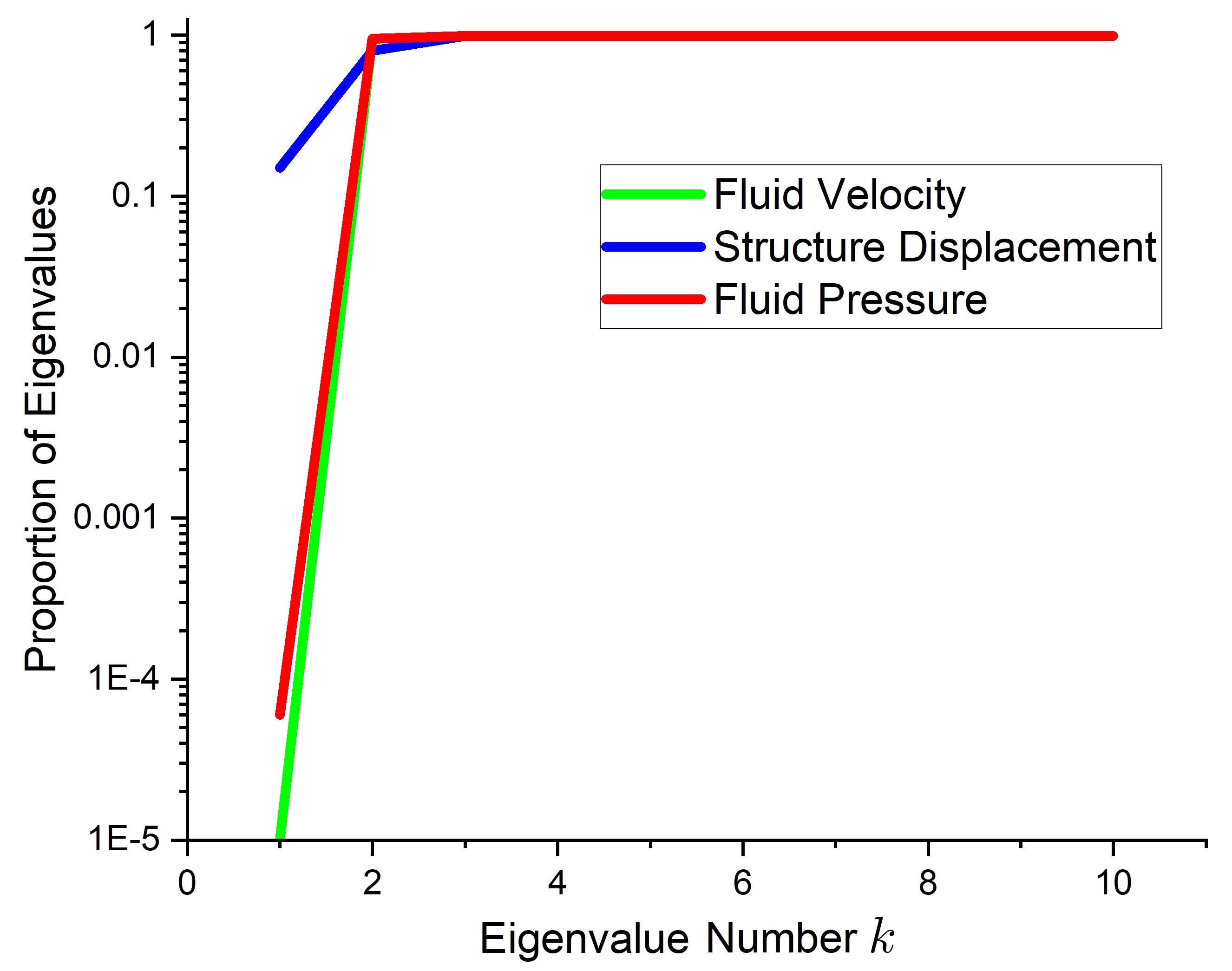}}
\subfigure[$10.0 \sim 10.1s$]{\includegraphics[width=3.6cm, height
=2.4cm]{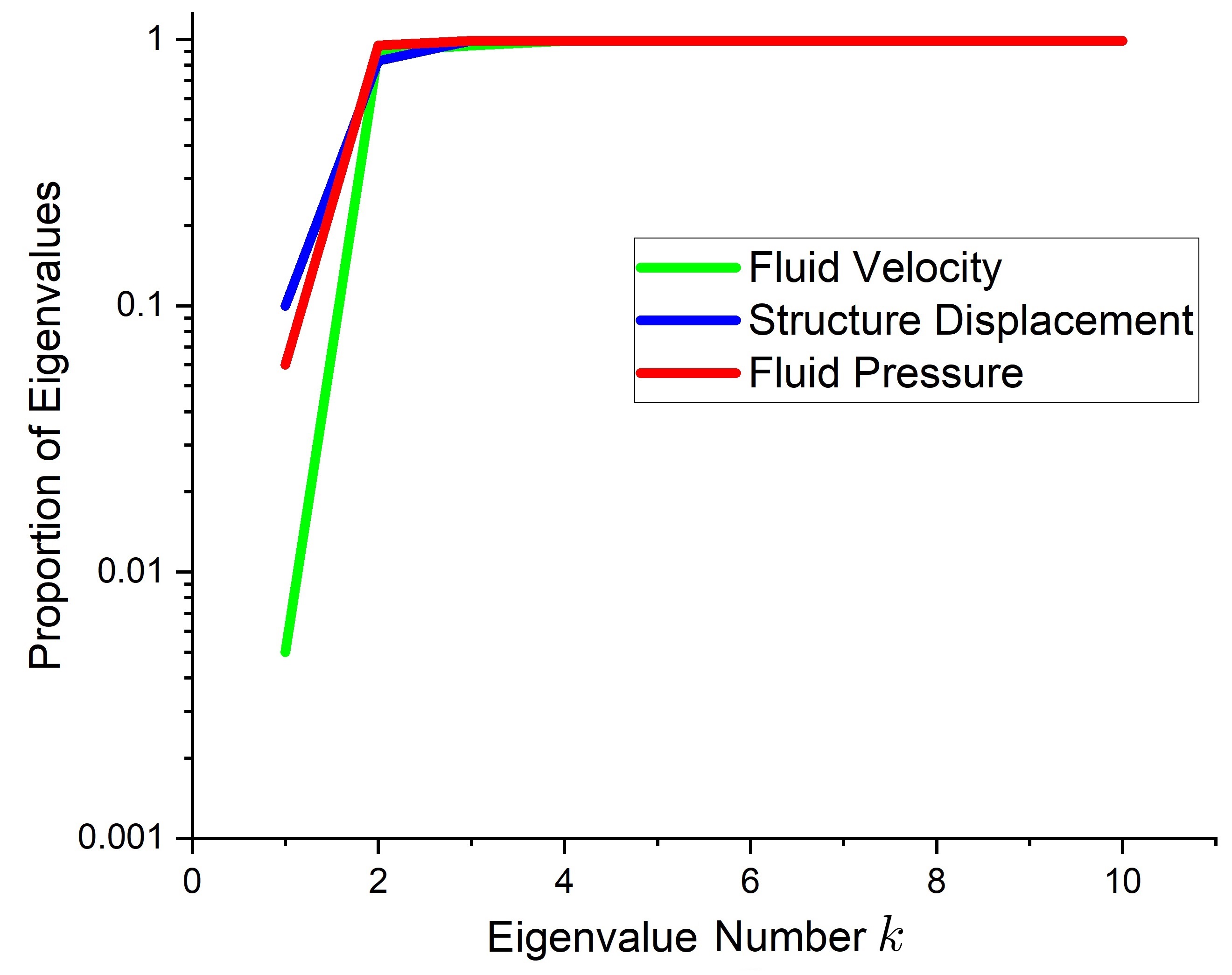}}
    \caption{History of proportion of eigenvalues within typical time segments}
\label{1e3Eigenvalueszhanbidecay}
\end{figure}

In the following, we introduce the proportion of eigenvalues,
$\frac{\sum_{k = 1}^{{N}_*}\lambda_{k}^{*}}{\sum_{k =
1}^{{N}_{T_g}}\lambda_{k}^{*}}$, where $* \in \{\bm{u}_{f}, p_{f},
\hat{\bm{d}}_s\}$ and $\lambda_{k}^{*}$ denotes the $k$-th
eigenvalue of $\mathcal{C}_*$ in $[T_{g},T_{g+1}]$, to demonstrate
the proportion of the sum of all selected eigenvalues amongst the
sum of all eigenvalues in each time segment. We further introduce
the following three error indicators that are used later to compare
numerical results between the FOM and ROM:
\begin{enumerate}
    \item The relative spatial $L^2$ error between the FOM solution
$\bm{U}_{g}^{n}$ and the ROM solution $\mathbf{u}_{h}^{n,rom}$,
$\frac{\|\bm{U}_{g}^{n} -
\mathbf{u}_{h}^{n,rom}\|_{L^2(\Omega)}}{\|\bm{U}_{g}^{n}\|_{L^2(\Omega)}}$;
    \item The relative spatio-temporal $L^2$ error between the FOM solution
$\bm{U}_{g}^{n}$ and the ROM solution $\mathbf{u}_{h}^{n,rom}$,
$\frac{\|\bm{U}_{g}^{n} -
\mathbf{u}_{h}^{n,rom}\|_{L^2(0,T;L^2(\Omega))}}{\|\bm{U}_{g}^{n}\|_{L^2(0,T;L^2(\Omega))}}$;
    \item The absolute error of $y$-displacement of point $A$ between the FOM
result $Dy_{fom}$ and the ROM result $Dy_{rom}$, expressed as
$(Dy_{fom} - Dy_{rom})$.
\end{enumerate}
Figure \ref{1e3Eigenvaluesdecay} shows the eigenvalue changes
associated with three variables: fluid velocity, fluid pressure and
structural displacement within several typical time segments.
Through Figure \ref{1e3Eigenvaluesdecay} we intend to reveal the
following fact that in each time segment $[T_g, T_{g+1}]$, these
$N_{T_g}$ vectors, $\left(\bm{U}_{g}^{1},
\bm{U}_{g}^{2},\cdots,\bm{U}_{g}^{N_{T_g}}\right)$, are linearly
dependent, leading to a low-rank snapshot matrix $\bm{U}_g$, which
explains why ROM works in this scenario
\cite{alfioquarteroni_2014_reduced}. Figure
\ref{1e3Eigenvalueszhanbidecay} illustrates proportions of
eigenvalues of the first 10 eigenvalues that are associated with
three variables within several typical time segments, which helps us
to determine how many POD bases need to be selected for each
variable, at least. In addition, from Figures
\ref{1e3Eigenvaluesdecay} and \ref{1e3Eigenvalueszhanbidecay} we
also observe the following facts:
\begin{figure}[H]
    \centering
   \subfigure[$2.0 \sim 2.1s$]{\includegraphics[width=3.6cm, height =2.4cm]{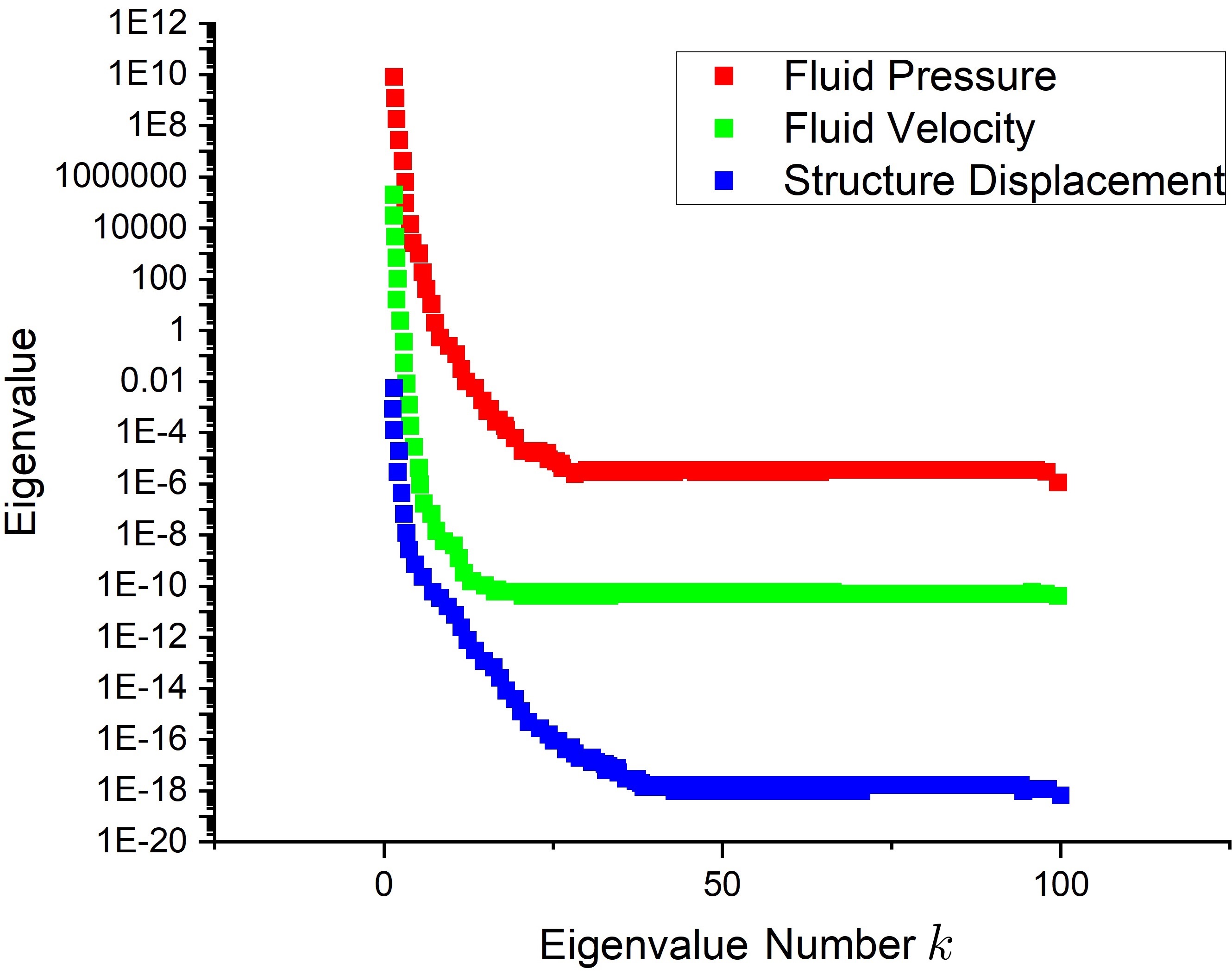}}
    \subfigure[$6.0 \sim 6.1s$]{\includegraphics[width=3.6cm, height =2.4cm]{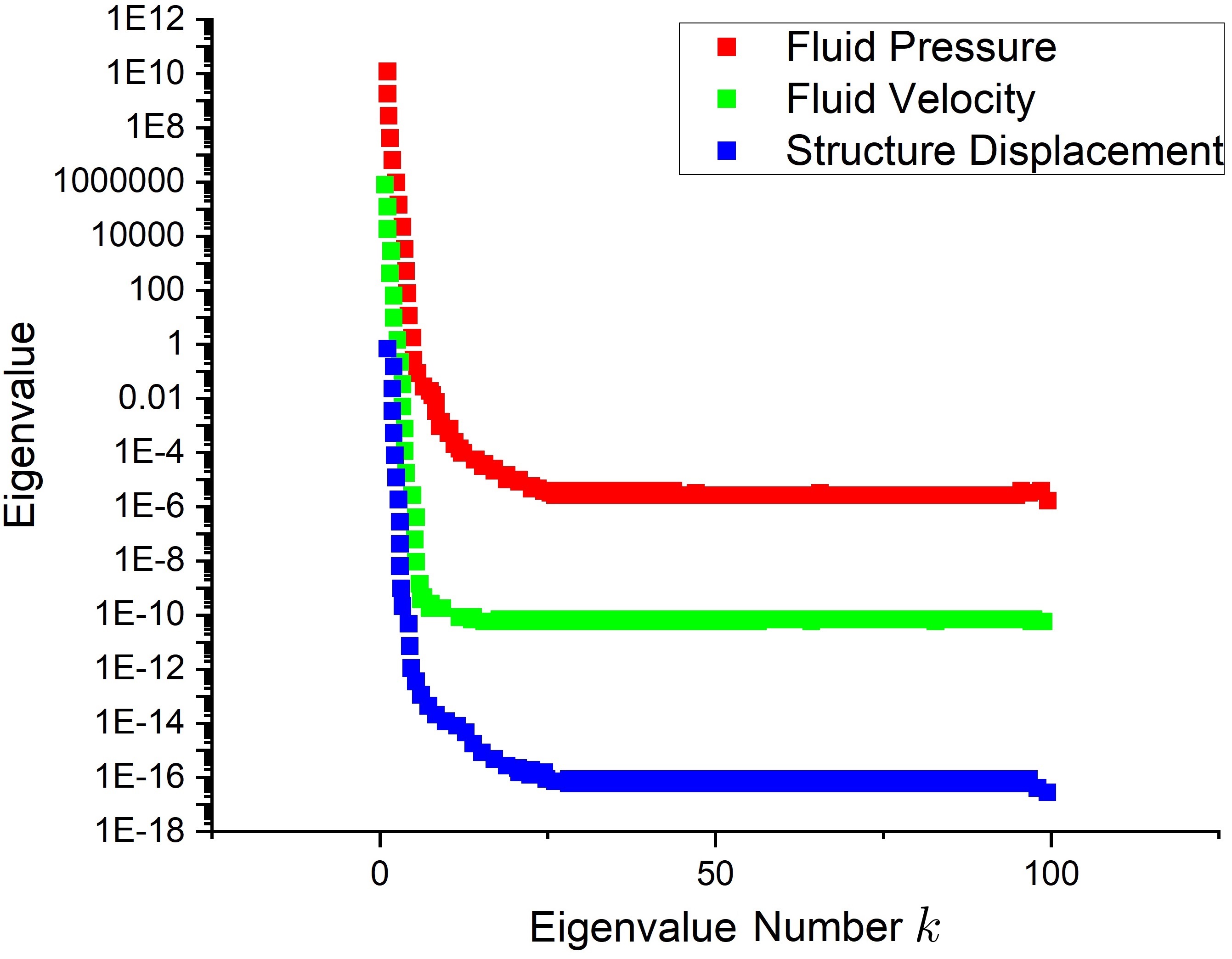}}
\subfigure[$10.0 \sim 10.1s$]{\includegraphics[width=3.6cm,
height=2.4cm]{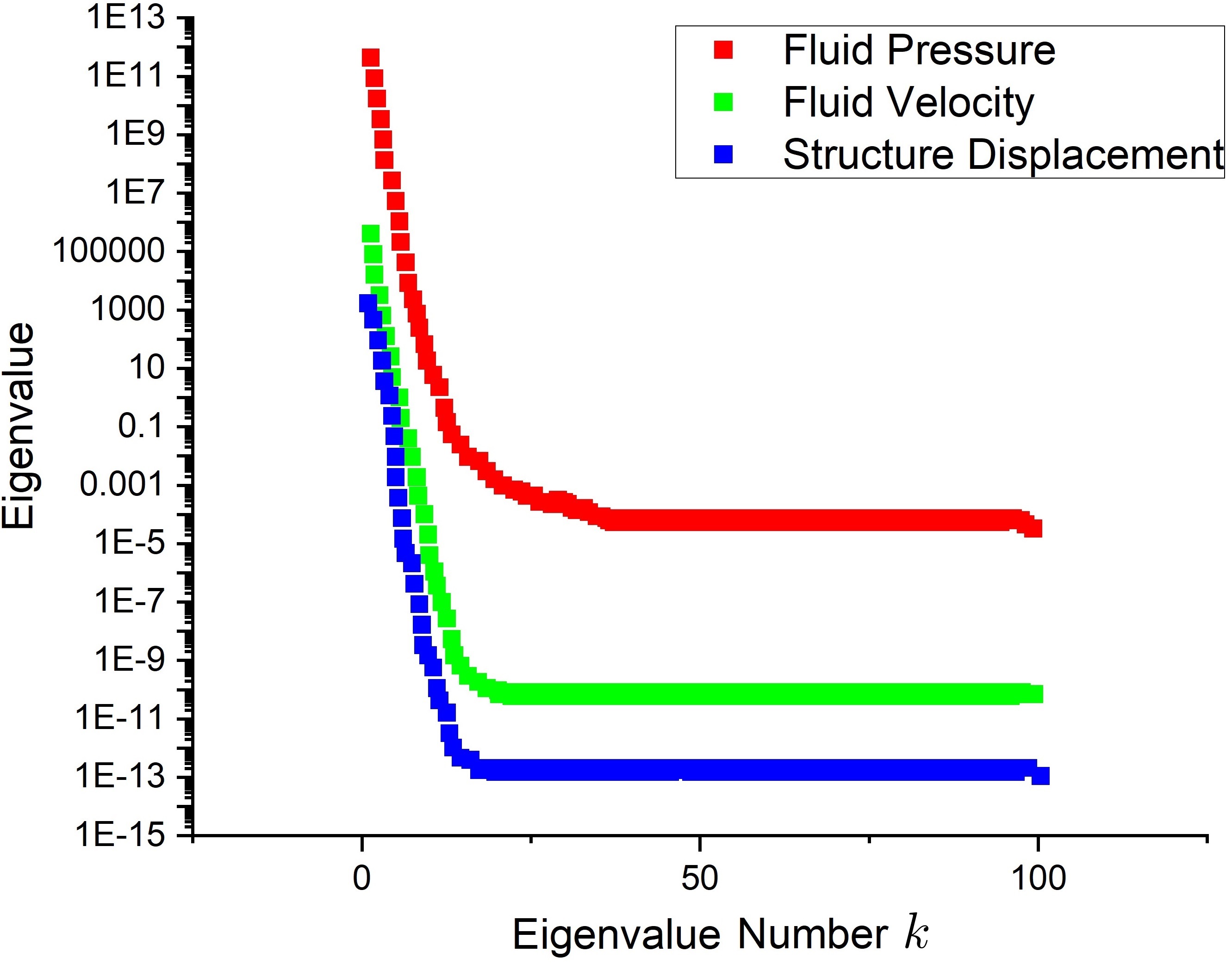}}
    \caption{History of eigenvalue decay within typical time segments}
\label{1e3Eigenvaluesdecay}
\end{figure}
\begin{itemize}
\item Within the same time period, the eigenvalue decay rates of both fluid
velocity and fluid pressure are closer to each other, while they are
significantly slower than that of the structural displacement;
\item The eigenvalue decay rates of all variables slow down when time marches (i.e., the eigenvalue number grows);
\item When the number of POD bases is chosen larger than 5,
proportions of eigenvalues for different variables are all close to
1;
\item The eigenvalue number with high proportion increases with time,
i.e., the bigger eigenvalue number, the higher proportion of
eigenvalues.
 \end{itemize}

Table \ref{details-S2} shows the comparison between the FOM and ROM
in terms of the number of DOFs and computational time during
different time periods, where the computational time is the time
taken to solve the linear algebraic system during that time period.
We can see that the ROM greatly saves the computational time for
almost $100\%$ corresponding to the FOM, and averagely speed up the
linear algebraic solver 8844 times, which is a huge improvement on
the computational efficiency.
\begin{table}[H]
   \centering
 \begin{tabular}{|c|c|c|c|}
 \hline
    Time subinterval & $2 \sim 6s$ & $6 \sim 10s$ & $10 \sim 15s$  \\
\hline
%
\# of DOFs of FOM & 18350 & 18350 & 18350 \\ \hline \# of DOFs of
ROM & 50 & 70 & 90 \\ \hline
Reduction Rate in \# of DOFs & 367:1 &  263:1  & 204:1  \\
\hline Comput. Time of FOM (s) & $5.52\times 10^{4}$& $5.52\times
10^{4}$ & $6.91\times 10^{4}$\\ \hline Comput. Time of ROM (s) &
5.62 &  6.25 & 8.77 \\ \hline Reduction Rate in Comput. Time&
99.990\% &  99.989\% & 99.987\% \\ \hline
Speedup & $9.822\times 10^3$ & $8.832\times 10^3$  & $7.879\times 10^3$ \\
    \hline
 \end{tabular}
  \caption{Numerical comparison between the FOM and ROM in terms of \# of DOFs
  and computational time during different time periods.}\label{details-S2}
 \end{table}

\begin{figure}[H]
    \centering
    \includegraphics[width = 7cm, height = 5cm]{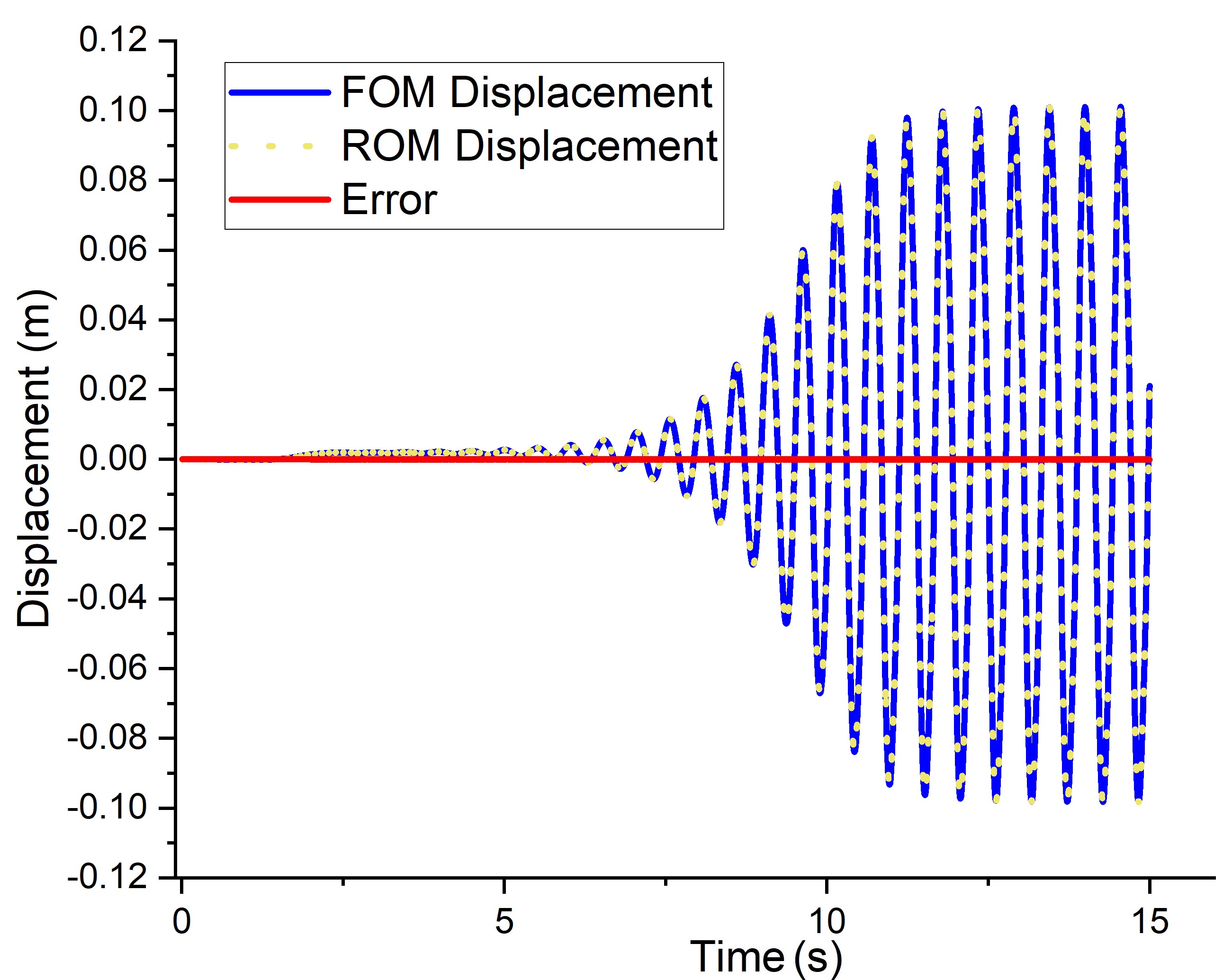}
    \caption{\textbf{Blue solid}: the $y$-displacement curve of FOM at point $A$;
    \textbf{Yellow dot}:the $y$-displacement curve of ROM at point $A$;
    \textbf{Red solid}: Error curve of $y$-displacement at point $A$ between FOM and ROM,
    $Dy_{fom} - Dy_{rom}$, over the entire time interval.}
    \label{ROMdispy18750error}
\end{figure}

\begin{figure}[H]
\centering
   \subfigure[Horizontal velocity of FOM]{\includegraphics[width=6cm, height =1.6cm]{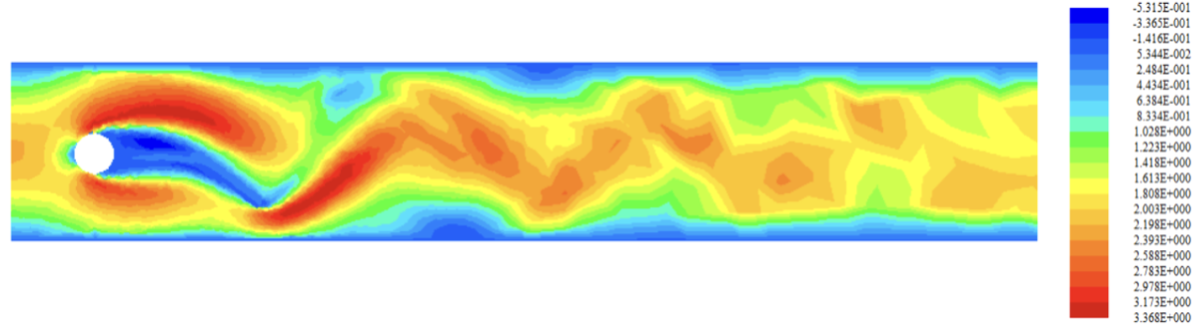}}
    \subfigure[Horizontal velocity of ROM]{\includegraphics[width=6cm, height =1.6cm]{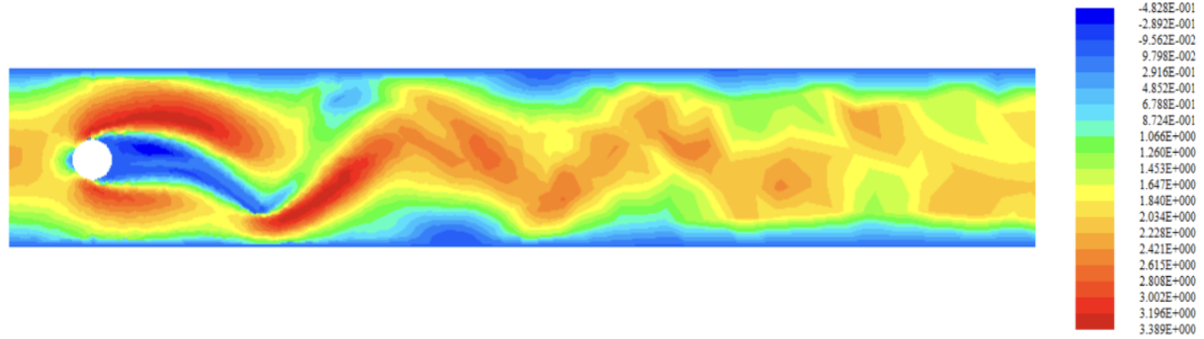}}
\subfigure[Vertical velocity of FOM]{\includegraphics[width=6cm,
height =1.6cm]{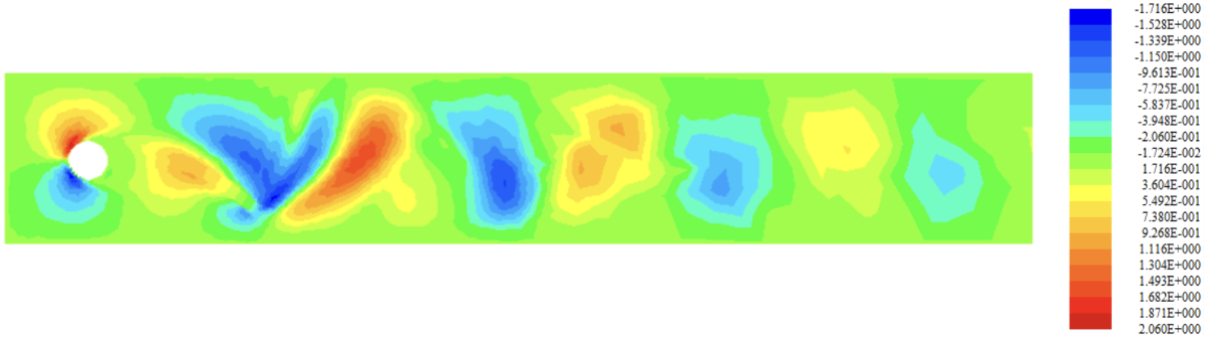}}
    \subfigure[Vertical velocity of ROM]{\includegraphics[width=6cm, height =1.6cm]{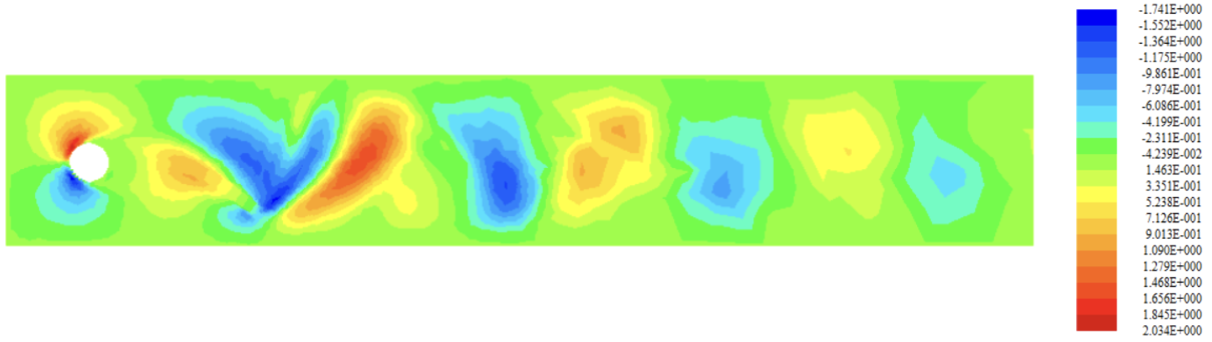}}
\subfigure[Fluid pressure of FOM]{\includegraphics[width=6cm, height
=1.8cm]{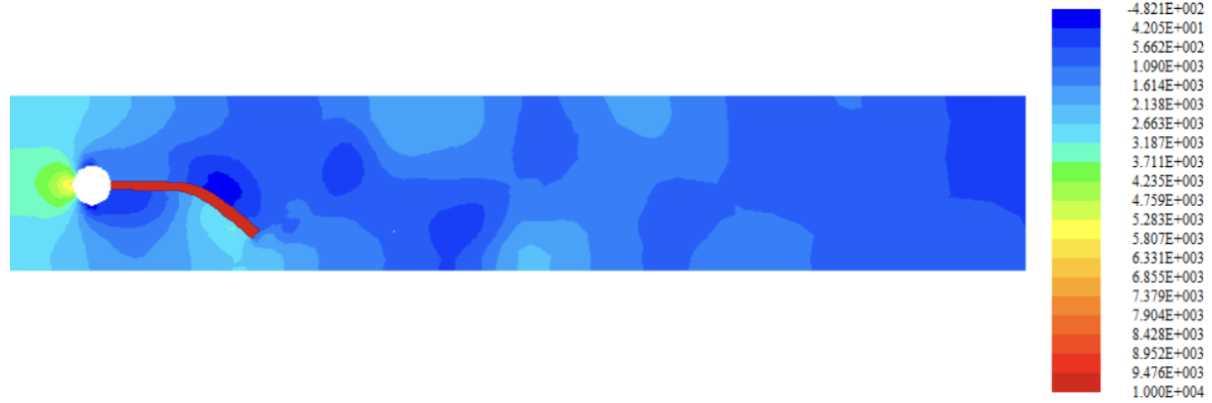}}
    \subfigure[Fluid pressure of ROM]{\includegraphics[width=6cm, height =1.8cm]{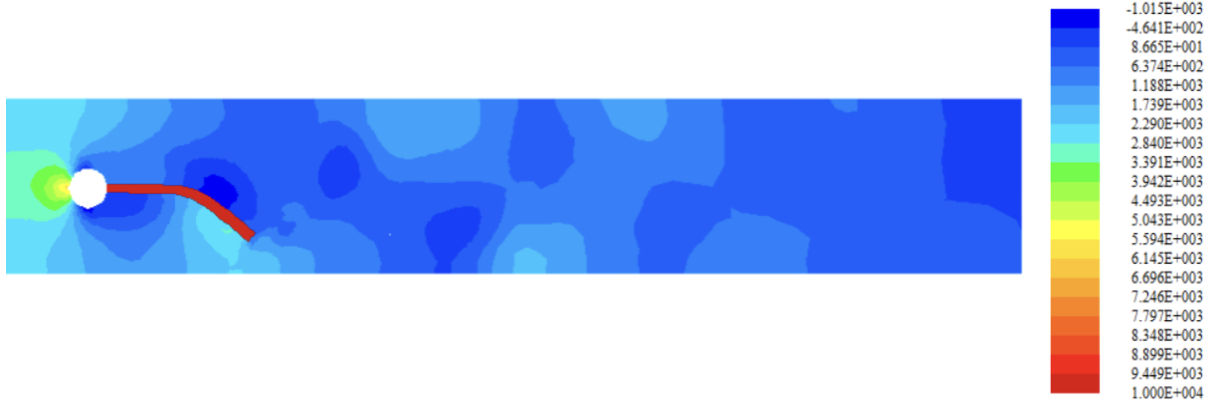}}
    \caption{FSI solutions between FOM and ROM at the terminal time $t = 15s$.}
    \label{Representativesolutions4751}
\end{figure}

\begin{figure}[H]
\centering
   \subfigure[Total error.]{\includegraphics[width=4.5cm, height =3cm]{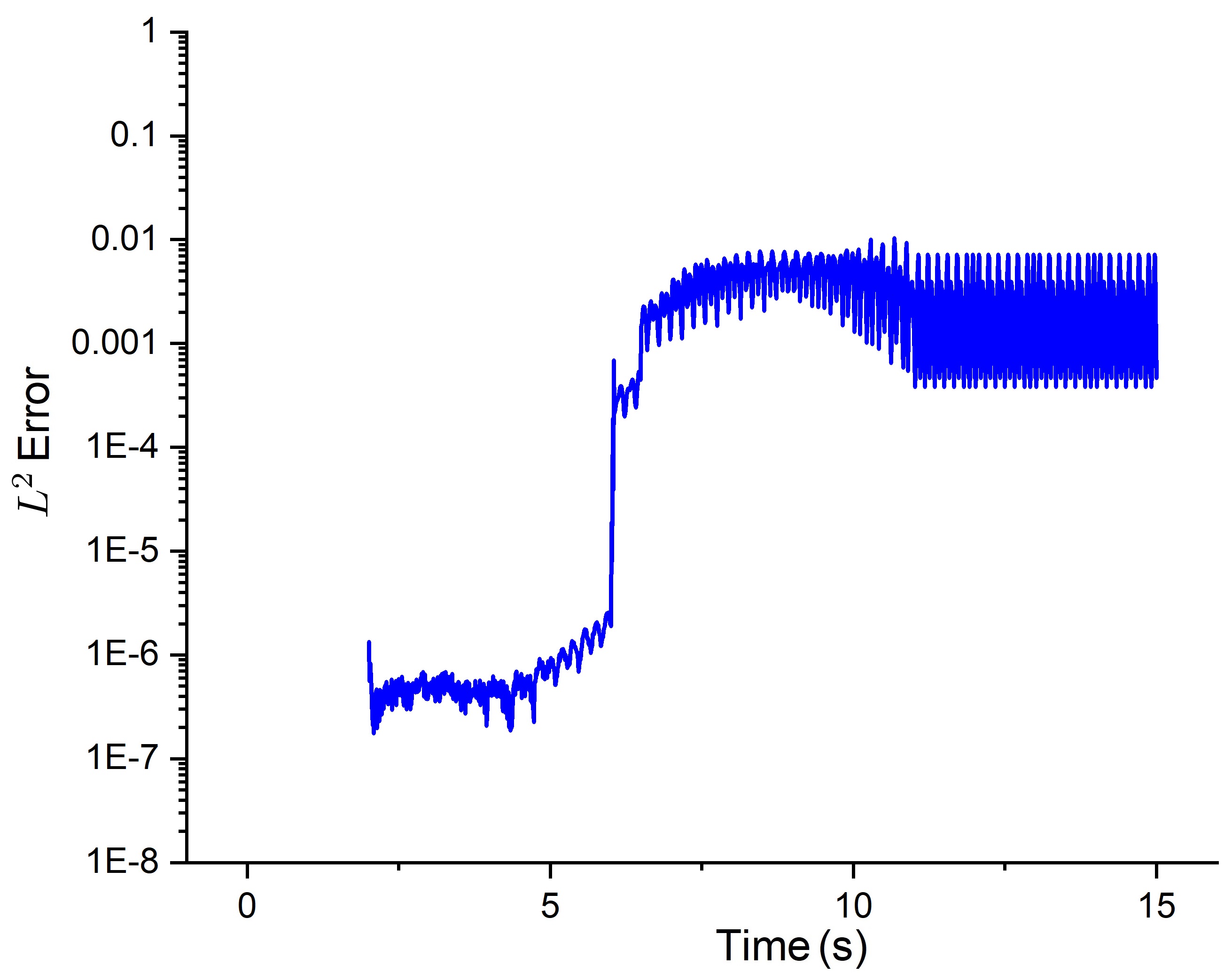}}
   \hspace{1cm}
    \subfigure[Errors of all individual variables combining with total error.]{\includegraphics[width=4.5cm, height =3cm]{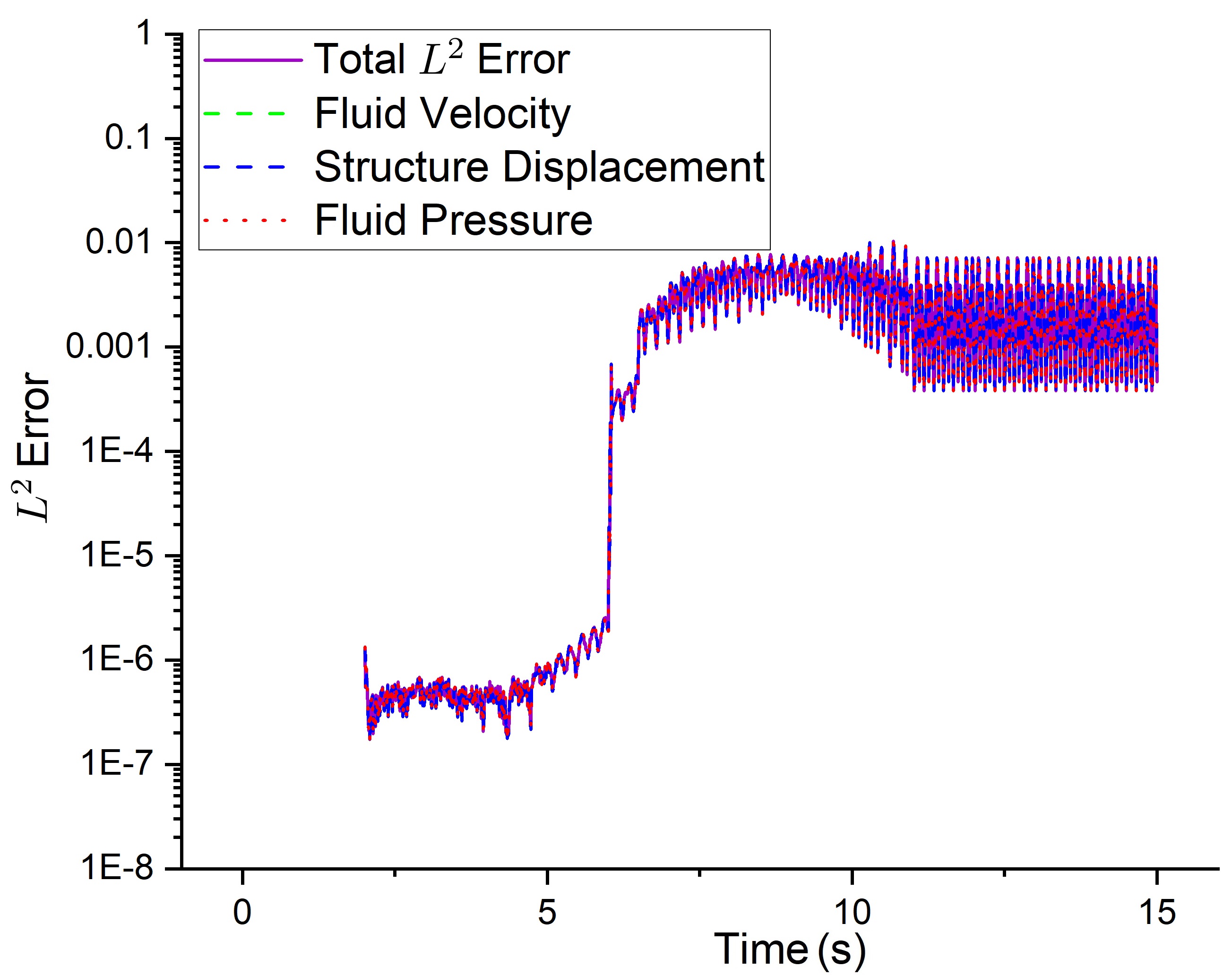}}
    \caption{Relative spatial $L^2$ errors between ROM and FOM over the entire time interval.}
    \label{errorrom18750}
\end{figure}

In regard to the numerical accuracy of ROM versus that of FOM, we
investigate solution errors between the ROM and FOM shown in Figures
\ref{ROMdispy18750error}-\ref{errorrom18750}, and observe the
following numerical phenomena:
\begin{itemize}
\item The vibration curve of the beam tail end, i.e., the $y$-displacement of
point $A$ with time, has a good match with that of the FOM, and the
error $(Dy_{fom} - Dy_{rom})$ is well controlled within $\pm0.002$;
\item The contour snapshots of velocity and fluid pressure at T=15s are
similar between the FOM and ROM. The total relative spatial $L^{2}$
error is well controlled within $0.01$;
\item The relative spatial $L^2$ error of fluid pressure is similar to the total
relative spatial $L^2$ error, while that of fluid velocity and
structural displacement are slightly smaller than the total relative
spatial $L^{2}$ error, meaning that the ROM approximation to the
fluid pressure is less accurate than that to the fluid velocity and
structural displacement.
\item As time marches, errors also grow.
But as the beam vibration stabilizes, the relative spatial $L^{2}$
errors for all variables finally stabilize;
\item The relative spatio-temporal $L^2$ error is $0.0056$.
 \end{itemize}

\begin{figure}[H]
    \centering
    \includegraphics[width = 7cm, height = 5cm]{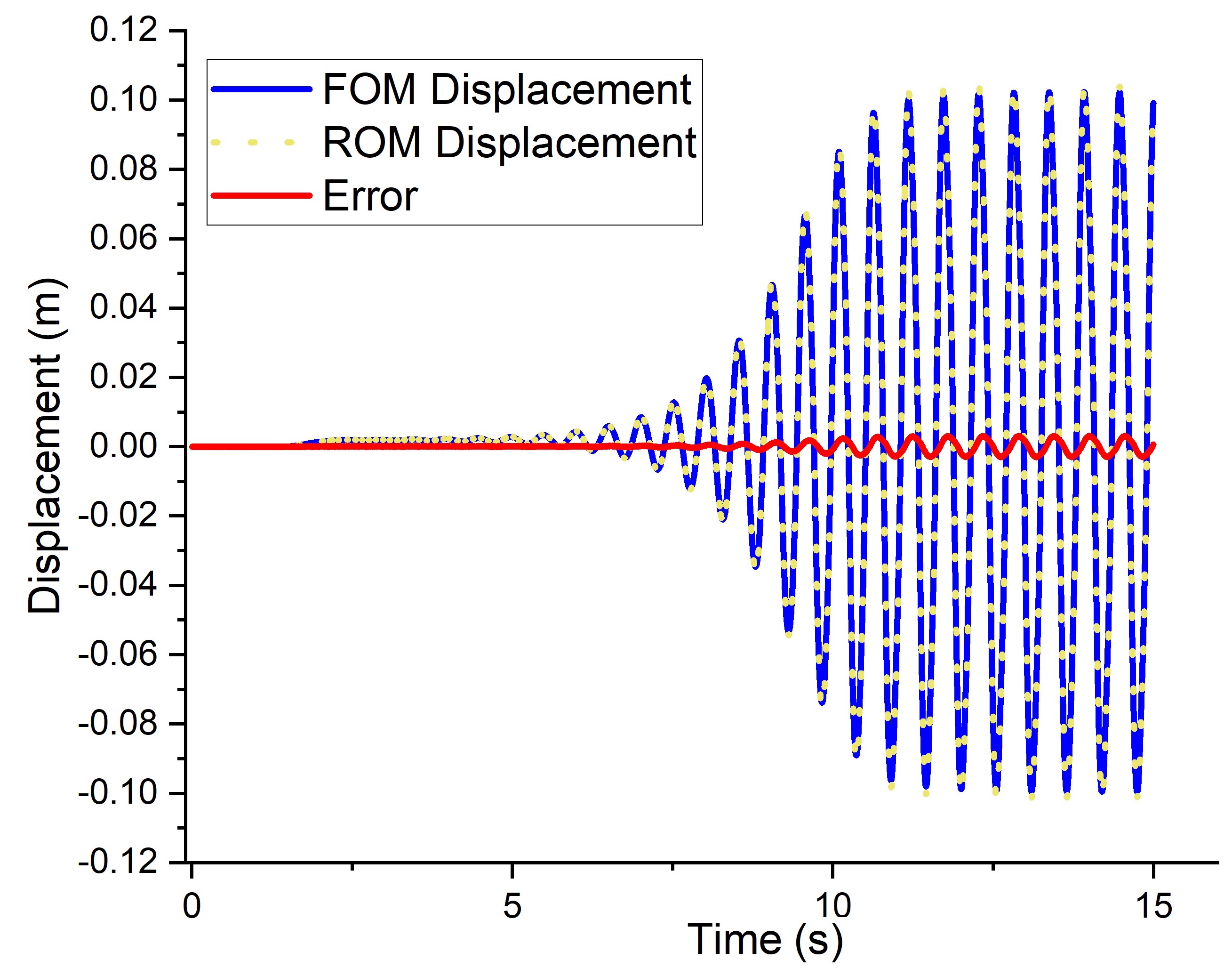}
    \caption{Error of $y$-displacement at point $A$ between FOM and ROM for Case $\hat{U}=1.52$.}
    \label{ROMdispy18750uin152error}
\end{figure}

\begin{figure}[H]
\centering
   \subfigure[Horizontal velocity of FOM]{\includegraphics[width=6cm, height =1.5cm]{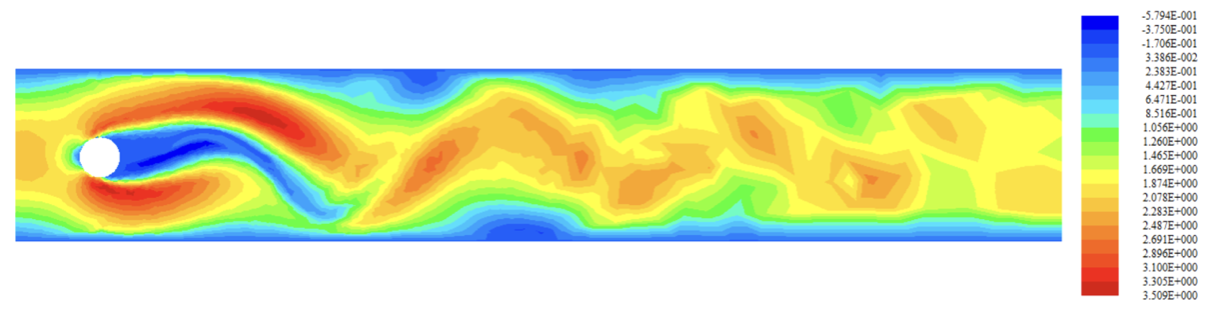}}
    \subfigure[Horizontal velocity of ROM]{\includegraphics[width=6cm, height =1.5cm]{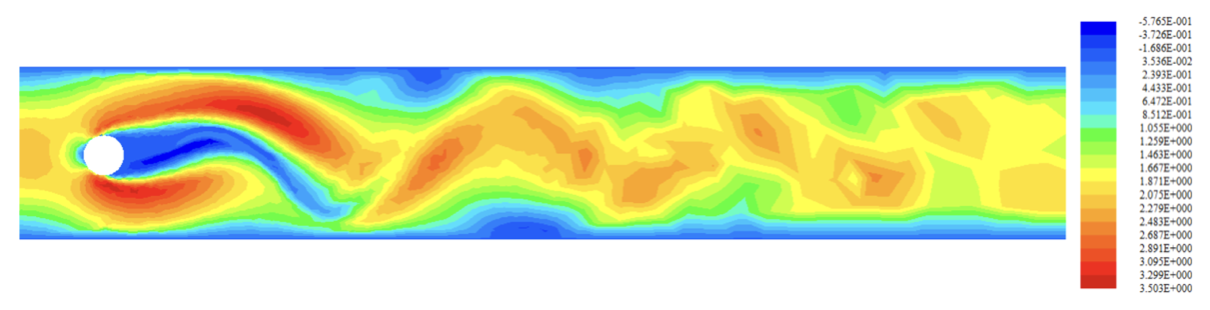}}
\subfigure[Vertical velocity of FOM]{\includegraphics[width=6cm,
height =1.5cm]{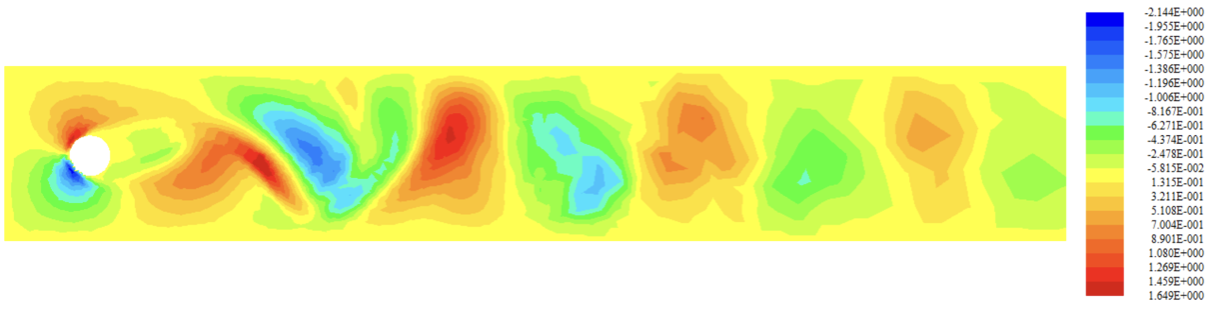}}
    \subfigure[Vertical velocity of ROM]{\includegraphics[width=6cm, height =1.5cm]{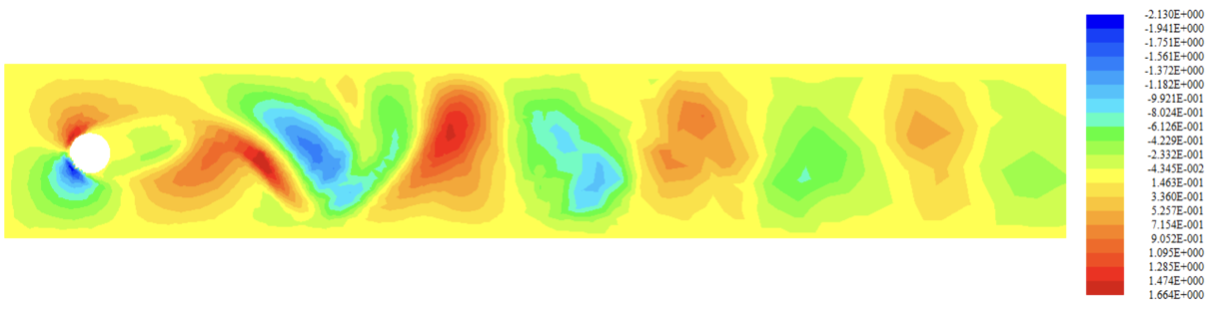}}
\subfigure[Fluid pressure of FOM]{\includegraphics[width=6cm, height
=1.5cm]{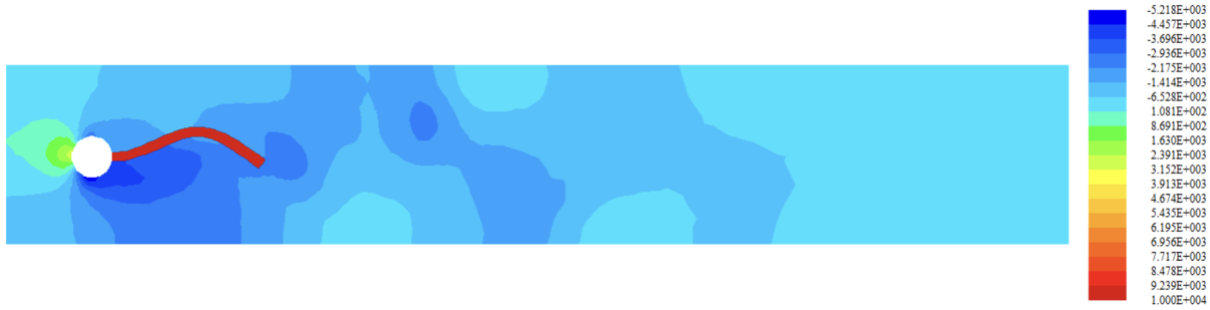}}
    \subfigure[Fluid pressure of ROM]{\includegraphics[width=6cm, height =1.5cm]{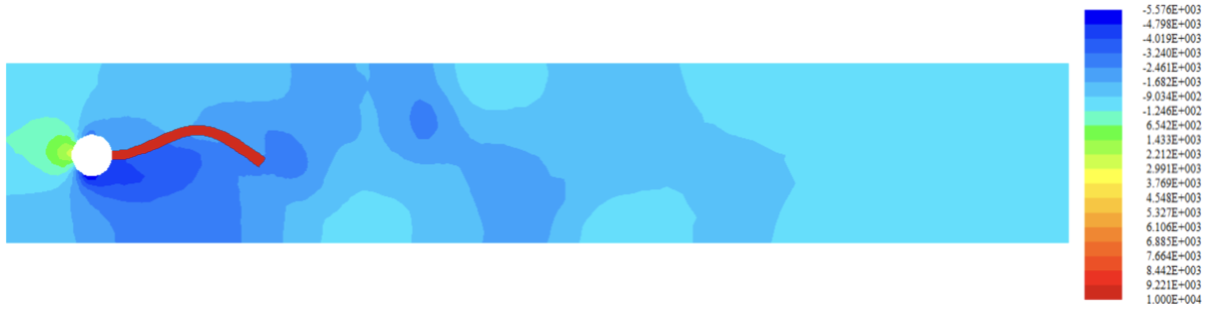}}
    \caption{FSI solutions between FOM and ROM at the finial time $t = 15s$ for Case $\hat{U}=1.52$.}
    \label{Representativesolutionsuin15218350}
\end{figure}

\begin{figure}[H]
\centering
   \subfigure[Total error.]{\includegraphics[width=4.5cm, height =3cm]{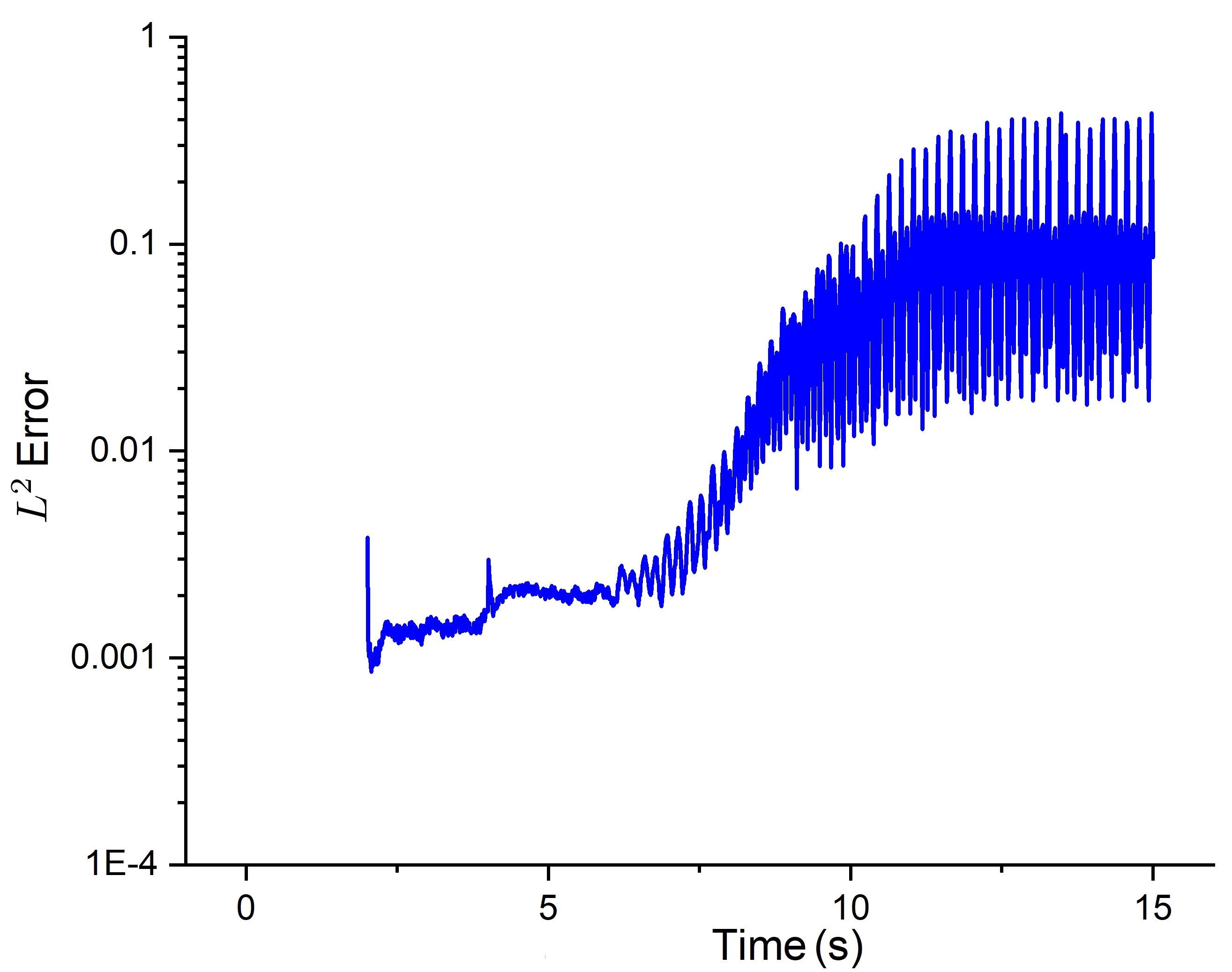}}
   \hspace{1cm}
    \subfigure[Errors of all individual variables combining with total error.]{\includegraphics[width=4.5cm, height =3cm]{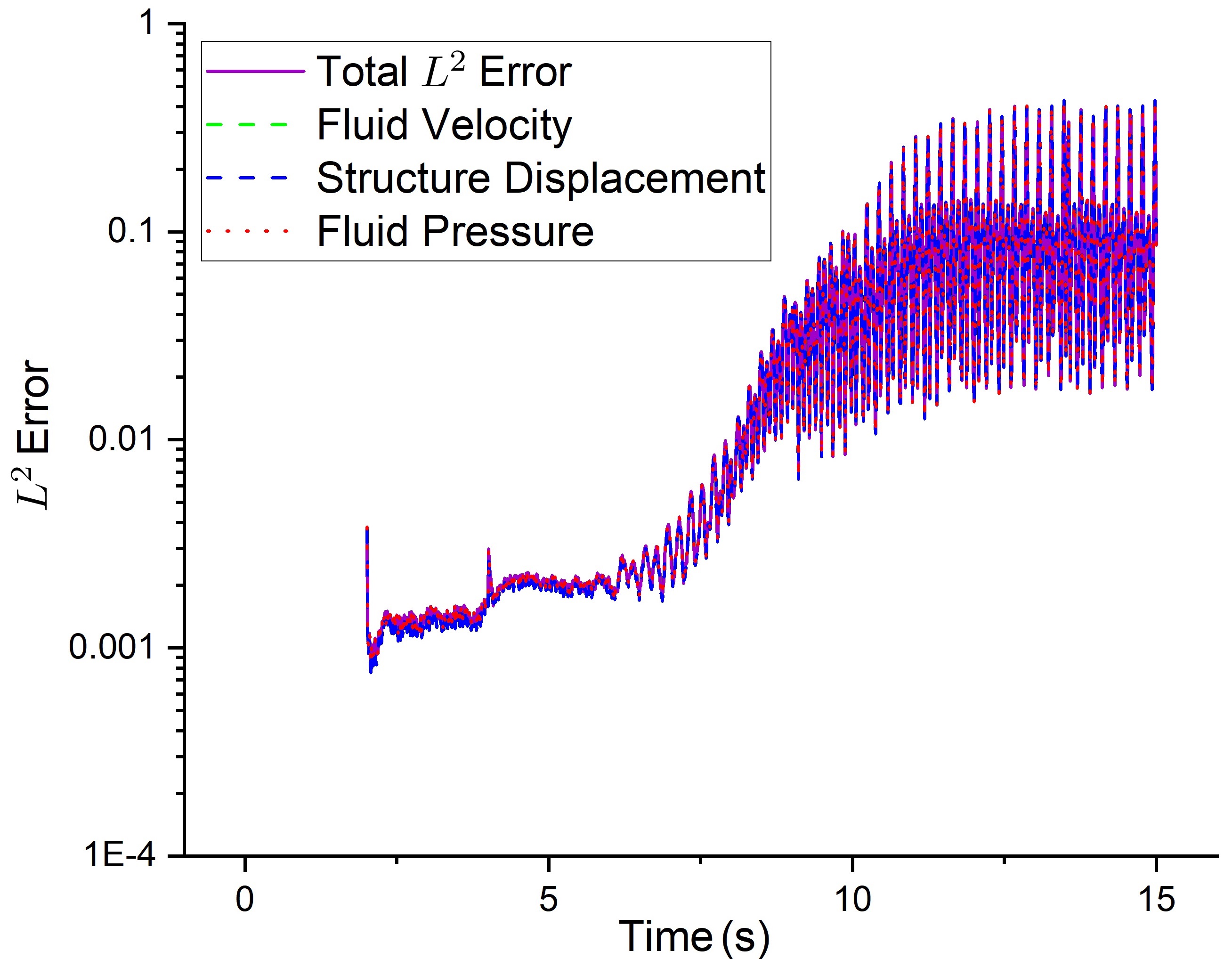}}
    \caption{Relative spatial  $L^2$ errors between ROM and FOM over the entire time interval for Case $\hat{U}=1.52$.}
    \label{errorrom18750uin152}
\end{figure}

\begin{figure}[H]
    \centering
    \includegraphics[width = 7cm, height = 5cm]{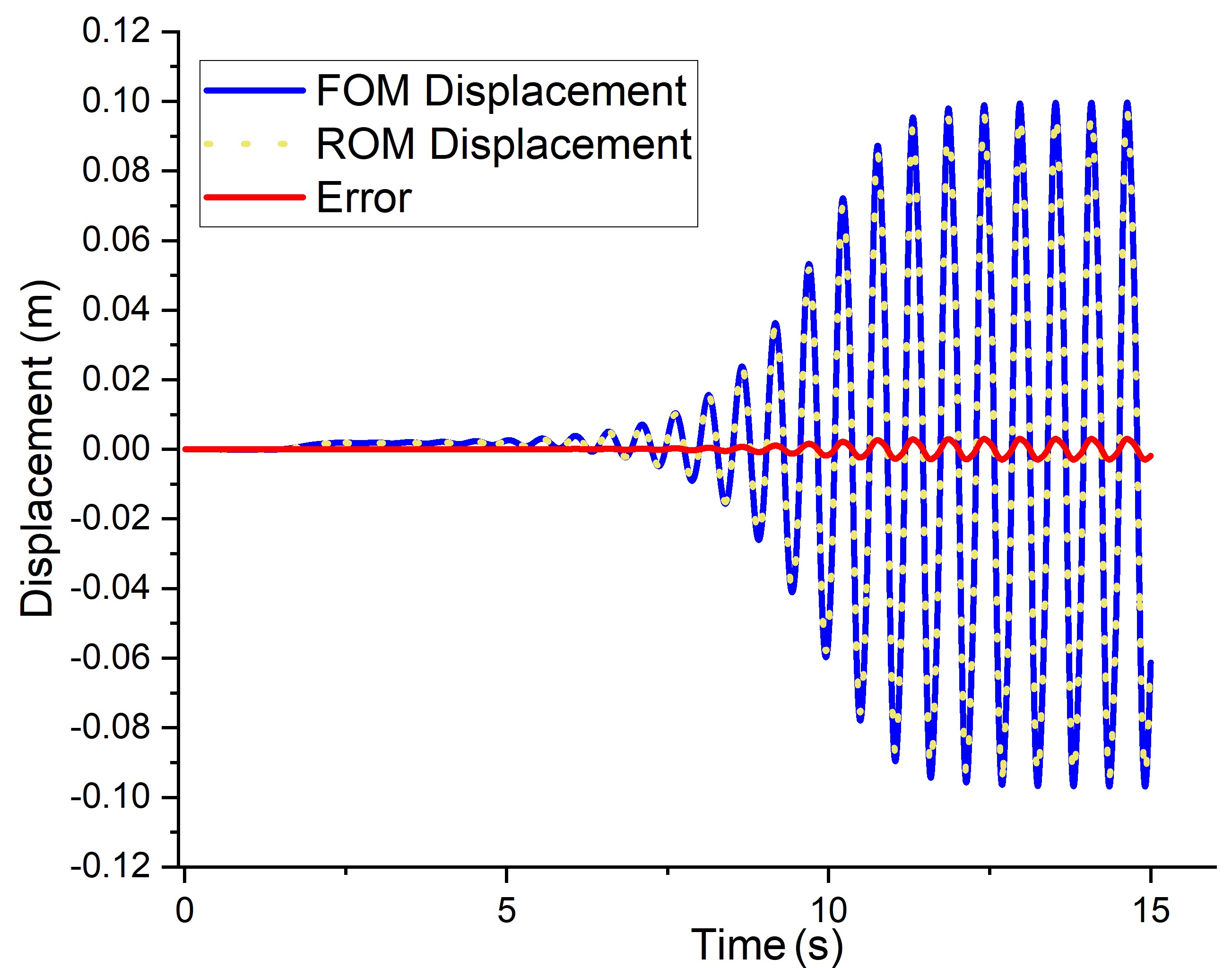}
    \caption{Error of $y$-displacement at point $A$ between FOM and ROM for Case $\hat{U}=1.48$.}
    \label{ROMdispy18750uin148error}
\end{figure}

\begin{figure}[H]
\centering
   \subfigure[Horizontal velocity of FOM]{\includegraphics[width=6cm, height =1.5cm]{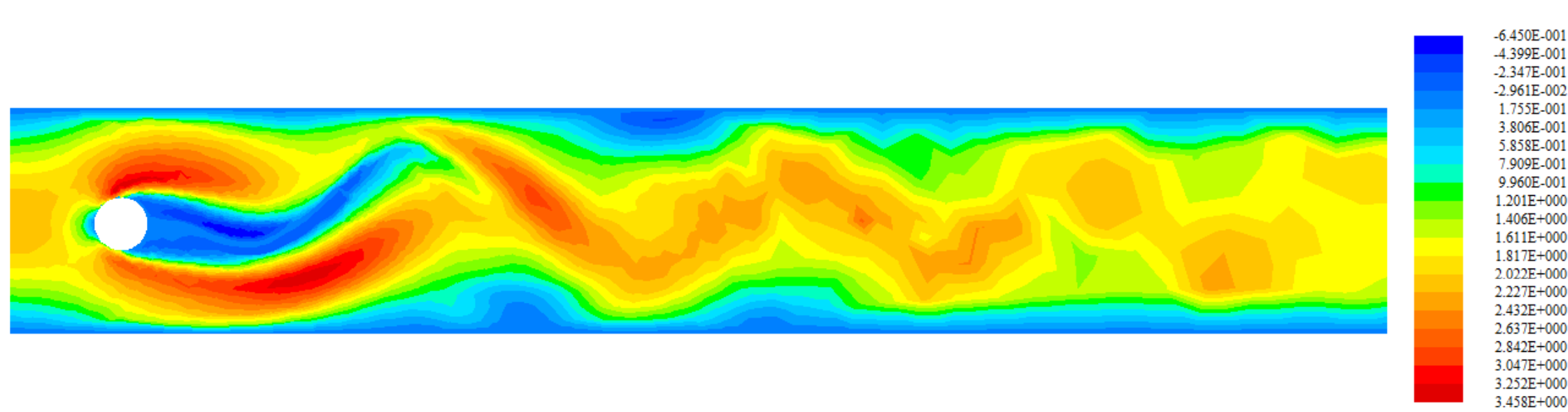}}
    \subfigure[Horizontal velocity of ROM]{\includegraphics[width=6cm, height =1.5cm]{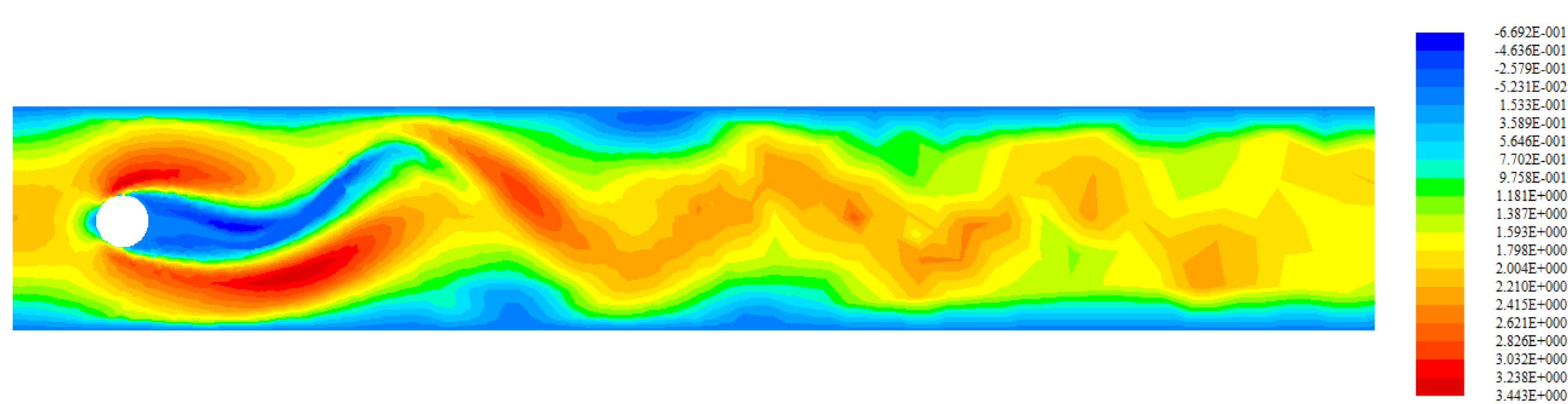}}
\subfigure[Vertical velocity of FOM]{\includegraphics[width=6cm,
height =1.5cm]{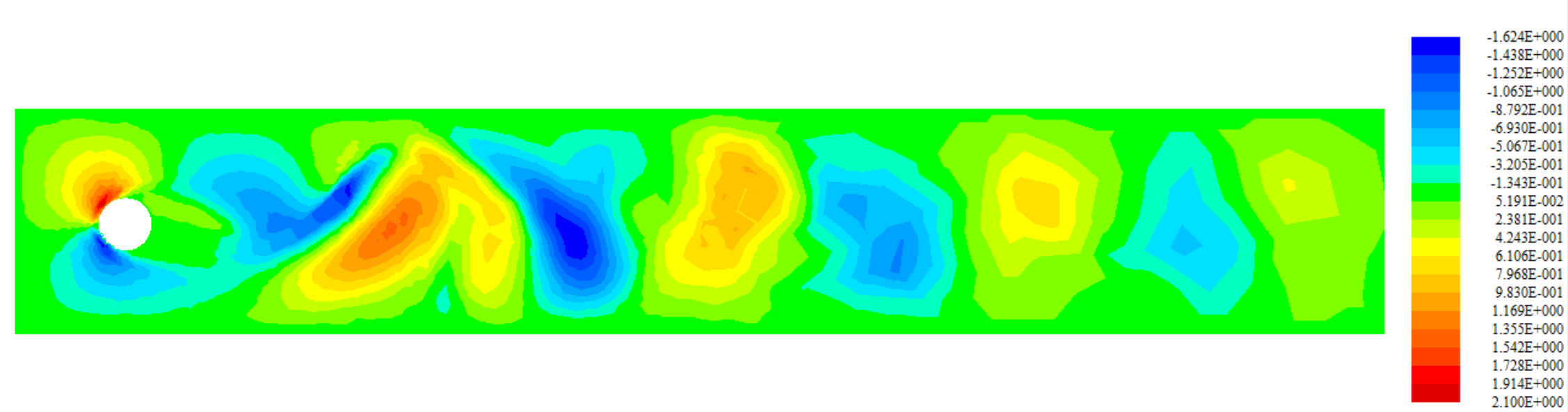}}
    \subfigure[Vertical velocity of ROM]{\includegraphics[width=6cm, height =1.5cm]{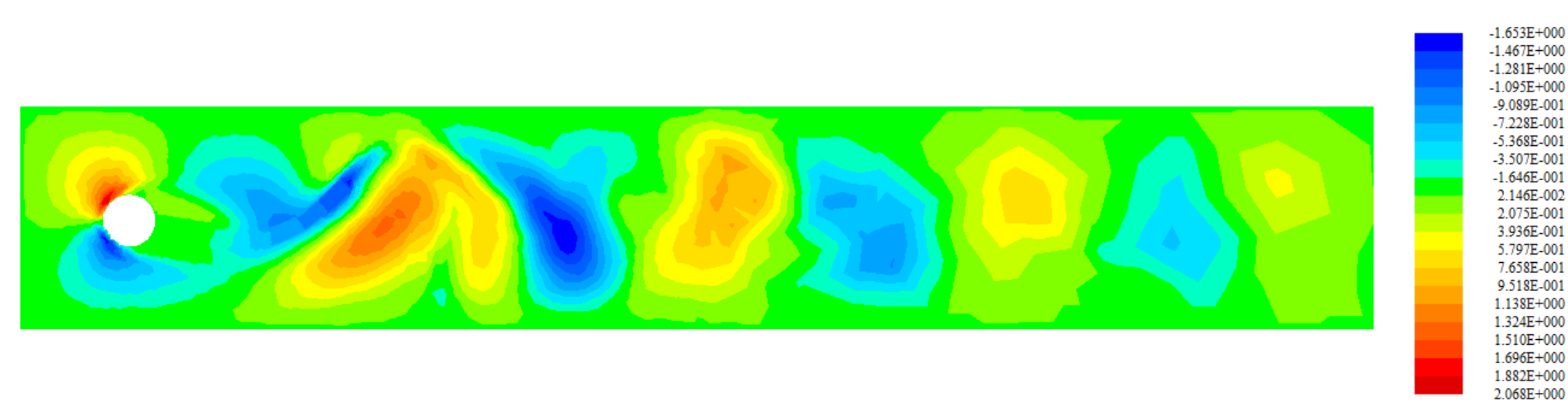}}
\subfigure[Fluid pressure of FOM]{\includegraphics[width=6cm, height
=1.5cm]{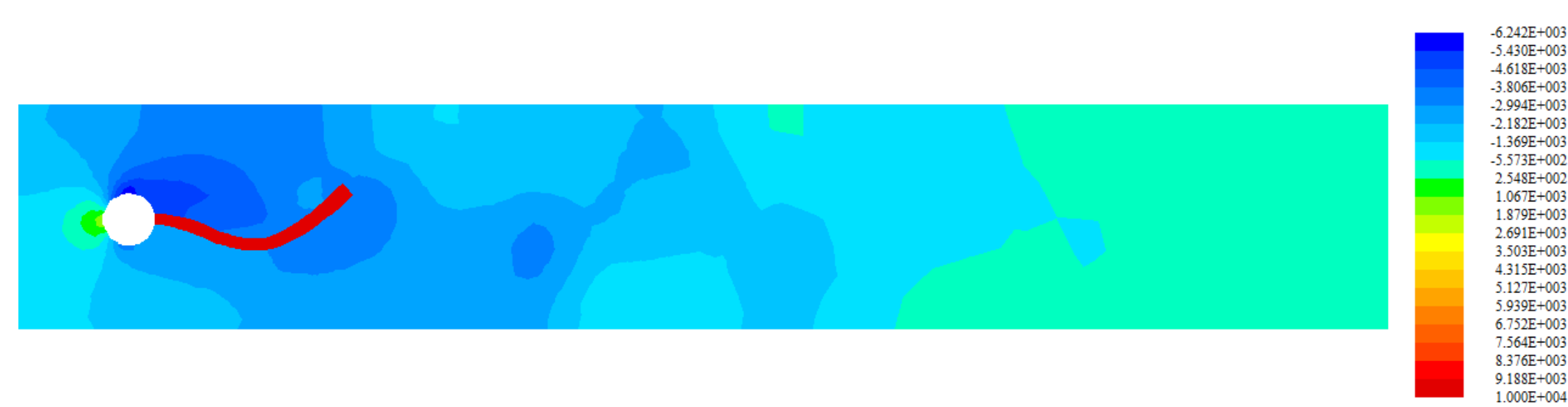}}
    \subfigure[Fluid pressure of ROM]{\includegraphics[width=6cm, height =1.5cm]{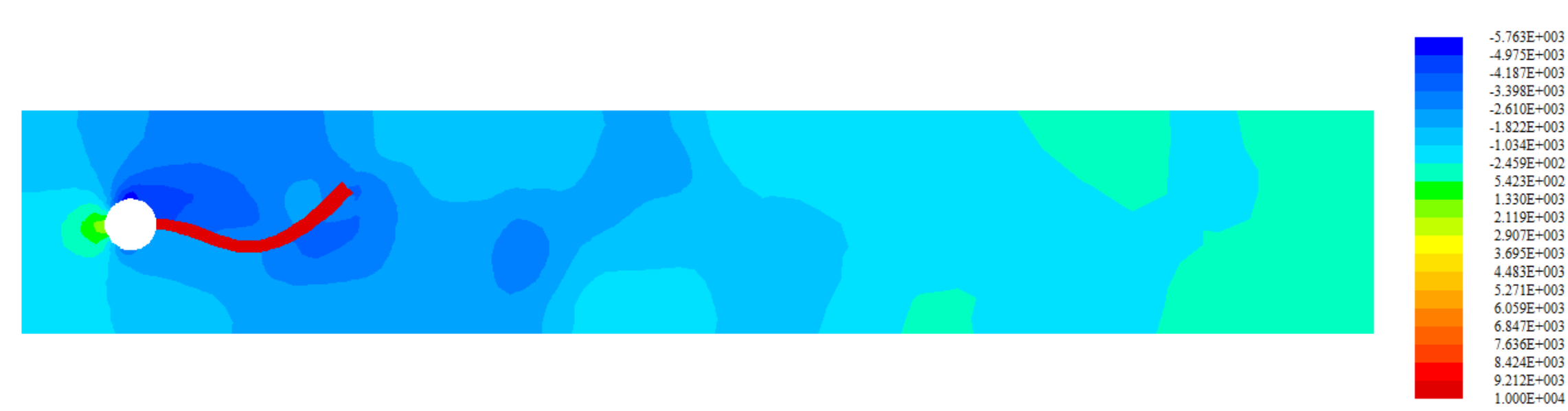}}
    \caption{FSI solutions between FOM and ROM at the finial time $t = 15s$ for Case $\hat{U}=1.48$.}
    \label{Representativesolutionsuin14818350}
\end{figure}

\begin{figure}[H]
\centering
   \subfigure[Total error.]{\includegraphics[width=4.5cm, height =3cm]{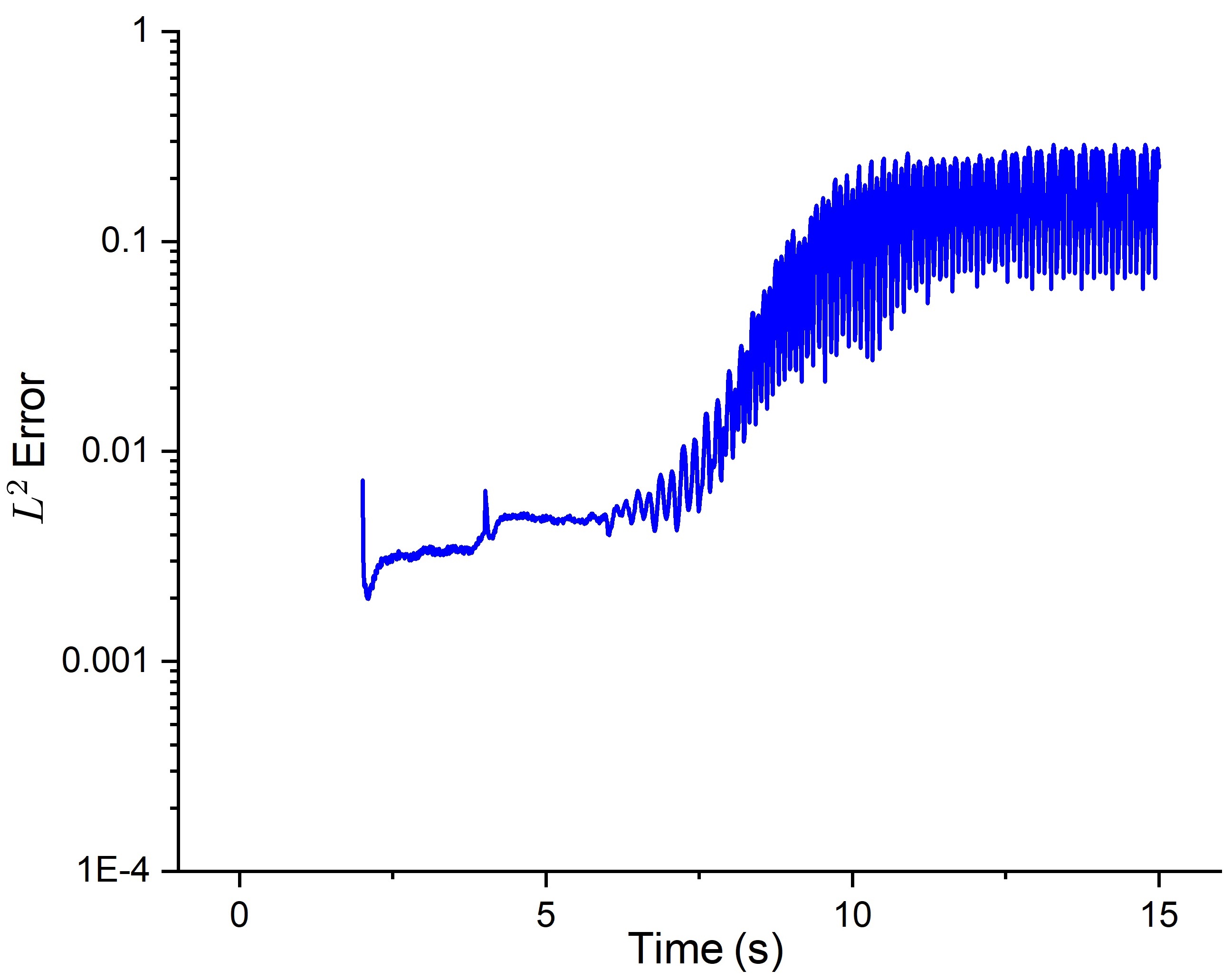}}
   \hspace{1cm}
    \subfigure[Errors of all individual variables combining with total error.]{\includegraphics[width=4.5cm, height =3cm]{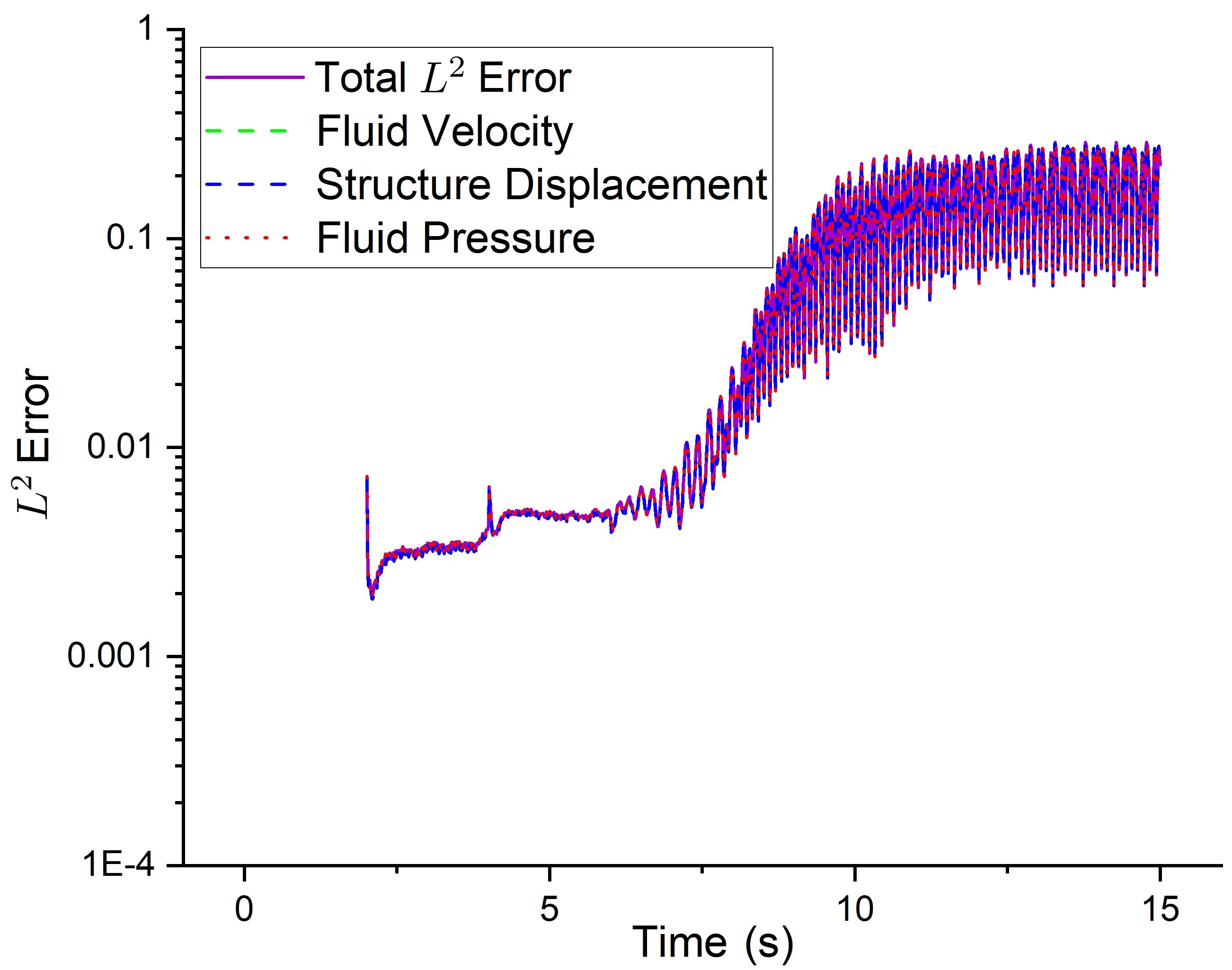}}
    \caption{Relative spatial  $L^2$ errors between ROM and FOM over the entire time interval for Case $\hat{U}=1.48$.}
    \label{errorrom18750uin148}
\end{figure}

\begin{figure}[H]
    \centering
    \includegraphics[width = 7cm, height = 5cm]{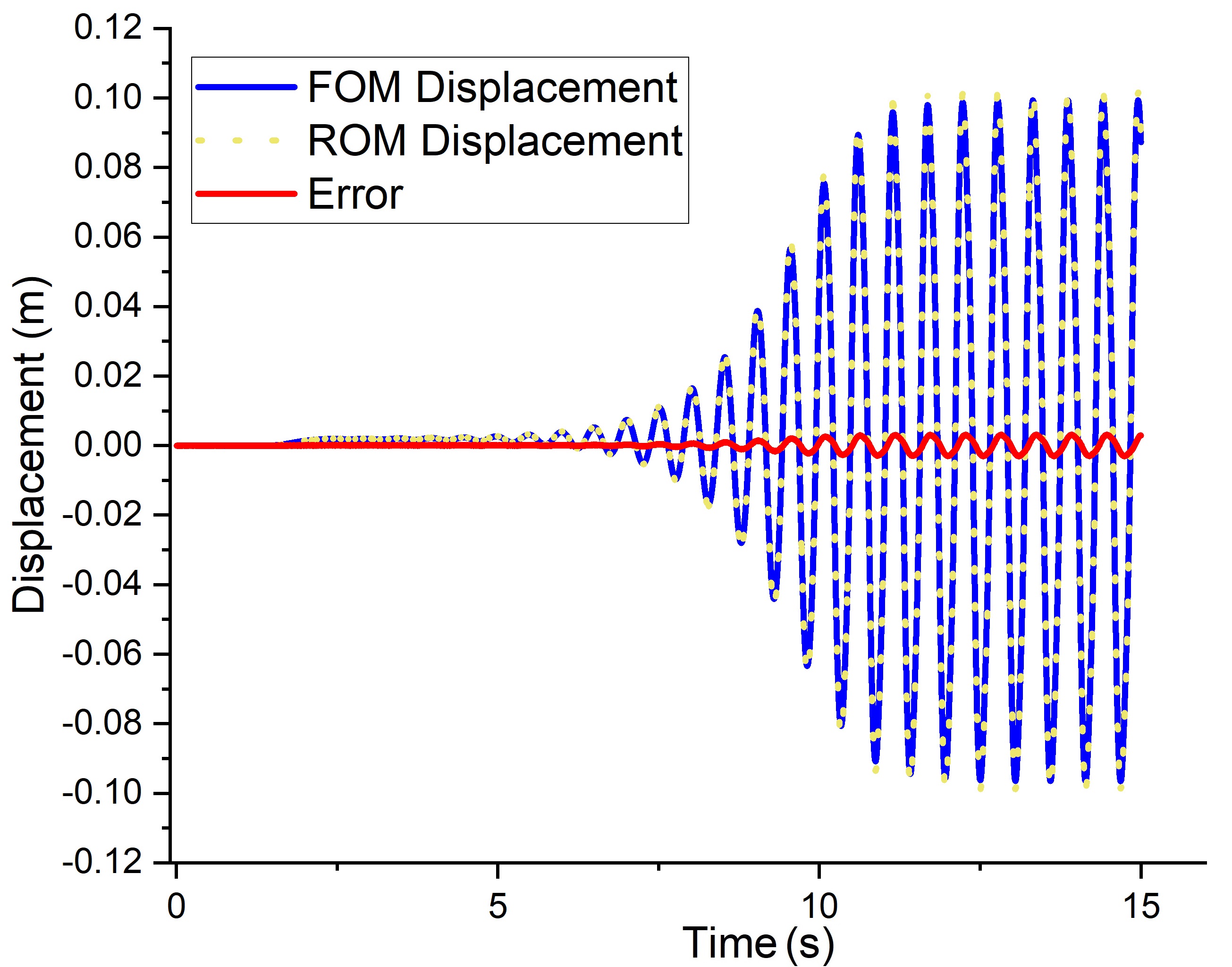}
    \caption{Error of $y$-displacement at point $A$ between FOM and ROM for Case $\mu_{s}=0.52 \times 10^{6}$.}
    \label{ROMdispy18750mu052error}
\end{figure}

\begin{figure}[H]
\centering
   \subfigure[Horizontal velocity of FOM]{\includegraphics[width=6cm, height =1.5cm]{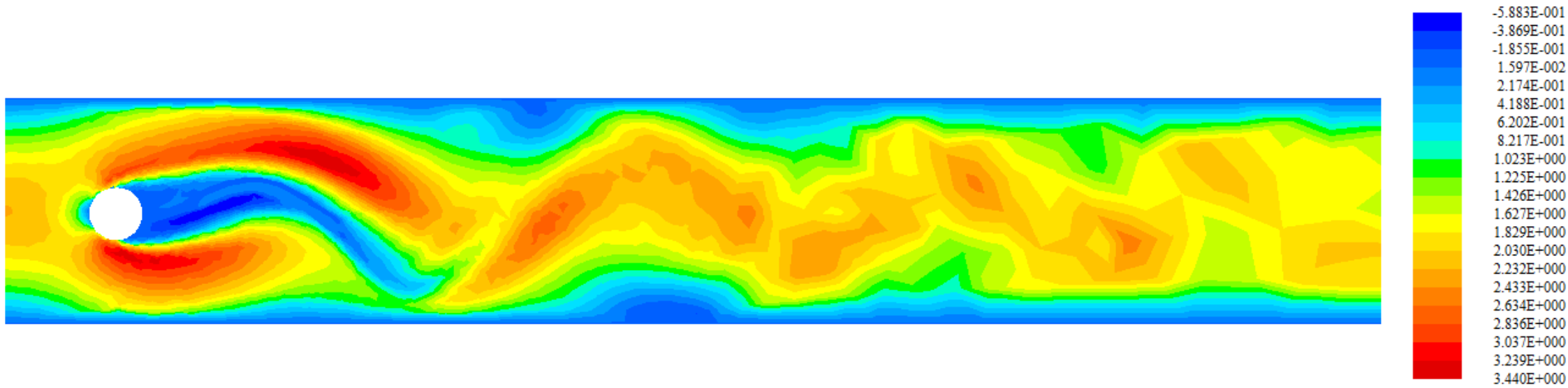}}
    \subfigure[Horizontal velocity of ROM]{\includegraphics[width=6cm, height =1.5cm]{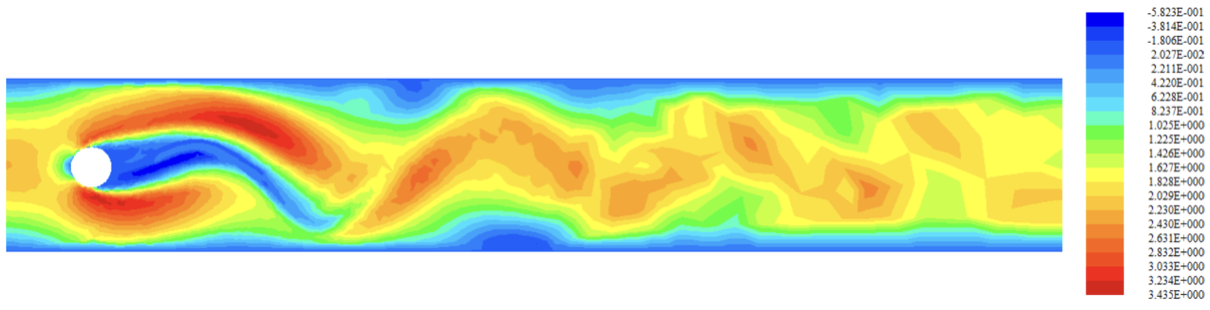}}
\subfigure[Vertical velocity of FOM]{\includegraphics[width=6cm,
height =1.5cm]{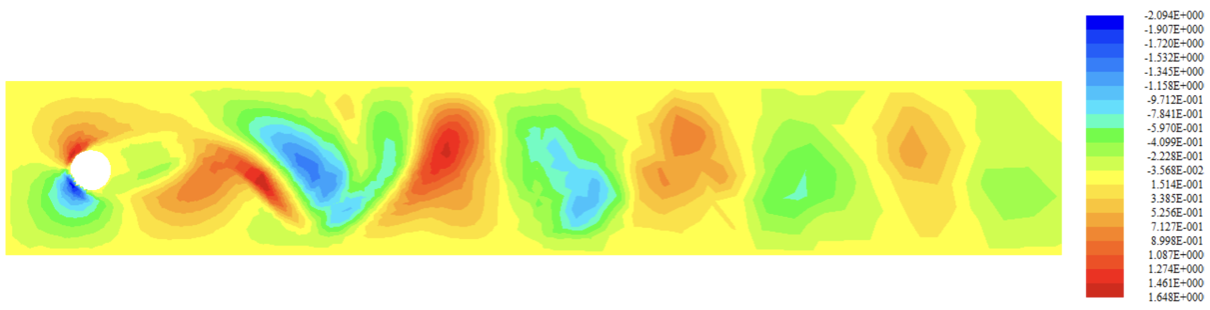}}
    \subfigure[Vertical velocity of ROM]{\includegraphics[width=6cm, height =1.5cm]{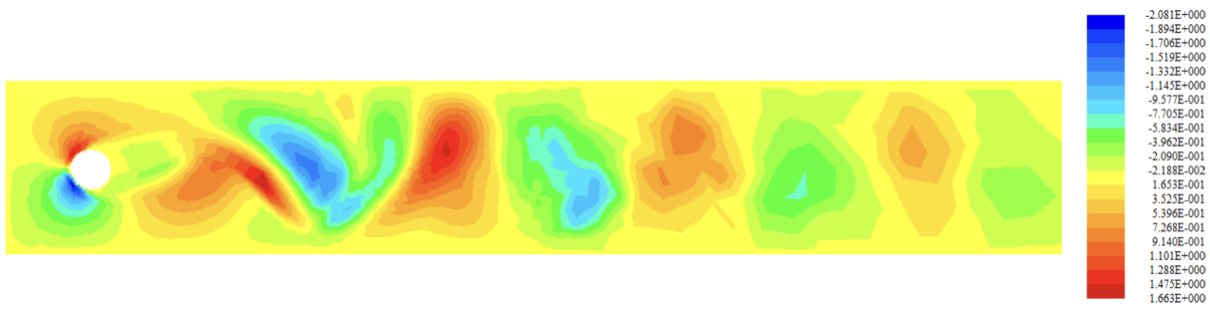}}
\subfigure[Fluid pressure of FOM]{\includegraphics[width=6cm, height
=1.5cm]{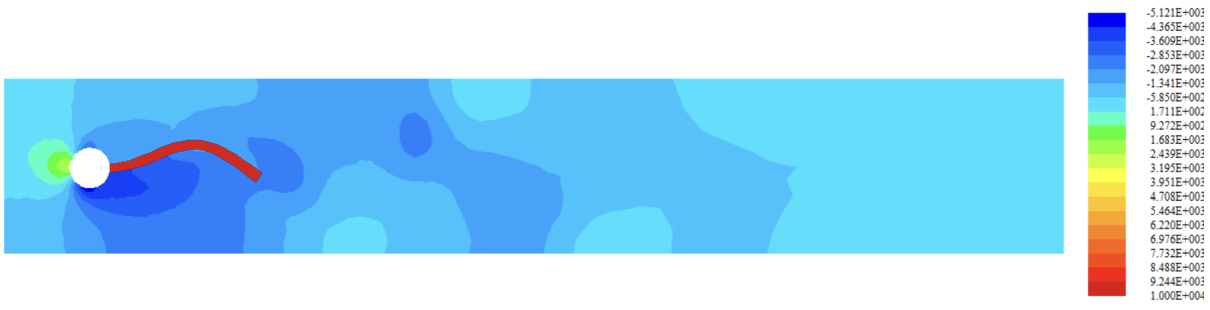}}
    \subfigure[Fluid pressure of ROM]{\includegraphics[width=6cm, height =1.5cm]{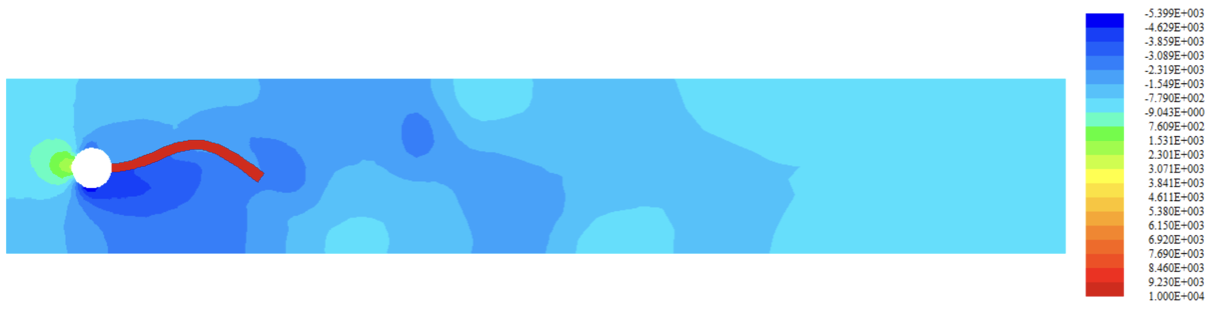}}
    \caption{FSI solutions between FOM and ROM at the finial time $t = 15s$ for Case $\mu_{s}=0.52 \times 10^{6}$.}
    \label{Representativesolutionsmu05218350}
\end{figure}

\begin{figure}[H]
\centering
   \subfigure[Total error.]{\includegraphics[width=4.5cm, height =3cm]{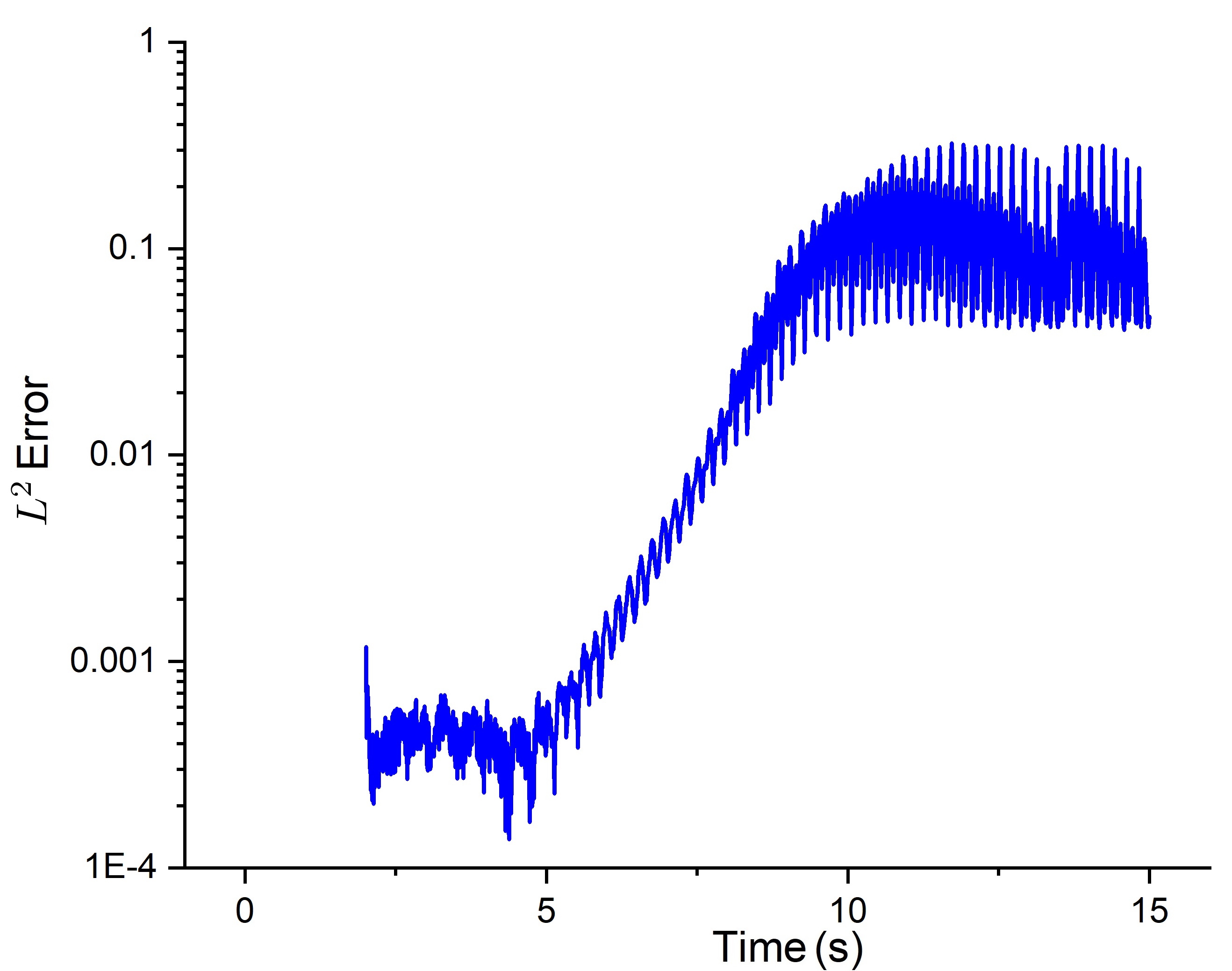}}
   \hspace{1cm}
    \subfigure[Errors of all individual variables combining with total error.]{\includegraphics[width=4.5cm, height =3cm]{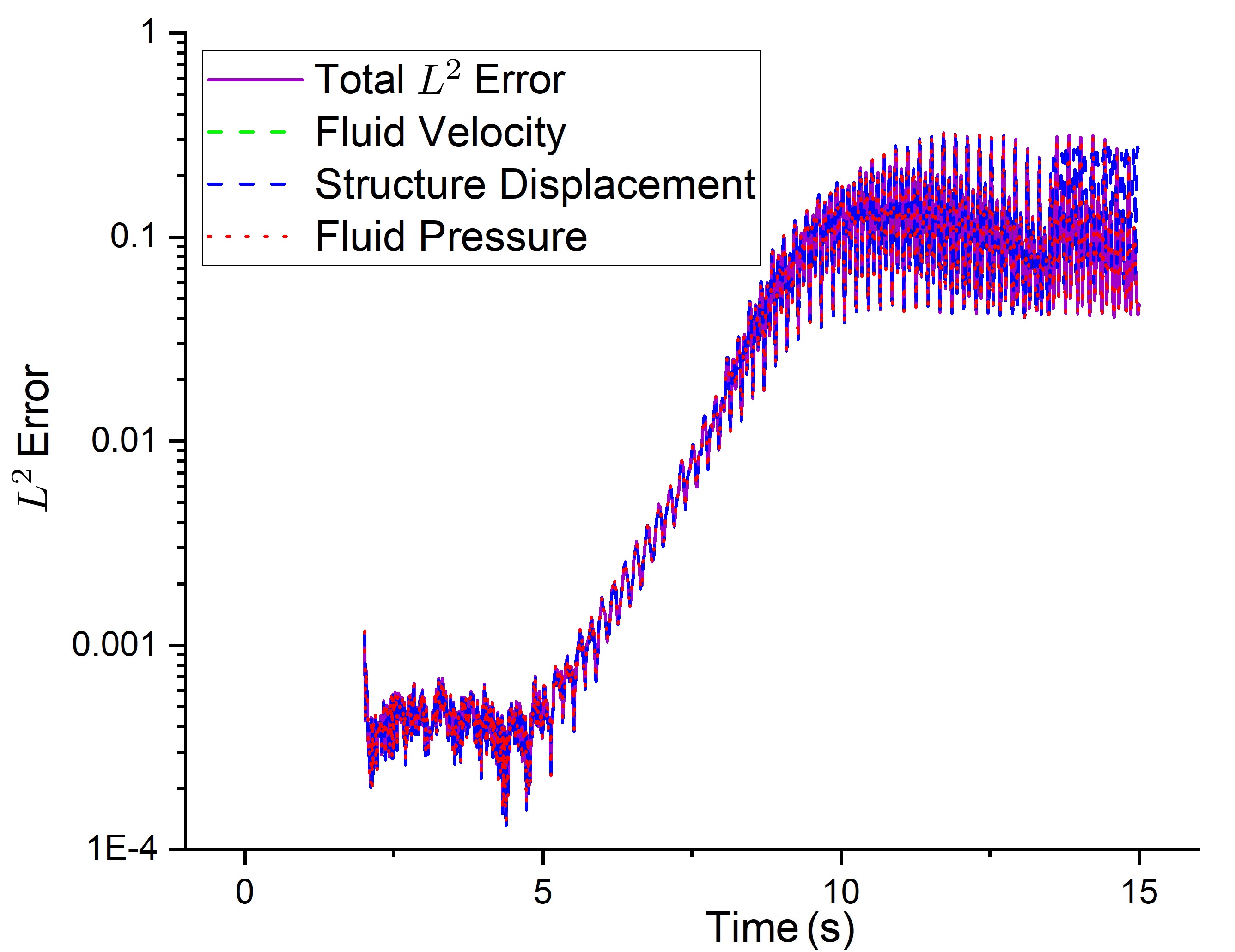}}
    \caption{Relative spatial  $L^2$ errors between ROM and FOM over the entire time interval for Case $\mu_{s}=0.52 \times 10^{6}$.}
    \label{errorrom18750mu52}
\end{figure}

\begin{figure}[H]
    \centering
    \includegraphics[width = 7cm, height = 5cm]{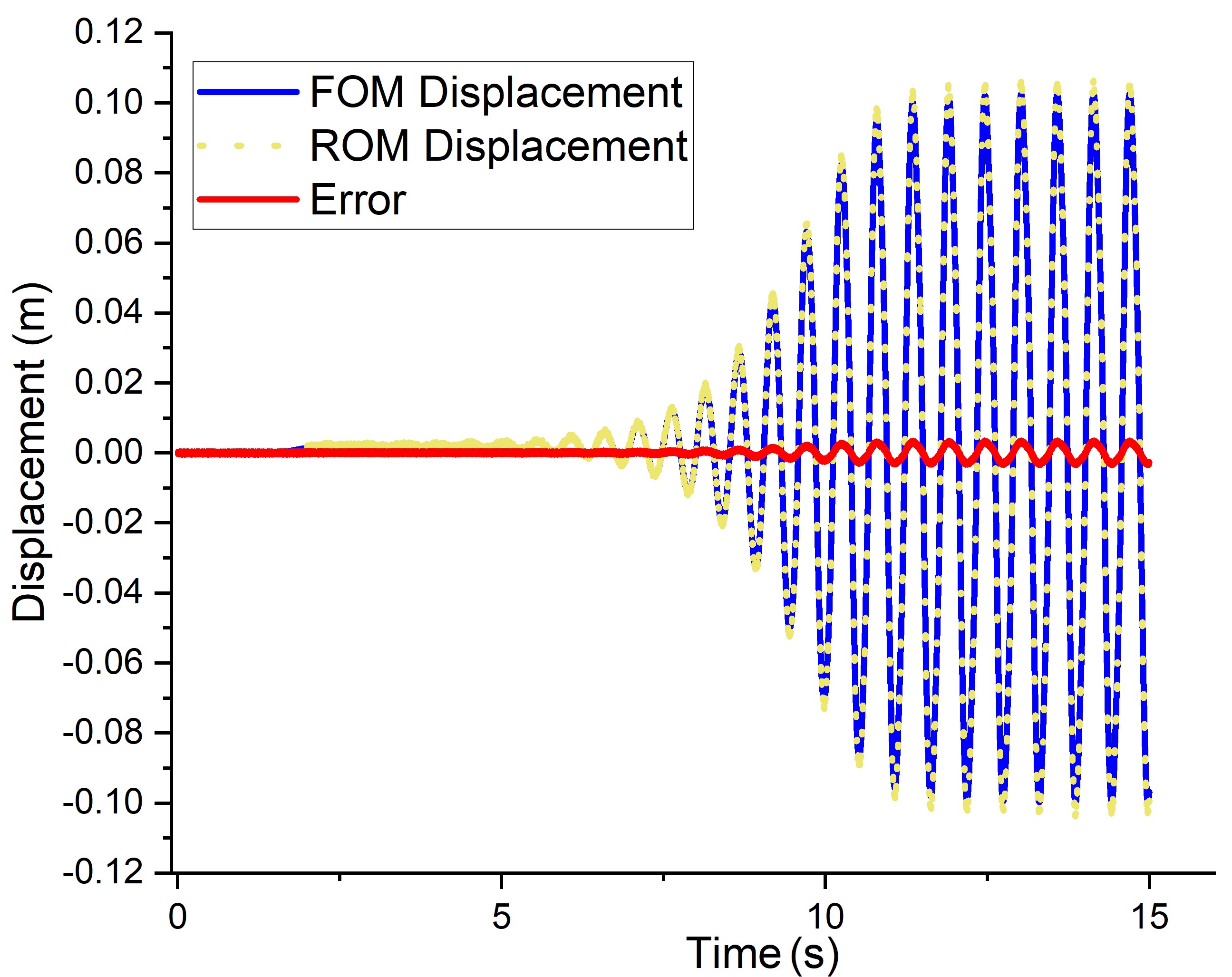}
    \caption{Error of $y$-displacement at point $A$ between FOM and ROM for Case $\mu_{s}=0.48 \times 10^{6}$.}
    \label{ROMdispy18750mu048error}
\end{figure}

\begin{figure}[H]
\centering
   \subfigure[Horizontal velocity of FOM]{\includegraphics[width=6cm, height =1.5cm]{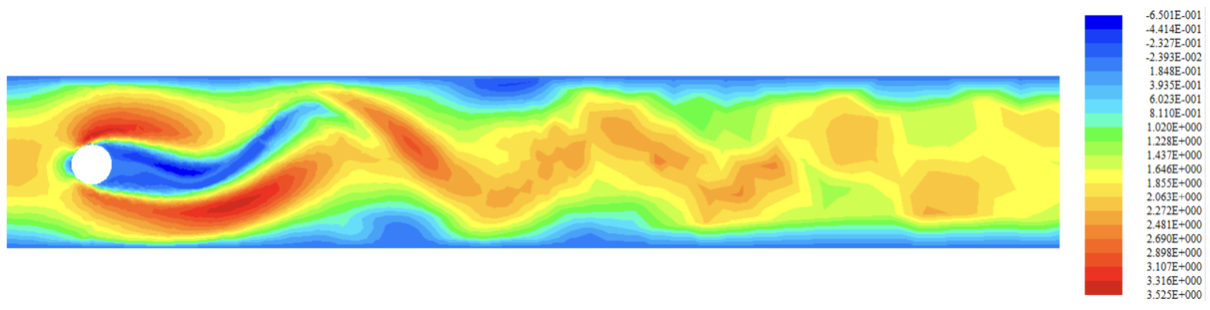}}
    \subfigure[Horizontal velocity of ROM]{\includegraphics[width=6cm, height =1.5cm]{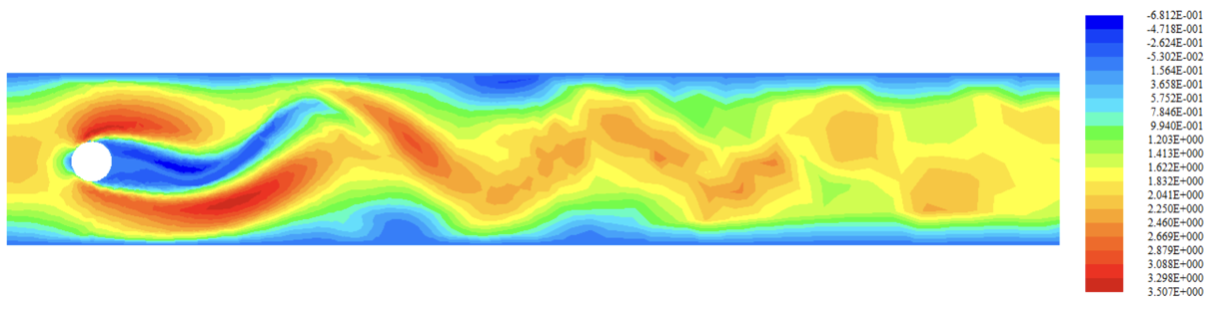}}
\subfigure[Vertical velocity of FOM]{\includegraphics[width=6cm,
height =1.5cm]{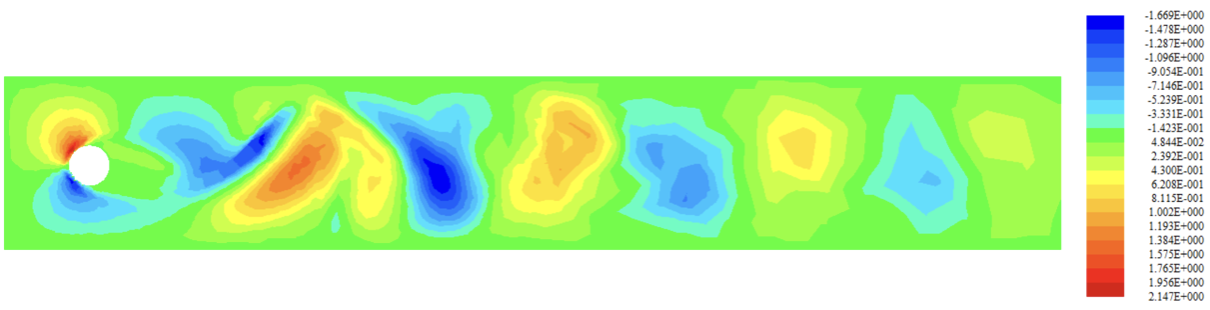}}
    \subfigure[Vertical velocity of ROM]{\includegraphics[width=6cm, height =1.5cm]{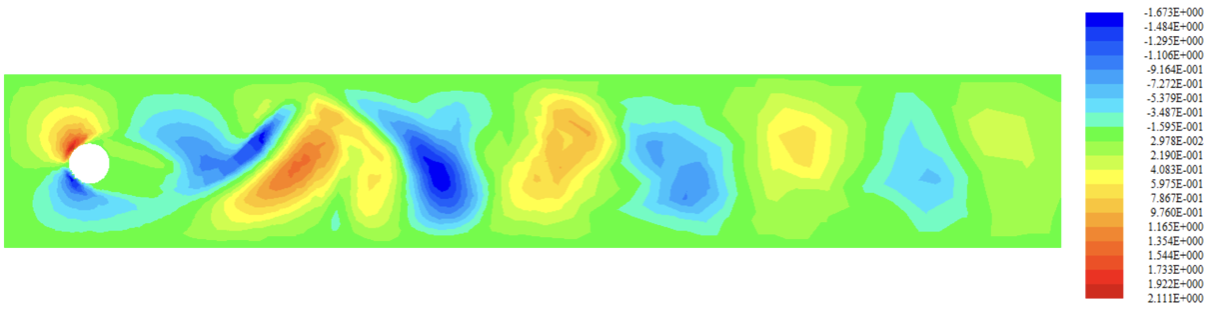}}
\subfigure[Fluid pressure of FOM]{\includegraphics[width=6cm, height
=1.5cm]{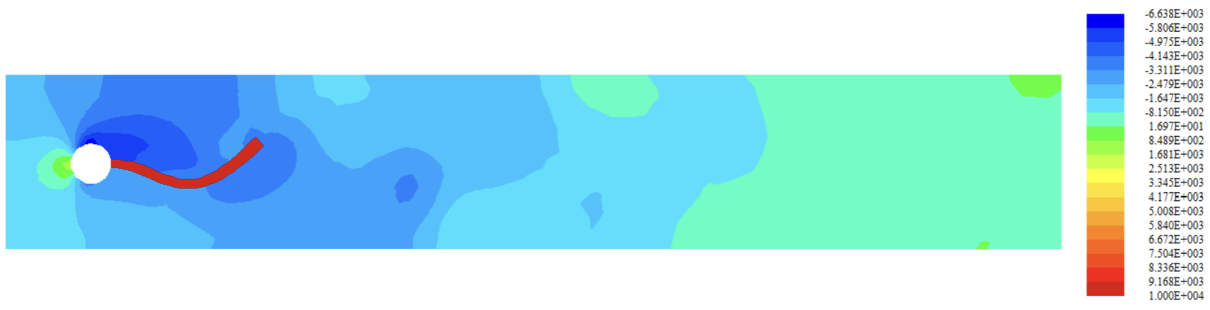}}
    \subfigure[Fluid pressure of ROM]{\includegraphics[width=6cm, height =1.5cm]{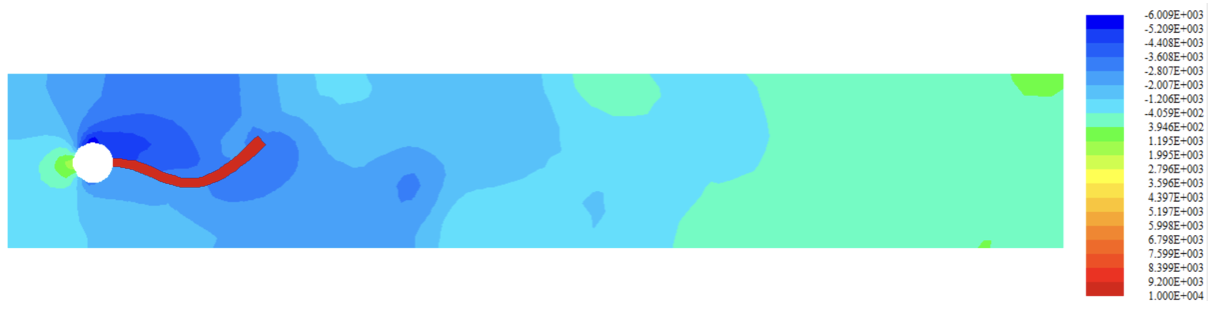}}
    \caption{FSI solutions between FOM and ROM at the finial time $t = 15s$ for Case $\mu_{s}=0.48 \times 10^{6}$.}
    \label{Representativesolutionsmu04818350}
\end{figure}

\begin{figure}[H]
\centering
   \subfigure[Total error.]{\includegraphics[width=4.5cm, height =3cm]{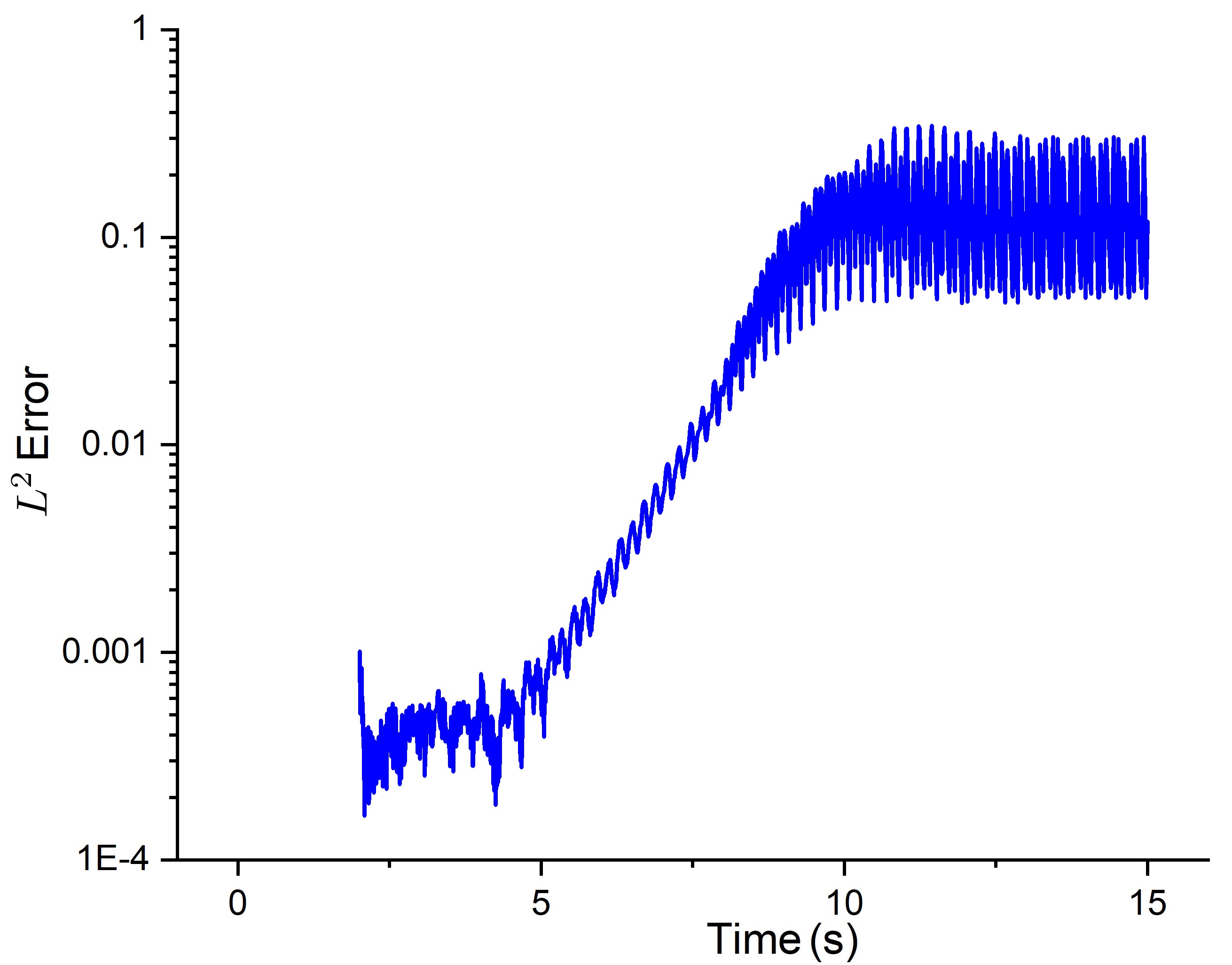}}
   \hspace{1cm}
    \subfigure[Errors of all individual variables combining with total error.]{\includegraphics[width=4.5cm, height =3cm]{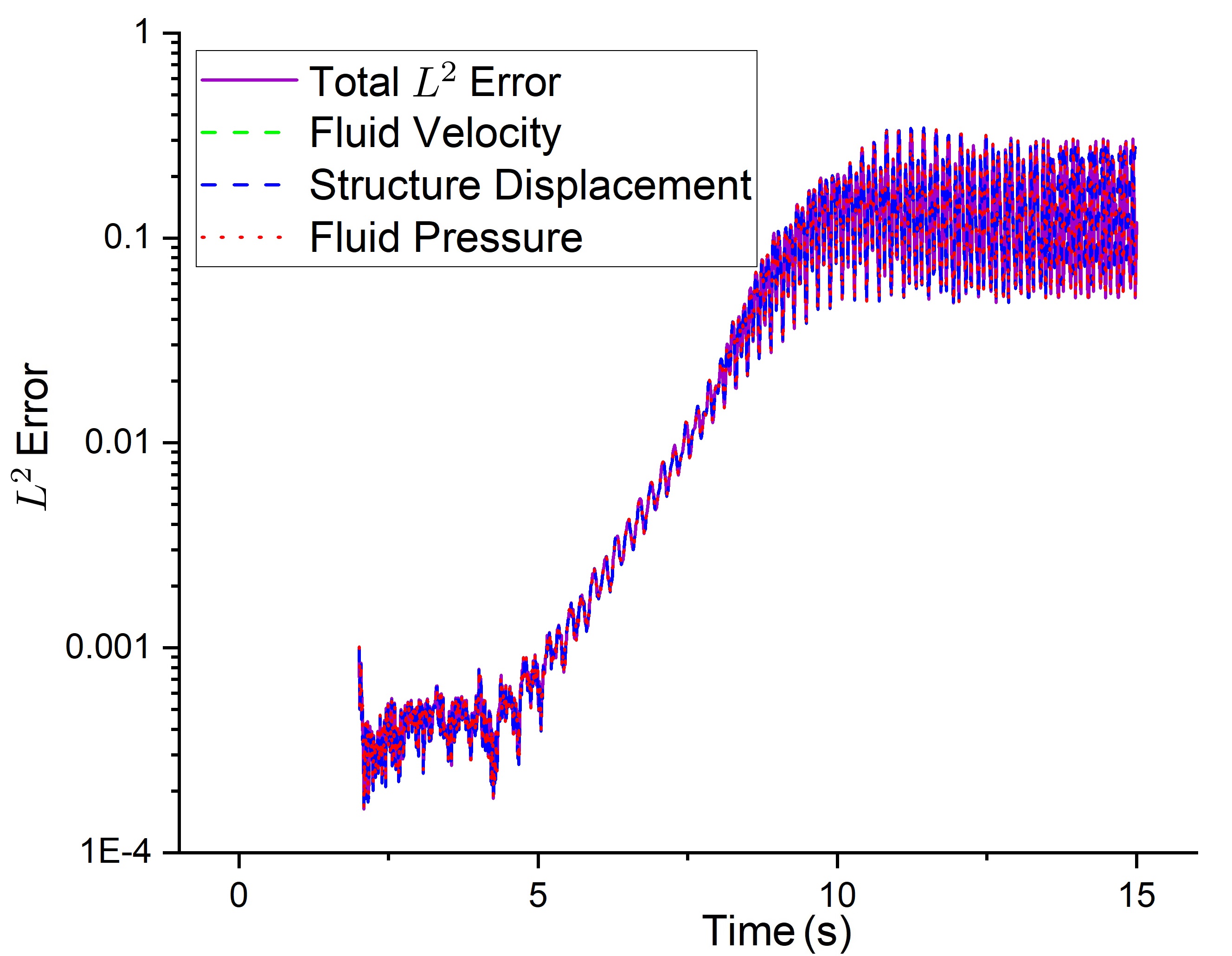}}
    \caption{Relative spatial  $L^2$ errors between ROM and FOM over the entire time interval for Case $\mu_{s}=0.48 \times 10^{6}$.}
    \label{errorrom18750mu048}
\end{figure}

Next, we test whether the constructed POD basis can let the
developed ROM have a good approximation to the FOM when physical
parameters are perturbed in a reasonable range. Without loss of
generality, we choose to change the maximum incoming fluid velocity,
$\hat{U}$, for the fluid part and the Young's modulus, $\mu_{s}$,
for the structure part. While keeping all the other parameters
unchanged, we perturb $\hat{U}=1.5$ by $\pm\, 0.02$ , or
$\mu_{s}=0.5\times 10^{6}$ by $\pm\, 0.02\times 10^{6}$, i.e., we
let $\hat{U}=1.52$ or $1.48$, $\mu_{s}=0.52\times 10^{6}$ or
$0.48\times 10^{6}$, respectively, resulting in four testing cases
in total. Figures
\ref{ROMdispy18750uin152error}-\ref{errorrom18750mu048} show
corresponding results of FOM and ROM, and of their comparison errors
under these four scenarios,
from which we can observe the following numerical phenomena:
\begin{itemize}
\item The POD bases generated at $\hat{U} = 1.5$ , $\mu_{s} = 0.5 \times 10^{6}$ have a certain degree of
generalization ability, and can reproduce relatively high-fidelity
solutions even if physical parameters are slightly changed in all
cases;
\item The error of $y$-displacement at point $A$, $(Dy_{fom} - Dy_{rom})$,
is well controlled within $\pm0.01$ and gradually stabilizes along
with the stabilization of beam vibration for all cases;
\item The contour snapshots of velocity and fluid pressure at T=15s are
similar between the FOM and ROM. The total relative spatial $L^2$
error is well controlled around $0.1$, and gradually stabilizes
along with the stabilization of vibration for all cases.
\item As time marches, errors also grow.
But as the beam vibration stabilizes, the relative spatial $L^{2}$
errors for all variables gradually stabilize for all cases.
\item  The relative spatio-temporal $L^2$ errors for all cases
are displayed below: (1) Case $\hat{U}=1.52$: 0.0632; (2) Case
$\hat{U}=1.48$: 0.0421; (3) Case $\mu_{s}=0.52 \times 10^{6}$:
0.0387; (4) Case $\mu_{s}=0.48 \times 10^{6}$: 0.0394. In summary,
all relative spatio-temporal $L^2$ errors are well controlled around
$0.0459$, on the average.
 \end{itemize}




{\subsection{Motivation of partitioning the spatial and temporal
dimension}\label{sec:notimedivide} As algorithmically discussed in
Section \ref{S3} and numerically illustrated in Section
\ref{sec::comfomrom}, our developed ROM approach involves
partitioning both the spatial and temporal dimensions, which has not
yet been seen in the existing classical ROM for time-dependent
problems including FSI. In this subsection, we illustrate why such a
detailed partition of spatial and temporal dimensions is necessary
by comparing results of the FOM and the classical ROM for the
presented benchmark problem on a coarse mesh and time partition:
4751 DOFs in total and $\Delta t=0.01s$, while keeping all the other
parameters unchanged.


First of all, we investigate the case of no partition for any
spatial and temporal dimensions, which means $G=1$ then $g=0$, only,
thus there is only one time segment: $[T_0,T_1]=[2s,15s]$, leading
to $N_{T_0} = 1301$, as well as $\mathcal{N}_h = 4751$. Therefore,
the correlation matrices (\ref {correlationmatrices}) turns out to
be ${\mathcal{C}}_{\bm{U}_0} := {\bm{U}_0}^{\top}\bm{U}_0$ that
belongs to $\mathbb{R}^{{N_{T_{0}}} \times {N_{T_{0}}}}$. As (\ref
{snapmatrix}) shows, we calculate all eigenvalues and corresponding
eigenvectors, then select sufficiently enough POD bases and perform
the online phase.
\begin{figure}[H]
    \centering
    \subfigure[Vibration curve at point $A$]{\includegraphics[width=5cm, height =4cm]{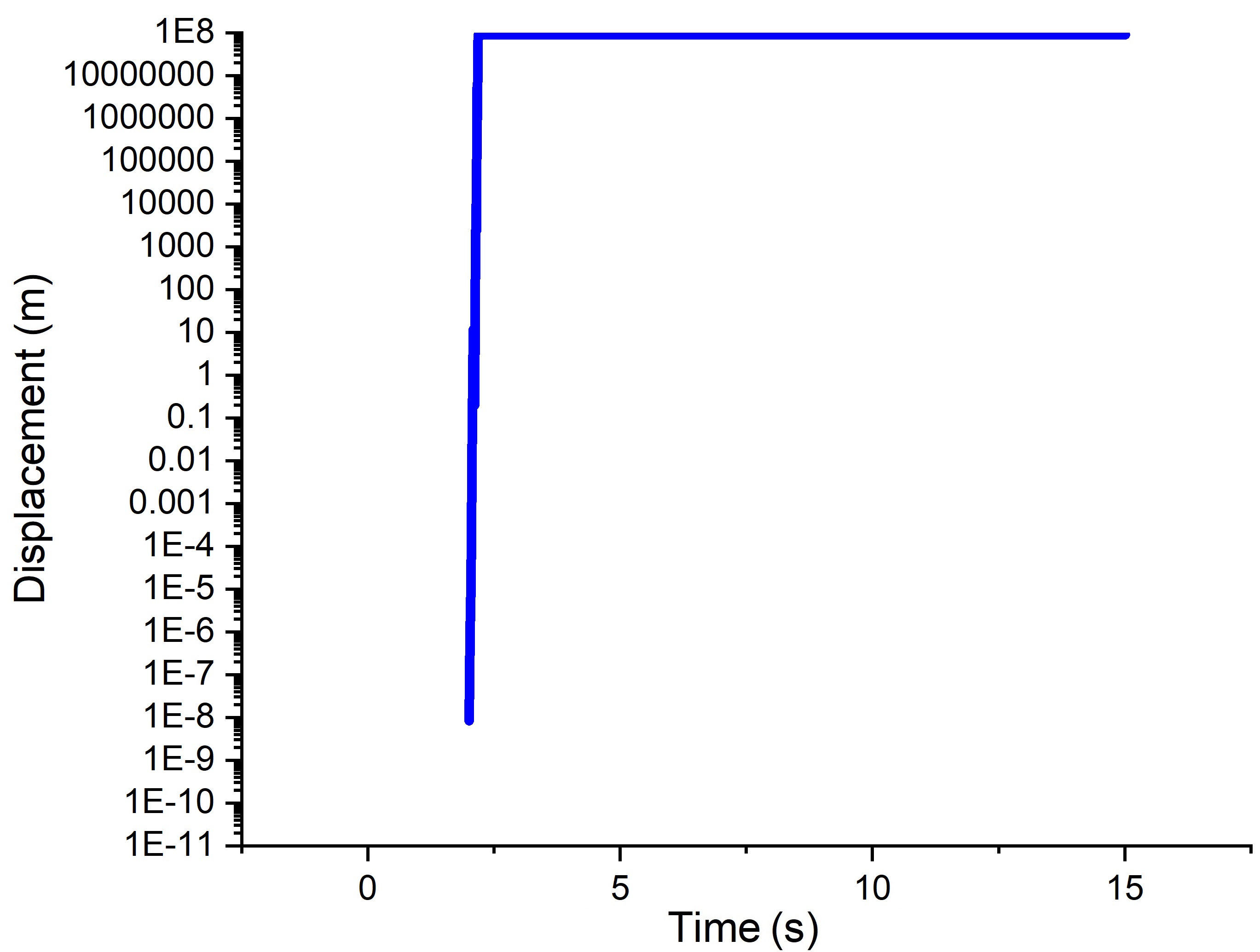}}
   \subfigure[Total relative spatial $L^2$ error]{\includegraphics[width=5cm, height =4cm]{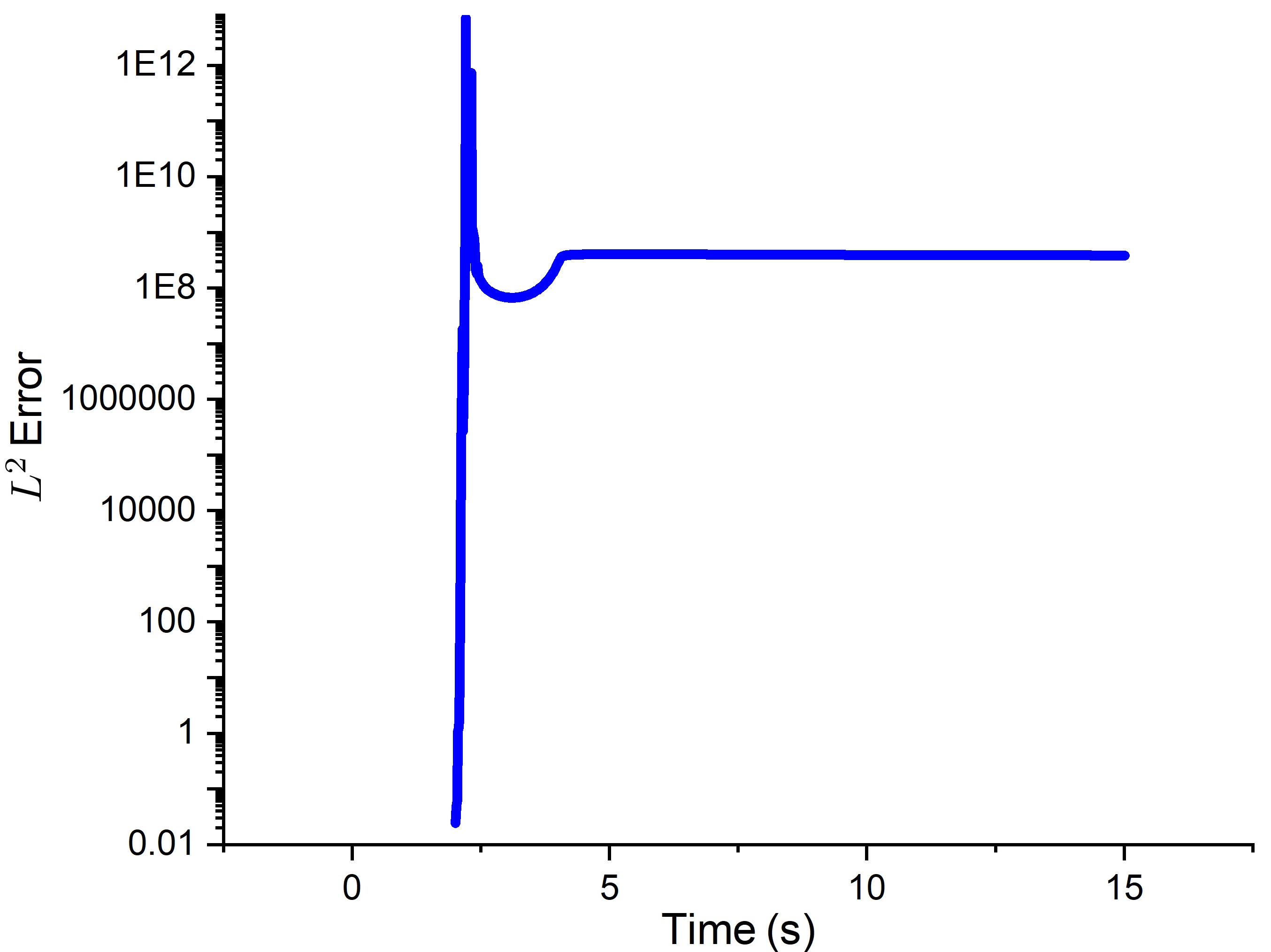}}
    \caption{ROM results without partition for spatial and temporal dimensions.}
    \label{wholeprocess}
\end{figure}
The obtained ROM results are shown in Figure \ref{wholeprocess},
where we can see that right after the ROM computation starts, the
total relative spatial $L^2$ error increases rapidly as high as
$10^{12}$, correspondingly, the vibration amplitude of point $A$
also increases rapidly to the value around $10^{8}$, which far
deviates from the FOM result, showing that such designed ROM
completely fails in this FSI problem.

\begin{figure}[H]
    \centering
    \subfigure[Vibration curve at point $A$]{\includegraphics[width=5cm, height =4cm]{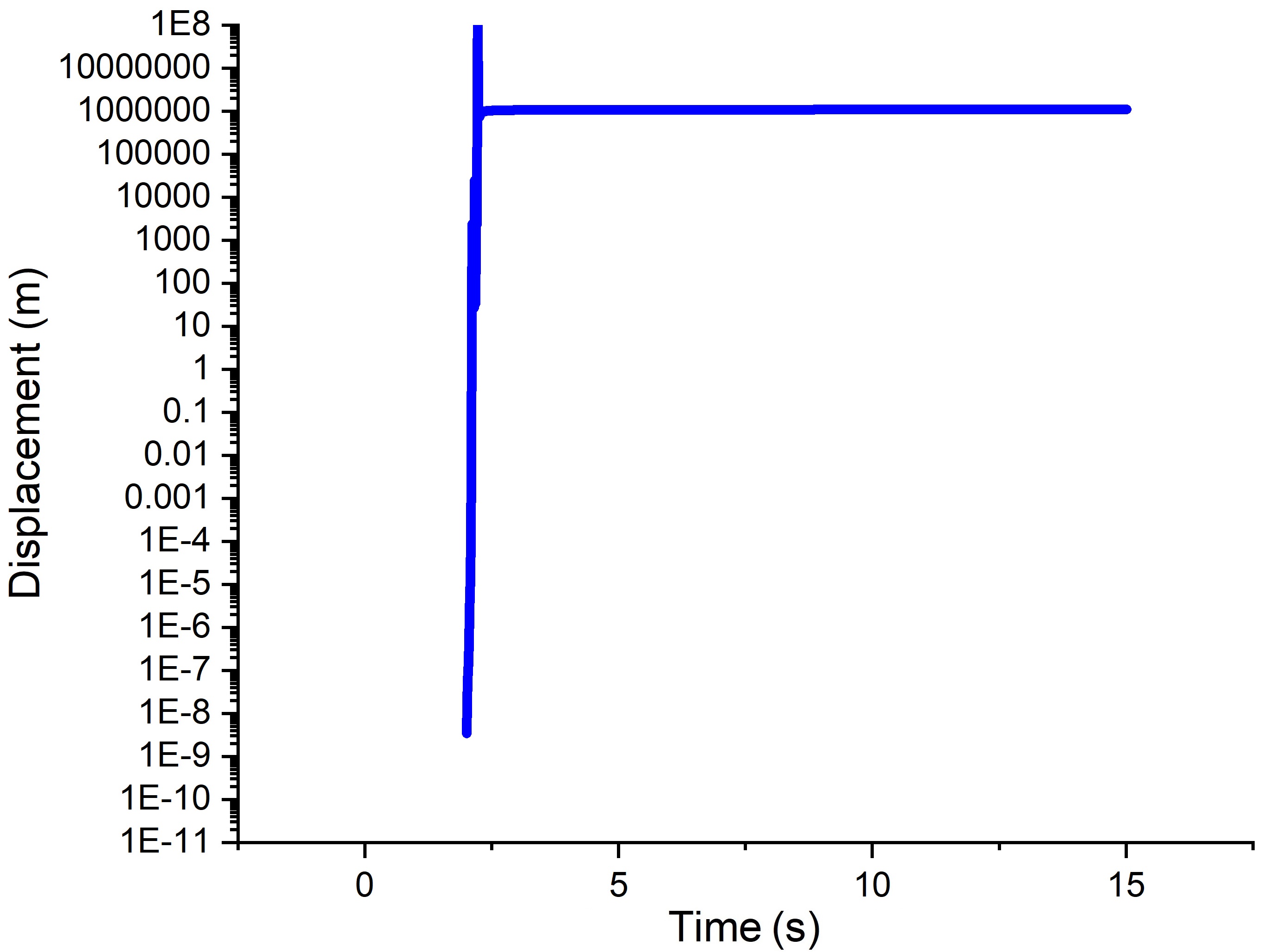}}
   \subfigure[Total relative spatial $L^2$ error]{\includegraphics[width=5cm, height =4cm]{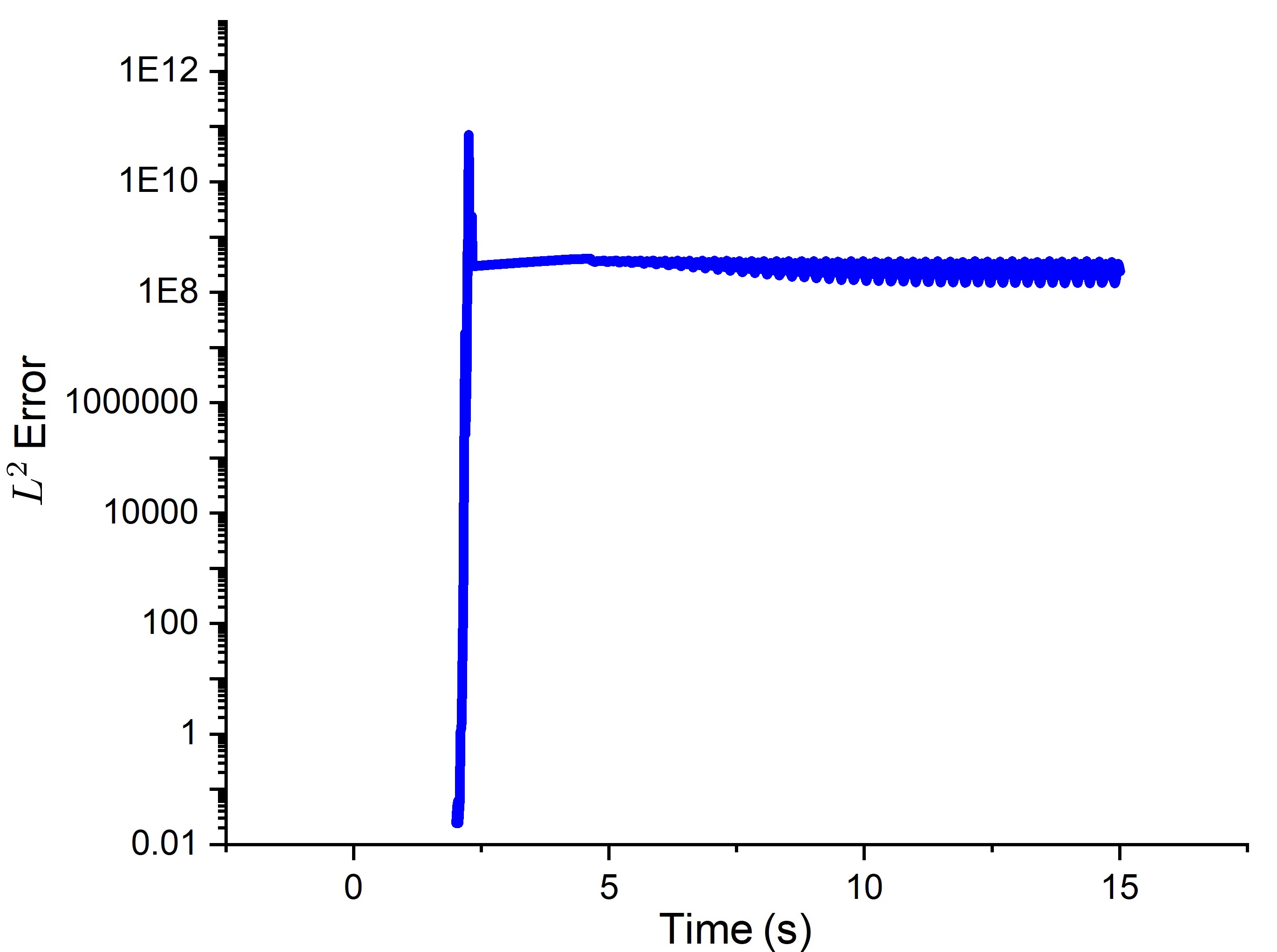}}
    \caption{ROM results without a partition of temporal dimension.}
\label{dividesthetime}
\end{figure}
In the second case, we only partition spatial dimensions in the same
way as described in Section \ref{sec:offline} but no partition for
the temporal dimension, i.e., we still keep $G=1$ and $N_{T_0} =
1301$. Then we select sufficiently enough POD bases in the online
phase, and perform the ROM computation whose results are shown in
Figure \ref{dividesthetime}. We can observe that although the total
relative spatial $L^2$ error decreases one magnitude in comparison
with Figure \ref{wholeprocess}, it is still as large as $10^{11}$.
In addition, the vibration curve of point $A$ also deviates quickly
from the FOM result, jumps to $10^{8}$ or so in a short time,
resulting in a failure again.

However, if we partition both spatial and temporal dimensions, then
just like what we do in Section \ref{sec::comfomrom}, we can obtain
a good match for the vibration curve of point $A$ between the ROM
and FOM on this coarse mesh and time partition, where the ROM result
holds a comparable accuracy to the FOM in terms of both spatial- and
spatio-temporal $L^2$ errors, relatively. Therefore, we conclude
that partitioning both spatial and temporal dimensions is greatly
important and crucial for FSI problems when the structure
significantly deforms with time.

\section{Conclusion}\label{S5}
In this work, we develop a novel reduced order model (ROM) approach
to solve fluid-structure interaction (FSI) problems in an efficient
fashion while the accuracy is still within a reasonable range in
comparison with the solution of full-order model (FOM) approach. The
key innovation is to treat time as a non-reduced variable while
dividing the time interval into some time segments, and within each
time segment we utilize the classical proper orthogonal
decomposition (POD) method to achieve the order reduction in the
offline phase. By selectively combining the time-segmented POD
models, we apply the proposed ROM approach to a FSI benchmark
problem and solve the issue of increasing error on the beam tail's
vibration amplitude over long-term simulations in the online phase.
This hybrid strategy maintains the approximation accuracy for
predicting the beam tail's vibration under the circumstance of what
the FOM requires, while speeding up the linear algebraic solver 8800
times more and saving the computational time for almost $100\%$
corresponding to the FOM, simultaneously, retaining the
generalizability of offline model when physical parameters are
perturbed for the entire system in the online application. Our
approach demonstrates both the computational efficiency and the
robustness for parameters' perturbation, and overcomes limitations
of standard POD reductions for FSI problems with long transient
responses while the structure significantly deforms with time.

\section{Acknowledgement}
Q. Zhai and X. Xie were partially supported by the National Natural
Science Foundation of China (12171340),
S. Zhang was supported by Natural Science Foundation of Sichuan
Province, China (2023NSFSC0075), and P. Sun was supported in part by
a grant from the Simons Foundation (MPS-706640).

\bibliographystyle{siam}
\bibliography{ref,FSI,publicationlist4Sun}
\end{document}